\documentclass[12pt]{article}
\usepackage{amsmath,amssymb,graphicx}
\batchmode

\newtheorem{lem}{Lemma}
\newtheorem{thm}{Theorem}

\newcommand{\pr}{\noindent{\bf Proof}. }
\newcommand{\re}{\noindent{\bf Remark}. }
\newcommand{\res}{\noindent{\bf Remarks}. }

\newcommand{\pa}{\partial}
\newcommand{\one}{\cO(1)}
\newcommand{\bpsi}{\bar \psi}
\newcommand{\bPsi}{\bar \Psi}

\newcommand{\bx}{\bar x}
\newcommand{\sq}{\square}
\newcommand{\st}{\stackrel{}}

\newcommand{\al}{\alpha}
\newcommand{\De}{\Delta}
\newcommand{\de}{\delta}
\newcommand{\ga}{\gamma}
\newcommand{\Ga}{\Gamma}
\newcommand{\ka}{\kappa}
\newcommand{\La}{\Lambda}
\newcommand{\la}{\lambda}
\newcommand{\Om}{\Omega}
\newcommand{\om}{\omega}

\newcommand{\ep}{\epsilon}
\newcommand{\si}{\sigma}

\newcommand{\Up}{\Upsilon}

\newcommand{\cA}{{\cal A}} 

\newcommand{\cC}{{\cal C}}

\newcommand{\cV}{{\cal V}}
\newcommand{\cO}{{\cal O}}

\newcommand{\cH}{{\cal H}}
\newcommand{\cS}{{\cal S}}
\newcommand{\cE}{{\cal E}}

\newcommand{\cT}{{\cal T}}

\newcommand{\cK}{{\cal K}}
\newcommand{\cG}{{\cal G}}

\newcommand{\cN}{{\cal N}}

\newcommand{\cQ}{{\cal Q}}
\newcommand{\cL}{{\cal L}}
\newcommand{\cP}{{\cal P}}
\newcommand{\cU}{{\cal U}}
\newcommand{\cJ}{{\cal J}}
\newcommand{\cZ}{{\cal Z}}

\newcommand{\bbR}{{\mathbb{R}}}
\newcommand{\bbZ}{{\mathbb{Z}}}

\newcommand{\bbT}{{\mathbb{T}}}

\begin{document}
  
\title{Quantum Electrodynamics on the 3-torus \\ \small{I.- First step}}
\author{ 
J. Dimock\thanks{Research supported by NSF Grant  PHY0070905}
\thanks{dimock@acsu.buffalo.edu}\\
Dept. of Mathematics \\
SUNY at Buffalo \\
Buffalo, NY 14260 }
\maketitle

\begin{abstract}
We study the ultraviolet problem for quantum electrodynamics  on a
three dimensional torus.    
We start with the lattice gauge theory on 
a toroidal lattice and seek to  control the singularities as the lattice spacing is taken 
to zero.  This is done by following the flow of a sequence of  renormalization 
group transformations. The analysis is facilitated by splitting the space of gauge fields into
into  large fields  and small fields at each step following Balaban. In this  paper we
explore the first step in detail.
\end{abstract} 

\newpage

\tableofcontents

\newpage

\section{Introduction}

Quantum electrodynamics in a four dimensional space time, (QED)$_4$, is the basic theory of
electrons  and photons.  
The theory is very singular at short distances.   Nevertheless
it does have a well defined perturbation theory
(expansion in the coupling constant) provided various renormalizations are carried 
out.   At this level it gives a  spectacularly precise description of nature.

One would like to have a rigorous non-perturbative construction
of the model.  To date  this has not been possible and success is not likely 
anytime soon.   However  the singularities are weaker in lower
dimensions and  there is the  possibility of progress.
In this sequence of  papers we give a construction of  (QED)$_3$.  To avoid 
long distance (infrared) problems, which are also present, 
we work on the three dimensional torus.

The theory is formally defined by functional integrals.
The program is to add cutoffs to take out the  
  short  distance  (ultraviolet)  singularities and make the
theory well-defined. Then one tries to remove the cutoffs.  The singularities
that develop are to be cancelled by adjusting
the bare parameters - this  is renormalization.
At the same time one must keep control throughout over 
the size of the functional integrals - this is the stability 
problem.

An
effective  way to deal with these difficulties
is to  regularize the theory by
putting it on a  lattice  and then study the continuum
limit.  \footnote{An alternative would be to work with a continuum
theory and put in explicit momentum cutoffs. For a discussion of this
approach to gauge field theories see Magnen, Rivasseau, and Seneor \cite{MRS93}}
This regularization  has the advantage of preserving gauge invariance.  This  
helps with the treatment of the singularities-
the counterterms required are minimal.
The stability problem  also seems more tractable in the lattice
approximation.

Our general  method  is the renormalization
group   (RG) technique. We first scale up to a unit lattice theory.
By block averaging  we  generate a  sequence of effective actions  corresponding to different length 
scales.   The effective coupling constant starts from near zero and grows  but remains small. 
Renormalization cancellations are made perturbatively at each level.
 The remainders must also be controlled
and for this we need control over the size of the gauge field.  This  is accomplished by breaking the
functional  integral   into large and small field regions at each level.  The contribution of the 
large field region is suppressed by the free measure, and in the small field
region we can do the perturbative analysis.  Stability emerges naturally
in this approach as well.

The scheme outlined above was originally developed by Balaban 
for a study of scalar (QED)$_3$  \cite{Bal82a}, \cite{Bal82b}, \cite{Bal83a},
 \cite{Bal83b}. See also  King  \cite{Kin86}.  The method was further developed 
 by    Balaban,  Brydges,  Imbrie, and
Jaffe  \cite{BBIJ84}, \cite{BIJ88} who used it in their
  study of the Higgs mechanism for this model.  
There is further exposition of this work in lecture notes by Imbrie  \cite{Imb86}.
 Balaban continued by proving  basic stability bounds
 for pure
Yang-Mills YM$_3$, YM$_4$   \cite{Bal87},\cite{Bal88}. 
 Balaban, O'Carroll, and Schor  \cite{BOS89}, \cite{BOS91} gave an analysis of 
the fermion propagators   that arise in a RG approach.
These   papers were the main  inspiration for the present work.

Let us also mention some  earlier related  results.
Weingarten and   Challifour \cite{WeCh79} and Weingarten  \cite{We80} gave a construction of (QED)$_2$.
For 
scalar electrodynamics there is extensive work for $d=2$ by Brydges, Fr\"{o}hlich,
and Seiler \cite{BFS79}.  Magnen and Seneor gave a treatment of 
Yukawa in $d=3$ \cite{MaSe80}.   For a heuristic discussion of (QED)$_3$ 
see   \cite{DJT82}.

The present work will not  directly generalize to   (QED)$_4$.  The problem 
is that this model lacks ultraviolet asymptotic freedom, the flow 
of the effective coupling constants away zero.  
However the present work is  progress toward a construction of quantum 
chromodynamics  (QCD)$_{3,4}$, the fundamental theory of the strong interactions.  
It seems likely that combining  Balaban's results on  Yang-Mills \cite{Bal87},\cite{Bal88}
with  the present work would be sufficient, albeit
still on a torus.
\bigskip

Acknowledgement:   I would like to thank T. Balaban for helpful conversations.

\section{Preliminaries}

\subsection{the model}  The basic torus is   $\bbT_{M}  =   (\bbR  / L^{M} \bbZ )^3$
where  $L$ is a fixed large positive number and $M \geq 0$ is a fixed nonnegative
integer.  It has volume  $L^{3M}$.  In this torus we consider  lattices  with spacing  $L^{-N}$
defined by  
 \begin{equation}
\bbT_{M}^{-N}  =   ( L^{-N}\bbZ  / L^{M} \bbZ )^3  
\end{equation}
Let   $\bpsi,\psi$ be fermi fields and  let  $\cA$ be an Abelian
gauge field on the lattice.
The lattice version of  Euclidean (QED)$_3$ in the Feynman gauge is defined by the density 
\begin{equation}  \label{def1}
\begin{split}
 \rho(\bpsi, \psi, \cA)  
=\exp \left( - \frac12  (\cA, (- \De + \mu^2) \cA) 
- ( \bpsi, ( D_e(\cA)+ m ) \psi)  - \de v - \de E  \right)
\end{split}
\end{equation}
Here    $D_e(A)$ is the lattice Dirac operator with potential $\cA$ and coupling constant  
$e \geq 0$,  and $\de v, \de E$ are mass and energy counterterms.
We are interested in integrals like  the normalization factor (partition function)
\begin{equation}
Z=\int  \rho( \bpsi,\psi, \cA)   \ d \bpsi  \ d \psi\ d\cA
\end{equation}
 and in correlation functions such as  the two point function 
\begin{equation}
<  \psi(x)  \bpsi(y)>  =  Z^{-1} \int \psi(x)   \bpsi(y) \ \rho(\bpsi, \psi, \cA) \ d \bpsi  \ d \psi \
d\cA
\end{equation}
The problem is to control the  $N \to \infty$ limit.

Let us explain the terms  in (\ref{def1}) is more detail.
The gauge field $\cA = \cA(xx')$ is a real valued function on bonds 
 $x,x'$ (nearest neighbors)
 in $\bbT_{M}^{-N}$  satisfying $\cA(xx') = -\cA(x'x)$.
If  $\{e_{\mu}\} =  \{e_1,e_2,e_3\}$ are oriented unit basis vectors we 
write   $\cA_{\mu}(x)   = \cA( x, x + L^{-N} e_{\mu})$.  The derivative of  $\cA_{\mu}$
in the direction $e_{\nu}$ is 
\begin{equation}  \label{lattice1}
(\pa_{\nu} \cA_{\mu})(x)   = ( \cA_{\mu}(x +L^{-N} e_{\nu}) - \cA_{\mu}(x))/ L^{-N}
\end{equation}
and $ (\cA, (- \De + \mu^2) \cA) $ is the quadratic form defined
by the lattice Laplacian:
\begin{equation}  \label{Laplacian}
 (\cA, (- \De +\mu^2 ) \cA)  =   \sum_{\mu, \nu,x } L^{-3N} |(\pa_{\nu} \cA_{\mu})(x)|^2
  +  \mu^2  \sum_{\mu,x } L^{-3N} |\cA_{\mu}(x)|^2 
\end{equation}

The  fermion fields  $\bar \psi_{\al}(x), 
\psi_{\al}(x) $ are  indexed by    $x \in  \bbT_{M}^{-N}$ 
and  $1 \leq \al  \leq  4$.  They  are the natural  basis for the space of functions
from the lattice to pairs of  four component spinors.
They provide a basis for the Grassman algebra generated by this 
space.  This is the space of fermions fields. Integration  over fermion 
fields means 
 projection onto the element of maximal degree. The
Dirac operator is given
by 
\begin{equation}  \label{Dirac}
   ( \bpsi,  D_e(\cA)\psi)= L^{-2N}\sum_{xx'}  \bpsi(x) \ga_{xx'} 
 e^{ieL^{-N} \cA(xx')} \psi(x')\   
 - \frac{3r}{2}  L^{-2N}\sum_x    \bpsi(x) \psi(x) 
\end{equation}
If
 $x' = x \pm  L^{-N} e_{\mu}$ then 
then 
 $ \ga_{xx'} = \frac12 ( r \pm \ga_{\mu} ) $
where  $\ga_{\mu}$ are the usual self-adjoint Dirac matrices satisfying 
 $\{\ga_{\mu}, \ga_{\nu} \} = 2 \de_{\mu \nu}$.
 This can also be written as
\begin{equation}    \label{lattice4}
\begin{split}
   ( \bpsi,  D_e(\cA)\psi)=&
 L^{-2N}\sum_{x,\mu}  \bpsi(x) ( \frac{ r + \ga_{\mu} }{2} )
e^{ieL^{-N} \cA(x,x+L^{-N} e_{\mu}) }\psi(x + L^{-N}e_{\mu})\\
+&
 L^{-2N}\sum_{x,\mu}  \bpsi(x)   ( \frac{ r - \ga_{\mu} }{2} )
e^{ieL^{-N} \cA(x,x-L^{-N} e_{\mu})} \psi(x - L^{-N}e_{\mu})\\
 -& \frac{3r}{2}  L^{-2N}\sum_x    \bpsi(x) \psi(x) \\
\end{split}
\end{equation}
Formally the $r$ terms disappear as $N \to \infty$ and we 
get the usual continuum action  
$\sum_{\mu} \int  \bpsi(x)  \ga_{\mu} 
(\pa_{\mu}  + ie \cA_{\mu}(x))\psi(x ) dx$.  The propagator for this action has
some well-known pathologies when $r=0$, so we take a fixed  $0 <r  \leq 1$.

 Actually we want a modification of 
the above corresponding to anti-periodic boundary conditions for
the fermions.   Instead of basis vectors  $ \bar \psi(x), \psi(x) $
indexed by the torus $ \bbT_{M}^{-N}$  we consider basis
vectors   $ \bar \psi(x), \psi(x) $ indexed by the infinite lattice
$(L^{-N} \bbZ)^3$, but with the identifications
\begin{equation}
\begin{split}
\psi(x+ L^{M}e_{\mu}) =& -  \psi(x)  \\
\bpsi(x+ L^{M}e_{\mu}) =& -  \bpsi(x)  \\
\end{split}
\end{equation}
Then   $j(x,x') \equiv \bpsi(x) \ga_{xx'} 
 e^{ieL^{-N} \cA(xx')} \psi(x')$ satisfies 
$j(x+ L^{M}e_{\mu},x'+ L^{M}e_{\mu}) =  
j(x,x')$.
Thus it is a function on the torus and so we can define sums like 
$\sum_{xx'} j(x,x')$  which appear in  (\ref{Dirac}).
The anti-periodic boundary conditions are necessary  for  Osterwalder-Schrader positivity,
that is for a Hilbert space structure \cite{BFS79}.   This is 
physically desirable and  will also be technically useful.

The above expression depends on three parameters: the coupling constant  $e$,
the fermion mass $m$ , and 
the photon mass $\mu$.    The parameters  $e,m$ are arbitrary. 
The photon mass is added because the Feynman gauge is not a complete
gauge.  We need it to suppress constant fields.  One would really
like to eliminate it, but this would mean picking another gauge or 
using the Wilson action.  In either case the treatment would 
be more complicated, but still feasible.  We regard the photon mass as 
a convenience, not  an essential modification of this UV problem.

The model has a charge conjugation symmetry.  Define the charge conjugation 
matrix $C$  to satisfy $ C^T \gamma_{\mu} C  =  - \gamma_{\mu}^T$  and we can
take  $C^T = C^{-1}  =  -C$.  Then we have  
$  C^T \ga_{xx'}  C  =   \ga_{x'x}^T  $  .  Under the transformation
\begin{equation}
\begin{split}
\bpsi  \rightarrow &  C \psi  \\  
\psi  \rightarrow &   -C \bpsi  \\  
\cA \rightarrow &  -\cA  \\
\end{split}
\end{equation}
the term   $(\bpsi , D ( \cA)  \psi )  $ is invariant.  In fact the entire density has this symmetry and
we  preserve it throughout our analysis.

The  term $(\bpsi , D ( \cA)  \psi )  $  is also  invariant under gauge transformations:
\begin{equation}
\begin{split}
\bar \psi  \rightarrow & e^{i e_0\la} \bar  \psi  \\
 \psi  \rightarrow &  e^{-i e_0\la} \psi   \\
 \cA \rightarrow  &  \cA -  \pa \la   \\
\end{split}
\end{equation}
The term  $ (\cA, (- \De + \mu^2) \cA) $ is of course not invariant.  Nevertheless 
we will be able to preserve a substantial amount of gauge invariance throughout 
the analysis
\bigskip

\noindent
\textbf{The scaled model}: 
Before proceeding we scale the problem up to a unit lattice
with large volume.  This changes our ultraviolet problems to 
infrared problems and puts us in  the natural home for the renormalization 
group.
The new   lattice  is  $\bbT_{N+M}^{0}$ with volume  $ L^{3(N+M)}$
and unit lattice spacing.  For fields  $\Psi,A$ on this lattice we 
define
$\rho_0(\bPsi,\Psi, A) = 
\rho(\bPsi_{L^{-N}},\Psi_{L^{-N}}, A_{L^{-N}})$ where 
\begin{equation}
\begin{split}
  (A_L)(xx')  = &   (\si_LA)(xx') = L^{-1/2}A(L^{-1}x,L^{-1}x') \\
(\Psi_L)(x)   = & (\si_L  \psi)(x) = L^{-1} \Psi(L^{-1} x ) \\    
\end{split}
\end{equation}
(The  scaling operator $\sigma_L$ is defined differently for bosons and  and fermions).
Then 
\begin{equation}  \label{def2}
\begin{split}
 \rho_0(\bPsi,\Psi , \cA)  
=\exp \left( - \frac12  (A, (- \De + \mu_0^2)A) 
- ( \bPsi, ( D_{e_0}(A)+ m_0 ) \Psi)  -\de v_0 - \de E_0 \right)
\end{split}
\end{equation}
Here the (real) inner products and operators are all on the unit lattice 
(i.e. put $N=0$ in (\ref{lattice1})-(\ref{lattice4})), and the new coupling
constants are all  very small:
\begin{equation}   e_0 = L^{-N/2} e  \ \ \ \ \ m_0  = L^{-N}m
\ \ \ \ \ \mu_0 = L^{-N}\mu
\end{equation}
The mass counterterm is taken to have the  Wick-ordered form
\begin{equation}
v_0  =  \sum_x  : \bPsi(x)  \de m_0  \Psi(x) :_{\hat S_0}
\equiv   \sum_x  \bPsi(x)  \de m_0  \Psi(x)    - tr ( \de m_0 \hat S_0(x,x))
\end{equation}
where  $\hat S_0 =  (D(0) +m_0)^{-1}$  and  
$\de m_0$ is to be specified.
In all the above we have used
 the subscript zero  (rather than $N$) because this 
is the starting point for the RG flow.
\bigskip

\re  The Dirac operator $ D_{e_0}(A)$  depends on the 
potential $A$ through  the phase   $e^{ie_0 A_{xx'}}$.
In analyzing the model it  is   tempting to replace   
$e^{ie_0 A_{xx'}}$ by    $ 1 + (e^{ie_0 A_{xx'}}-1)$
right from the start.
The first term gives a  kinetic term and the second a 
potential.  This does break gauge invariance,  but  the
split is natural for a perturbative analysis
and the associated RG transformations are clean.
We refrain from making the split   because  we do not have good 
control over the size of $A$. The potential
is not necessarily  small even though  $e_0$ is small.
(The term    $\exp(-\mu_0^2 \sum A^2)$  in the density
 does not    give any significant suppression
of large  $A$  because the 
the coefficient  $\mu^2_0$  is too small.)  Instead we wait and carry out the split in each fluctuation 
step.  By declining to do it all at once we are going 
to have background fields in  our fermion propagators.
This will prove to be a nuisance.

\subsection{counterterms}

Renormalization means adjusting  bare parameters to cancel the 
singularities in the theory.  Since our model is super-renormalizable it 
suffices to   choose the counterterms to cancel  singularities in low 
order perturbation theory.  Suppose one  splits off a potential 
as above and computes various quantities to second order in $e_0$.  One finds  an apparent  
logarithmic divergence in the fermion mass  and a quadratic divergence in the  vacuum energy as $N \to
\infty$. (We ignore an apparent linear divergence in the photon mass.  This will be ruled out by gauge
invariance).

\begin{figure} 
\includegraphics{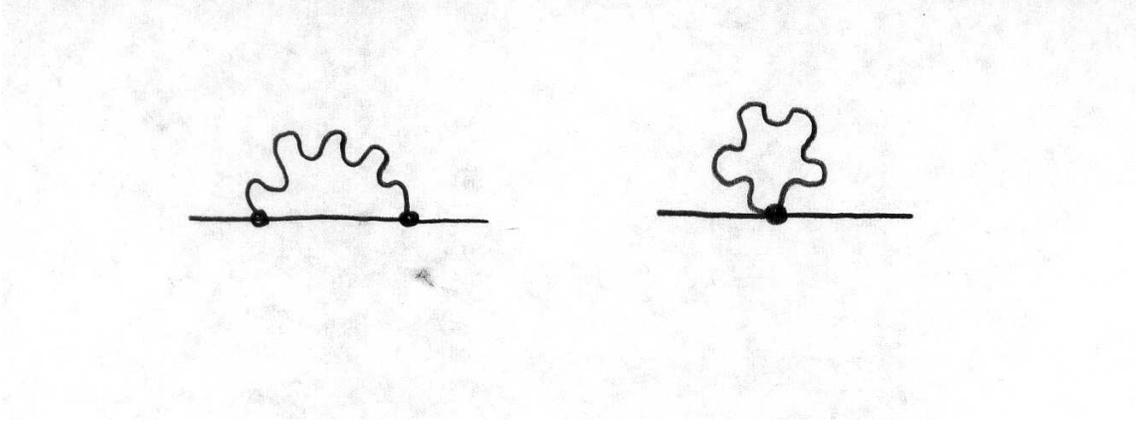}
\caption{ The fermion self-energy  $\Sigma_0(x,y)$     }  \label{sigma}
\end{figure}

Consider first  the fermion mass.  In the two point function $<\Psi(x) \bPsi(y)>$
for the scaled model
look at the amputated one-particle irreducible part.
The second order contribution is given by certain Gaussian integrals and computing them
we find
the  expression $ \de m_0 \ \de_{x,y} + \Sigma_0 (x,y) $ where 
\begin{equation}  \label{sigmaa}
\begin{split}
 \Sigma_0(x,y)  = &  \sum_{x'y'zw}
 \cV^{(1)}_{e_0,\mu}(z,x,x') (D(0)+m_0)^{-1}(x',y')
  \cV^{(1)}_{e_0,\mu}(w,y',y)  (-\De + \mu_0^2)^{-1}(z,w)  \\
 & + \sum_{z,z }\cV^{(2)}_{e_0,\mu, \mu}(z,z,x,y)  (\De+\mu_0^2)^{-1}(z,z)   
    \\ 
\end{split}
\end{equation} 
Here we have defined  vertices 
\begin{equation}
\cV^{(n)}_{e_0,\mu_1, \dots, \mu_n}(z_1, \cdots z_n,x,x')=
[\frac{ \pa^n  D_{e_0}(A,x,x')}{\pa A_{\mu_1}(z_1)\cdots \pa A_{\mu_n}(z_n)}]_{A=0}  = \cO(e_0^n)
\end{equation}
They vanish unless  $z_1 = \cdots = z_n$ and  $\mu_1 = \cdots =\mu_n$.
The vertex  $\cV^{(1)}_{e_0,\mu}$ is the lattice version  of $ie_0 \gamma_{\mu}$.
The sums in (\ref{sigmaa})  are restricted to nearest neighbors of $x,y$ in  $\bbT^0_{N+M}$.
The expression  
  corresponds to the diagrams shown in figure \ref{sigma}.
Terms in which the fermion fields in the potential contract to each other
do not occur.   \footnote{
 For example such  a term is 
\[
  \sum_{\tilde z z ww'}   \cV^{(1)}_{e_0,\mu}(\tilde z,x,y)  (-\De + \mu_0^2)^{-1}(\tilde z,z) \ 
 \textrm{tr}\left(  \cV^{(1)}_{e_0,\mu}( z,w,w')  (D(0)+m_0)^{-1}(w',w)  \right)  
\]
The trace vanishes by charge conjugation invariance  for we have that 
$C^{-1}  \cV^{(1)}_{e_0,\mu}( z,w,w')  C  = -[ \cV^{(1)}_{e_0,\mu}( z,w',w)]^T $
and  $C^{-1}  (D(0)+m_0)^{-1}(w',w) C = [(D(0)+m_0)^{-1}(w',w)]^T$.
Wick ordering removes self-contraction in the counterterm.
} 

The  singularity comes from  the non-summability of $\Sigma_0(x,y)$.
As $d(x,y)  \to \infty$ the best estimates uniform in $N$ are    $ (-\De + \mu_0^2)^{-1}(xx',yy')  = 
\cO( d(x,y)^{-1})$ for bosons  and    $ (-D(0) + m_0)^{-1}(x,y)  = 
\cO( d(x,y)^{-2})$ for fermions  and hence   $\Sigma_0(x,y)  = \cO( d(x,y)^{-3})$.
This gives the   logarithmic divergence  when summed. 
The counterterm is chosen to compensate and we take
\begin{equation}
\de m_0  = -  \sum_{y \in  \bbT^{0}_{N+M}} \Sigma_0(x , y)
\end{equation}
This is independent of $x$ by translation invariance.

When the cutoffs are removed the divergent part of  $\de m_0$ 
formally  vanishes by symmetry considerations.
This suggests that perhaps we do not really   need the mass counterterm.
We include it nevertheless as a convenient way to cancel dangerous looking terms in 
our effective actions.
Along these lines we mention also  that   $\de m_0$ is formally a Dirac scalar
when the cutoffs are removed.  We do not claim it is true as it stands.

The vacuum energy counterterm can also be chosen from looking 
at the divergences in perturbation theory.  But now one would have
to include all diagrams up to sixth order.  We do not really want to go
to the trouble of keeping track of all these diagrams, particularly 
since the vacuum energy does not contribute at all to the correlation
functions.  Instead we pick a counterterm by a method more suited 
to our analysis.
The counterterm is taken to have the form
\begin{equation}
  \de E_0 =  \sum_{j=0}^{N-1}  \delta \cE_j \ L^{3(N+M-j)} 
\end{equation}
where the $j^{th}$ term is chosen to remove dangerous terms
in the $(j+1)^{st}$  RG step.   The factor  $L^{3(N+M-j)}$ is 
the volume in which we  will be working at that point.  The energy densities 
$\de \cE_j$ are to be specified, but we will have   $\delta  \cE_j = \cO(e_j^2)$
where  $e_j =  L^{-(N-j)/2}$ is a running coupling constant.
Note the quadratic
divergence  $ \de E_0 = \cO(L^{2N}) $.

\subsection{the RG transformation}

The renormalization  group (RG) is a series of transformations which average
out the short distance features of the model, leaving only the
the long distance properties in which we are interested (now that 
we have scaled the model).
Here we  define a  simple  version of  the first RG transformation.
Our purpose is to explain the general idea and establish some notation.
This simple version will not be suitable for iteration which is the 
eventual goal.  \textit{Counterterms are omitted}.

First we define averaging operators.
These take functions  on  $\bbT_{N+M}^0$ to functions  on  $\bbT_{N+M}^1$
and are defined  for spinors and vector fields respectively by  
\begin{equation}
\begin{split}
(Q_{e_0}(A)f)(y) &   =  L^{-3}  \sum_{x \in B(y)}\exp \left( ie_0 
 A(\Ga_{yx}) \right) f(x)   \\
(Q h)_{\mu}(y) &   =  L^{-3}  \sum_{x \in B(y)} h_{ \mu}(x)  \\
\end{split}
\end{equation}
Here for  $y \in \bbT_{N+M}^1$  we have defined
\begin{equation} 
 B(y) =  \{ x \in \bbT_{N+M}^0:  |x-y|  \leq L/2\}
\end{equation}
The distance  is  $|x-y| = \sup_{\mu} |x_{\mu} -y_{\mu}|$ so this 
in an  $L$-cube centered on  $y$. 
We assume $L$ is odd so the $B(y)$ form a partition. 
Also  $\Ga_{yx}$ is a rectilinear path from $x$ to $y$ obtained by 
successively changing each of the three components,
and   $A(\Ga)  =  \sum_{x'x \in \Ga}  A_{x'x}$.
The spinor   averaging operator is chosen to be  gauge covariant:
\begin{equation}
Q_{e_0}(A - \pa \la) =  e^{-ie_0 \la }Q(A)   e^{ie_0 \la}
\end{equation}

Let    $Q_{e_0}(A)^T, Q^T$ denote the transpose operators which take functions
on  $\bbT_{N+M}^1$ to functions  on  $\bbT_{N+M}^0$.  These are 
defined  with respect to the natural inner product on    $\bbT_{N+M}^1$, i.e. sums
are weighted by $L^3$.   They are computed to be 
\begin{equation}
\begin{split}
(Q_{e_0}(A)^Tf)(x) &   = \exp \left( ie_0  A(\Ga_{yx}) \right) f(y)   \\
(Q ^Th)_{\mu}(x) &   = h_{ \mu}(y)  \\
\end{split}
\end{equation}
where $y$ is the unique point so  $x \in B(y)$.
We have   $Q_{e_0} (-A) Q_{e_0}(A)^T = I$ and  $QQ^T=I$ while  
$ Q_{e_0}(-A)^TQ_{e_0} (A) $ and $Q^T Q$ are projection operators.
\bigskip

 Now  starting with the density   $ \rho_{0}$  on  $\bbT^0_{N+M}$ then we define a
transformed density on   $\bbT^1_{N+M} $ by 
\footnote{We frequently  allow $\Psi$ to stand for the pair  $(\bPsi, \Psi)$.}
\begin{equation}
\begin{split}  \label{first}
 \tilde  \rho_{1} (\Psi, A)  = &c_0 \int \exp \left(-  \frac{a}{2L^2} 
|A-QA_0|^2 \right)   
\exp \left(- \frac{a}{L}| \Psi -Q_{e_0}(A_0) \Psi_0|^2\right) \\  & \rho_{0} (\Psi_0, A_0) \  d\Psi_0
 \ dA_0 \\
\end{split}
\end{equation}
We have passed to a larger Grassman algebra generated by  $\Psi, \Psi_0$ and are
now integrating over the $\Psi_0$ part of it.
 We have also  introduced the 
(somewhat abusive) notation
\begin{equation}
| \Psi -Q(A_0) \Psi_0|^2 =  ( \bPsi - Q_{e_0}(-A_0)  \bPsi_0,   \Psi -Q_{e_0}(A_0) \Psi_0) 
\end{equation}
The positive constant  $a$ is arbitrary and 
the  constant  $c_0$ is chosen so that   
\begin{equation}
\int \tilde   \rho_{1} (\Psi, A)\ d \Psi  \ dA =
\int   \rho_{0} (\Psi_0, A_0) \ d \Psi_0   \ dA_0
\end{equation}

Now insert the expression for  $\rho_0$ into (\ref{first}).
The quadratic form in  $(A, A_0)$ is 
\begin{equation}  \label{bquad}
  \frac{a}{2L^2}  |A-QA_0|^2 + \frac12  (A_0,   (-\De  +\mu_0^2 ) A_0  ) 
\end{equation}
and we would like to diagonalize it.
To this end we introduce
 \begin{equation}  \label{sharp}
 \De^\# \equiv -\De  +\mu_0^2 +  \frac{a}{L^2}\ Q^TQ 
\end{equation} 
This is positive and so we can also define
\begin{equation}
\begin{split}
C =&  (\De^\# )^{-1} \\
H_1 =&    \frac{a}{L^2} \ (\De^\# )^{-1}Q^T \\
\tilde \De_1  =&   \frac{a}{L^2}  -  \frac{a^2}{L^4}\ Q ( \De^\# )^{-1}Q^T \\
\end{split}
\end{equation}
Here  $C$ is an operator on  $\bbT^0_{N+M}$  (more precisely an operator on functions on 
bonds in  $\bbT^0_{N+M}$),  $H_1$ is an
operator from $\bbT^1_{N+M}$ to $\bbT^0_{N+M}$ , and  $\tilde \De_1$
is an operator on  $\bbT^1_{N+M}$.
The   quadratic form  is diagonalized by the 
change of variables  $A_0 \to A_0 + H_1A$.
Under this transformation the quadratic form becomes
\begin{equation}
 \frac12 (A,\tilde \De_{1}A) + \frac12 (A_0,   \De^{\#}  A_0 )   
\end{equation}
We have split it into  a piece for the background field  $A$, 
and a piece for the fluctuation field  $A_0$.  Now identify  the Gaussian measure
\begin{equation}
d \mu_{\st{ C_0}}(A_0)  = Z_1^{-1}  \exp\left(  - \frac12   (A_0,  \De^{\#}  A_0 ) \right) dA_0 
\end{equation}
where $Z_1$ is a normalization factor.
Defining $\tilde A =H_1A$, and omitting counterterms for simplicity our transformed 
density has become
\begin{equation}
\begin{split} 
& \tilde  \rho_{1} (\Psi, A)  = c_0
\exp \left(- \frac12 (A,\tilde \De_{1}A) \right) \
Z_1
\\
 &\int  \exp \left(  -  \frac{a}{L}| \Psi -Q_{e_0}(\tilde A + A_0 ) \Psi_0)|^2 -
 (\bPsi_0 ,( D_{e_0}(\tilde A  +A_0) + m_0 ) \Psi_0 ) \right)    d\Psi_0 \  d\mu_{C_0}(A_0)\\
\end{split}
\end{equation}

Next we define a potential  $V_0$ by  
\begin{equation}  \label{pot1}
\begin{split}
&\frac{a}{L}| \Psi -Q_{e_0}(\tilde A + A_0 ) \Psi_0|^2 +
 (\bPsi_0 ,( D_{e_0}(\tilde A  +A_0) + m_0 ) \Psi_0 ) \\
=  & \frac{a}{L}| \Psi -Q_{e_0}(\tilde A )
\Psi_0)|^2 +
 (\bPsi_0 ,( D_{e_0}(\tilde A ) + m_0 ) \Psi_0 )  +   V_0(\Psi,\Psi_0,\tilde A, A_0)
\end{split}
\end{equation}
Thus $V_0$ is the part of the fermion quadratic form depending on 
the fluctuation field  $A_0$.  The fluctuation field is massive due to 
 the  $Q^TQ$ in $\De^{\#}$.  Hence large  $A_0$ is  suppressed and
we can expect that  $V_0$ is small.  
The most important contribution to $V_0$ is 
from the term 
\begin{equation}  \label{pot2}
   ( \bPsi_0,  D_{e_0}(\tilde A + A_0)\Psi_0)- ( \bPsi_0,  D_{e_0}(\tilde A)\Psi_0)= \sum_{xx'} 
\bPsi_0(x) \ga_{xx'} 
 e^{ie_0  \tilde A(xx')} ( e^{ie_0 A_0(xx')}-1)\Psi_0(x')
\end{equation}
The lowest order term  in $e_0$ gives the classical interaction vertex, but with the background $ e^{ie_0 
\tilde A}$. Higher order terms give multiphoton vertices.  There are also vertices coming
from the $Q$ terms.

Now consider the fermion quadratic form with background field, i.e. the first two 
terms on the right side of (\ref{pot1}).
We  want to diagonalize in  $\Psi, \Psi_0$.  This involves
the inverse of the operator   
\begin{equation}
D^\#(\tilde A) \equiv  D_{e_0}(\tilde A)  + m_0 +  \frac{a}{L} \ Q_{e_0}(- \tilde A)^T Q_{e_0}(\tilde A)
\end{equation}
(We could write  $  Q_{e_0}( -\tilde A)^T$   as  $Q_{e_0}( \tilde A)^*$   where the adjoint
refers to  a complex inner product).
However we cannot control the inverse unless with have control over
the field strength for $\tilde A$ .   This is an issue we address in the 
the rest of the paper.  \textit{  For now we just assume  that  $e_0 |\pa \tilde A|$
is sufficiently small.} In this case we can define 
\begin{equation}  \label{hhh}
\begin{split}
\Ga (\tilde A) =&  D^\#(\tilde A)^{-1} \\
H_1(\tilde A) =& 
\begin{cases}    
 \frac{a}{L} \  D^\#(\tilde A) ^{-1}   Q_{e_0}( -\tilde A)^T  &   \textrm{ on } \Psi \\
 \frac{a}{L} \  [ D^\#(\tilde A)^{-1}]^T   Q_{e_0}( \tilde A)^T  &   \textrm{ on } \bPsi \\
\end{cases}
 \\
\tilde D_1(\tilde A)   =& \ 
  \frac{a}{L}   -   \frac{a^2}{L^2} \ Q_{e_0}(\tilde A)  D^\#(\tilde A) ^{-1}  Q_{e_0}( -\tilde A)^T  \\
\end{split}
\end{equation}
Now  make the change of variables  $\Psi_0 \to \Psi_0 + H_1(\tilde A) \Psi$ and 
 $\bPsi_0 \to \bPsi_0 + H_1(\tilde A) \bPsi$. 
The quadratic form  becomes  
\begin{equation}
(\bPsi,\tilde D_{1}(\tilde A)\Psi) +  (\bPsi_0,  D^{\#}(\tilde A) 
\Psi_0 )
\end{equation}
We identify the fermion Gaussian measure 
\begin{equation}
d \mu_{\st{ \Ga(\tilde A)}}(\Psi_0)  =\tilde Z_1(\tilde A)^{-1} 
 \exp \left( -  (\bPsi_0,  D^{\#}(\tilde A) 
\Psi_0 )  \right)  d \Psi_0
\end{equation}
where $\tilde Z_1(\tilde A)$ is the normalization factor.
With $   \tilde \Psi(\tilde A) = H_1(\tilde A) \Psi$ 
the complete fluctuation integral is now   the Gaussian integral
\begin{equation}
\tilde \Xi_1(\Psi, \tilde A) \equiv  
\int  \exp \left( - V_0(\Psi,\Psi_0 + \tilde \Psi(\tilde A),\tilde A, A_0 ) \right)   
d\mu_{\st{\Ga(\tilde A)}}( \Psi_0 )  d \mu_{\st{C}}(A_0)
\end{equation}
and the overall density is 
\begin{equation}  \label{rep3}
 \tilde  \rho_{1} (\Psi, A)  =  c_0
\exp \left(- \frac12 (A,\tilde \De_{1}A) )-(\bPsi,\tilde D_{1}(\tilde A) \Psi ) \right)
 Z_1 \tilde Z_1(\tilde A)\ \tilde \Xi_1(\Psi,  \tilde A)
\end{equation}
We note that  $(\bPsi,\tilde D_{1}(\tilde A) \Psi )$ and  $ \tilde Z_1(\tilde A)$ and
$\tilde \Xi_1(\Psi,  \tilde A)$  are all invariant under gauge transformations
in the variables  $(\bPsi, \Psi, \tilde A)$.  (They  would not be invariant in the fundamental variables
 $(\bPsi, \Psi, A$).)

Finally we scale back to fields on a unit lattice.   For   $\Psi_1, A_1$ on  
$\bbT^{0}_{N+M-1}$ by  we  define 
\begin{equation}
 \rho_{1} (\Psi_1,A_1)  = \tilde  \rho_{1}(\Psi_{1,L}, A_{1,L}) 
\end{equation}
For the boson parts of this we define  
   $\cQ =  \si_L^{-1} Q \si_L   $  which is the natural averaging operator  mapping functions on 
$\bbT^{-1}_{N+M-1}$ (an $L^{-1}$ lattice ) to functions on  $\bbT^0_{N+M-1}$.
Then we define
\begin{equation}
\begin{split}
G_1 =& L^{-2} \si_L^{-1}C_0 \ \si_L =  \left(-\De  +\mu_1^2 + \cQ^T\cQ  \right)^{-1} \\
\cH_1 =& \si_L^{-1}H_1 \si_L =  a \ G_1 \cQ^T \\
\De_1  =& L^{2} \si_L^{-1} \tilde \De_1 \si_L  = a I - a^2 \cQ G_1 \cQ^T \\
\end{split}
\end{equation}
Here $G_1$ is on operator on  on  $\bbT^{-1}_{N+M-1}$ , $ \cH_1 $ is an 
operator from $\bbT^0_{N+M-1}$  to $\bbT^{-1}_{N+M-1}$, 
and   $\De_1$ is an operator on   $\bbT^0_{N+M-1}$.  The background field $\tilde A = H_1A$
scales to 
 $ H_1 A_{1,L}$ on $\bbT^0_{N+M}$, and it  can  now  be written   $\cA_{1,L} $
 where  $\cA_1 = \cH_1 A_1$ is a field on  $\bbT^{-1}_{N+M-1}$

Now for fermions we define   $\cQ_{e_1}(\cA_1) =  \si_L^{-1} Q_{e_0}(\cA_{1,L}) \si_L $ 
which is the natural averaging operator on $\bbT^{-1}_{N+M-1}$ with   $e_1,\cA_1$.
Then we define  
\begin{equation}
\begin{split}
S_1(\cA_1) =  &L^{-1}  \si_L^{-1}\Ga_0(\cA_{1,L}) \ \si_L =
  \left(D_{e_1}(\cA_1)  +  m_1 +\cQ_{e_1}(-\cA_1)^T\cQ_{e_1}(\cA_1)  \right)^{-1}
 \\
\cH_1(\cA_1) =& \si_L^{-1}H_1(\cA_{1,L}) \si_L =
\begin{cases}    
  a  S_1(\cA_1) \cQ_{e_1}(-\cA_1)^T  & \textrm{ on } \Psi \\
 a  S_1(\cA_1)^T \cQ_{e_1}(\cA_1)^T  &   \textrm{ on } \bPsi \\
\end{cases}
\\
D_1(\cA_1)  =& L\ \si_L^{-1}
\tilde D_1(\cA_{1,L}) \si_L  = a I - a^2 \cQ_{e_1}(\cA_1) S_1(\cA_1)
\cQ_{e_1}(-\cA_1)^T
 \\
\end{split}
\end{equation}
  The fermi field  $\tilde \Psi(\tilde A)$ scales to  $ H_1(\cA_{1,L}) \Psi_{1,L}$
 which can be written   $( \psi_1(\cA_1) )_{L}$
where $ \psi_1(\cA_1)    \equiv  \cH_1(\cA_{1}) \Psi_{1}$. This  field appears
in the scaled fluctuation integral   
\begin{equation}
\Xi_1 (\Psi_1,  \cA_1)=\tilde \Xi_1 (\Psi_{1,L} ,\cA_{1,L})
\end{equation}
 We also define $Z_1(\cA_1)  =  \tilde  Z_1(  \cA_{1,L})$.
Then we have for the new density (still under a small field  assumption)
\begin{equation}
  \rho_{1} (\Psi_1,A_1)  = c_0  
\exp \left(- \frac12 (A_1, \De_{1}A_1) )-(\bPsi_1, D_{1}(\cA_1) \Psi_1) \right)
Z_1 Z_1(\cA_1) \Xi_1 (\Psi_1,\cA_1)
\end{equation}

This expression shows a nice split into a kinetic part  (the exponential) and an effective interaction
(the rest).   Next one could compute the leading contributions to  $Z_1(\cA_1)$  and  
 $\Xi_1 (\Psi_1,\cA_1)$ in perturbation theory.  However we leave this for the full treatment.
\bigskip

\noindent  \textbf{Another representation}:
 Suppose we   undo some of the  steps that got us here.  Go back to the expression  (\ref{rep3})
for   $\tilde \rho_1(\Psi, A)$.
Express   $ \tilde Z_1(\tilde A)$  and  $Z_1$   as integrals over  $\Psi_0$ and $ A_0$ 
and then make the inverse transformations   $\Psi_0 \to \Psi_0 - \tilde \Psi(\tilde A)$  
and  $A_0 \to A_0 - \tilde A$.     This yields
\begin{equation}
\begin{split}  \label{fifth}
 \tilde  \rho_{1} (\Psi, A)  =&  c_0 \
\tilde  \Xi_1(\Psi,\tilde  A) 
\int \exp \left(-  \frac{a}{2L^2} 
|A-QA_0|^2 \right)   
\exp \left(- \frac{a}{L}| \Psi -Q(\tilde A) \Psi_0|^2\right) \\
&
   \exp \left( - \frac12  (A_0,   (-\De  +\mu_0^2 ) A_0  )  -  
 (\bPsi_0 ,( D_{e_0}(\tilde A) + m_0  ) \Psi_0 )  \right)  d\Psi_0\ dA_0
 \\
\end{split}
\end{equation}
This can also be scaled to get another representation for  $ \rho_{1} (\Psi_1,A_1)$.   
The expression  (\ref{fifth}) is  close to what we started with in (\ref{first}). 
 The change is that the  gauge field
$A_0$ has been replaced with a background field $\tilde A$ with fewer degrees of freedom, and there
is a corresponding correction factor $\tilde  \Xi_1(\Psi,\tilde  A) $.
Variations of this representation will be useful for iteration.

\subsection{some tools}

We introduce some tools that we need for subsequent developments.   These are roughly
 ordered by length scale.
\bigskip

\textbf{A}.  We use random walk expansions for the basic propagators  $C$ and  $\Ga( A)$
assuming $e_0 |\pa A|$ is  sufficiently small.
These are developed on a scale $M_0$  which is a  fixed power of $L$.  We consider
the $M_0$ lattice  $\bbT^{M_0}_{N+M}$
A path  
is a sequence  $\om  =(j_0,j_1,\dots, j_n)$  of points in this lattice   which are neighbors
 in the sense that $|j_{\al}  - j_{\al+1}|= 0  \textrm{ or }M_0$.  
Associated with a path  $\om$ is a sequence $  ( \cO_{j_0}, \cO_{j_1}, \cdots, \cO_{j_n})$ 
of \textit{overlapping} $2M_0$ cubes $\cO_j$  centered on the points  $j$. 
We write  $\om: x \to y$ if   $x \in \cO_{j_0}$ and $y \in \cO_{j_n}$.

 The random walk expansion has  the
form
\begin{equation}
\begin{split}
\Ga(A,x,y)  =&  \sum_{\om:x \to y}  \Ga_{\om}(A,x,y)\\
C(x,y) =&  \sum_{\om: x \to y}  C_{\om}(x,y)\\
\end{split}
\end{equation}
The  individual terms are  $(\cO(1)M_0^{-1})^{\ell(\om)}$  where $\ell(\om)$=number of steps in $\om$, and this 
 is sufficient for convergence   if  $M_0$ is large enough.  From the condition  $\om: x \to y$ we can 
also extract some exponential decay  and obtain for  $\beta = \cO(M_0^{-1})$
\begin{equation}  \label{expdecay}
|\Ga(A,x,y)  |, \ \  |C(x,y)| \leq C  \exp( - \beta d(x,y) )  
\end{equation}
We also note that 
 $\Ga_{\La,\om}(A)$  depends on  $A$  only  in 
$ \cO_{j_0}  \cup \cO_{j_1} \cup \cdots \cup \cO_{j_n}$.

See  appendix  A for details of this construction and the original references.
\bigskip

\textbf{B}.   The fluctuation integrals will be processed by a cluster expansion.     
This means all terms  must be localized.  This will be done on a  scale $M_1$,
also a fixed power of $L$ but larger than $M_0$.
$\De$ will denote a $M_1$ cube centered on the  $\bbT^{M_1}_{N+M}$ lattice.
These partition the lattice.
  A  union    of such  $\De$  will be denoted  by $X,Y,Z, \dots$.
Such a set is connected and called a polymer if for any two  $\De, \De'   \subset X$ there is 
a sequence  $ \De, \De_1, \dots  \De_n, \De'$ in $X$ such that consecutive
blocks touch in any dimension.  
The terms in our effective potentials will be polymer sums   $\sum_X E(X)$
where  $E(X)$ only depends on fields in   $X$.
\bigskip

\textbf{C.} We  introduce a  scale beyond which correlations 
are completely negligible.   This is defined by
\begin{equation}
r(e_0)  =  (\log e_0^{-1} )^r   =  \left(\frac N2  + \log e^{-1}\right)^r
\end{equation}
for some small positive integer $r$.  If $N$ is large then  $e_0$ is small and  $r(e_0)$ is 
large; we assume it is larger than   $M_1$. 
Let $R_0 = \inf_m \{ L^m: L^m \geq r(e_0) \} $ be an associated power of  $L$.  We will
 consider blocks
$\square$ of size $R_0$ centered on points in $\bbT^{R_0}_{N+M}$  and collections of such blocks denoted $\Om ,
\La, ...$.

We also  will use approximate   propagators which do not couple beyond this scale.
These are defined in terms of the random walk expansion. 
 Let $\chi_{\om}(x,y)$ is the characteristic function of   paths which stay within  $r(e_0)/2$ 
of both $x$ and $y$. Then we define
\begin{equation} \label{localprop}
\begin{split}
C^{loc}(x,y)=& \sum_{\om:x \to y}C_{\om}(x,y)\chi_{\om} (x,y) \\
\Ga^{loc}(A,x,y)=& \sum_{\om:x \to y}\Ga_{\om}(A,x,y)\chi_{\om} (x,y) \\
\end{split}
\end{equation}
These  vanish for   $d(x,y) \geq r(e_0)/2$ and still satisfy
  (\ref{expdecay}).
\bigskip

\textbf{D.}  We introduce a scale which distinguishes between  large and 
small values of   $|\pa A|$.      This is defined  by  
\begin{equation}
p(e_0)  =  (\log e_0^{-1} )^p = \left(\frac N2  + \log e^{-1}\right)^p 
\end{equation}
for some integer $p$ larger than $r$.
``Small fields'' will   satisfy  conditions like  $|\pa A|  \leq  \cO(p(e_0))$ and may actually be rather
large.
Still    $e_0|\pa A|$ would be  
small as required for the existence of $\Ga(A)$. 
\bigskip

\textbf{F}.  Finally we discuss introduce some norms for fermions.
 Let  $\xi$ stand for   $(0, \al, x)$ or $(1, \al , x)$ with  $x \in \bbT^{0}_{N+M}$
and $ 1 \leq  \al \leq 4 $.  Then  define
$\Psi(0, \al, x) = \Psi_{\al}(x)$ and  $\Psi(1, \al, x) = \bPsi_{\al}(x)$.  
We consider elements of the Grassmann algebra of the form:
\begin{equation}  \label{element}
 K(\Psi) = \sum_{n=0}^{\infty}  1/n! \sum_{\xi_1,...,\xi_n} k_n(\xi_1,..., \xi_n) \Psi
(\xi_1) ...
\Psi (\xi_n)  \end{equation}
The kernel  $k_n$ may also depend on the gauge field  $A$.
A family of norms is defined by 
\begin{equation} \| K \| _h = \sum_{n=0}^{\infty} \frac {h^n} { n!}\| k_n \|_1 
\end{equation} 
where $\| k_n \|_1$ is the $\ell_1$ norm and  $h= \cO(1)$ is an 
adjustable parameter.  These norms satisfy   $\| K L \|_h  \leq  \|K \|_h \|L\|_h$. 
Properties and variations  are discussed in  appendix  B.

We will also want to consider functions which only depend on  dressed fields
like  $\tilde \Psi = \tilde \Psi(\tilde A) = H_1(\tilde A) \Psi$.  These would have 
the form    
\begin{equation}  \label{dressed}
 K(\tilde \Psi) = \sum_{n=0}^{\infty}  1/n! \sum_{\xi_1,...,\xi_n} k_n(\xi_1,..., \xi_n)\tilde  \Psi
(\xi_1) ...
\tilde \Psi (\xi_n)  \end{equation}
We define  a  \textit{kernel norm} by  
\begin{equation} \label{kernelnorm}
 | K | _h = \sum_{n=0}^{\infty} \frac {h^n} { n!}\| k_n \|_1 
\end{equation} 
This still satisfies   $| K L |_h  \leq  |K |_h |L|_h$.

\newpage

\section{The first step}

We now give another version of the first RG transformation and this 
time it is something we can iterate.  The idea is to break up 
the  RG integral over the gauge field into a large fields  and 
 small fields.   This is done at each point in space
time   we generate a sum over regions where the field is large and 
regions where the field is small.   In the small field region 
we can carry out the  analysis of effective action  that we have sketched. 
Furthermore we can make a perturbation expansion   with good control over the remainder.
The contribution of the large field region makes a tiny contribution
due to tiny factors from the free action.   These small factors are 
enough to control the sum over the regions.   We now enter into the details
which follow especially  \cite{Bal82b}, \cite{BIJ88}.

\subsection{the split}
We analyze (\ref{first}), but now splitting into large and small field regions. 
The small field region will be a set on which the following  inequalities  hold 
\begin{equation}  \label{bounds1}
\begin{split}
   |\pa A_0 | \leq  & \ \  p(e_0)\\
  | A_0 | \leq  & \ \   \mu_0^{-1} p(e_0)\\
|A - Q A_0 | \leq & \ \   p(e_0)  \\
\end{split}
\end{equation}
These correspond to  various terms in the action.  Fields which violate them 
will be  exponentially suppressed.

Specifically we insert 
under the integral sign the expression
\begin{equation}
 1 =  \sum_{ \Om}  \zeta_{0}( \Om^c, A,A_0) 
\chi_{0}( \Om,A,A_0) 
 \end{equation}
Here  $\Om$  is an arbitrary collection of blocks  $\sq$ of size  $R_0$.
The function  $\chi_{0}( \Om,A,A_0) $ enforces the inequalities   (\ref{bounds1})
on $\Om$ and we take it to have the form 
\begin{equation}
\begin{split}
&\chi_{0}( \Om,A,A_0)\\
 =&  \prod_{x \in \Om}  \chi( |\pa A_0(x)| / p(e_0))
 \prod_{x \in \Om}  \chi( | A_0(x)| /\mu_0^{-1} p(e_0))
 \prod_{y \in \Om}  \chi( |(A-QA_0)(y)| / p(e_0)) \\
\end{split}
\end{equation}
where $\chi(\al)$ is a smooth positive function interpolating between  $\chi=1$ 
for   $|\al|  \leq   \frac12  $ and $\chi =  0$ for   $|\al|  \geq 1$.  We assume 
the derivatives satisfy   $|\chi^{(n)}|  \leq  \cO(n!^2)$.   
The function   $\zeta_{0}( \Om^c, A,A_0) $ is also smooth and  enforces that at least one 
of the inequalities  (\ref{bounds1}) (with a factor $\frac12$) is  violated at some point in 
every block in  $\Om^c$.   For the explicit formula see \cite{Bal82b}, \cite{BIJ88}.

Now  we have 
\begin{equation}  \label{rep1}
\begin{split}
& \tilde  \rho_{1}(\Psi, A)  =c_0e^{ \de E_0}  \sum_{\Om}  
\int d \Psi_0 d A_0 \   \zeta_{0}(\Om^c, A,A_0)\  \chi_{0}( \Om,A,A_0) \\
&\exp \left(- \frac{a}{2L^2} |A- QA_0|^2  - \frac12  ( A_0,(-\De + \mu_0^2) A_0) 
 \right) \\
&
\exp \left(- \frac{a}{L}| \Psi -Q_{e_0}(A_0) \Psi_0|^2
+ (\bPsi_0 ,( D_{e_0}(A_0) + m_0 ) \Psi_0 ) -v(\de m_0)
\right) \
\\
\end{split}
\end{equation}

\subsection{boson translation}
Again we seek to diagonalize the boson  quadratic form, but now only in the small
field region.
We  define a local version of    $H_1$ by   
\begin{equation}
H_1^{loc}  = \frac{a}{L^2} C^{loc} Q^T 
\end{equation}
We approximately diagonalize by making  the transformation
\begin{equation}  \label{bloc}
A_0 \to A_0 +  [  H_1^{loc} A ]_{\Om'}  \equiv   A_0 +  \tilde A_{\Om'}
\end{equation}   
 Here  $\Om'$ is $\Om$ shrunk by a corridor of  $R_0$ blocks so that   
$d(\Om^c, \Om') = R_0 \geq r(e_0)$.  Since $H_1^{loc}$ has range
$r(e_0)/2$  the expression $ [  H_1^{loc} A ]_{\Om'} $ only depends on $A$ in $\Om$
where we have some control.

\begin{lem} \label{wb}  Under the transformation (\ref{bloc})  the boson  quadratic form  (\ref{bquad}) becomes 
\begin{equation}   \label{bosontrans}
 \frac12  (A, [\tilde  \De_1]_{\Om}A) +\frac12(A_0, \De^\# A_0)  
+ F^b+ W^b
\end{equation}
Here each   $F^b= F^b(A,A_0)$ is localized in   $(\Om'')^c$  and  $W^b= W^b(A,A_0)$ has the form 
$ W^b =  \sum_{X \subset \Om} W^b(X) $ where the sum is over polymers
$X$.  The function  $W^b(X) =W^b(X,A,A_0)$  only
depends on
$A,A_0$ in $X$.   Assuming the bounds  (\ref{bounds1})  we have  with $\beta = M_0^{-1}$
\begin{equation}   
|W^b(X,A,A_0)| \leq  \cO(1)e^{-\frac12 \beta  r(e_0)}e^{- \frac12 \beta M_1|X|_1 }  
\end{equation}
\end{lem}
\bigskip

\res   Here are some notational conventions
\begin{enumerate}
\item   If $f$ is a function then  $f_{\Om} = \chi_{\Om} f$ is the restriction to 
$\Om$.  If $T$ is an operator the   $T_{\Om}  = \chi_{\Om} T \chi_{\Om}$ and  
 $T_{\Om\Om'}  = \chi_{\Om} T \chi_{\Om'}$. 
\item   
 The general  convention is that constants denoted  $\cO(1)$ may depend
on the parameters $L, M_0, M_1,\mu$,  but not on  $e,m$ or on the fundamental cutoffs  $N,M$.
A constant which is independent of all parameters is 
designated as universal.
\item   The symbol  $|X|_1$ stands for the number of elementary  $M_1$ blocks in  $X$.
Thus the volume is  $|X| = M_1^3|X|_1$.   The quantity $M_1|X|_1$ is the linear size of 
$|X|$.  In fact if $\cL(X)$ is the length of the shortest tree joining the centers of 
the blocks in $X$ then  $\cL(X)  = M_1(|X|_1 -1)$. We could have written the bound with
$\cL(X)$.  Since  $\beta M_1 = \cO(M_0^{-1} M_1)$ is assumed large  the factor 
$e^{- \frac12 \beta M_1|X|_1 } $ controls the sum over $X$.  The relevant  standard bound is 
that for  $\ka$ large enough 
\begin{equation}
\sum_{X \supset \De} e^{- \ka |X|_1}  \leq  1
\end{equation}
Or we could replace the condition  $X \supset \De$ by $X$ touches $\De$.
\item
Note  that    $e^{-r(e_0)} = \cO(e_0^n)$ for any $n$;  it is extremely small.
\end{enumerate}
\bigskip

\pr
Making the translation the term quadratic in $A_0$ remains
$\frac12(A_0, \De^\# A_0)$   where   $\De^\#$ is defined in  ( \ref{sharp}).

The cross term is    
\begin{equation}  \label{cross}
 ( A_0, \De^\#   [  H_1^{loc} A ]_{\Om'} ) - \frac{a}{L^2} (QA_0, A)
\end{equation}
We replace  $A_0$ by  $A_{0,\Om}$.  In the first term there is no change and 
the difference for the second term  is localized in  $\Om^c$ and
contributes to  $F^b$.  Thus it suffices to consider   (\ref{cross}) with
$A_{0,\Om}$.   Next we write 
\begin{equation}  \label{shift}
[H_1^{loc}A]_{\Om'}=[H_1^{loc}A_{\Om}]_{\Om'}  = H_1^{loc}A_{\Om} -[H_1^{loc}A_{\Om}]_{(\Om')^c} 
\end{equation}
The term  $[H_1^{loc}A_{\Om}]_{(\Om')^c}$ has $A$ dependence in $\Om \cap (\Om'')^{c}$, and its 
contribution also has  $A_0$ dependence in this set.   Thus  these terms contribute to $F^b$.
We are left with
\begin{equation} 
( A_{0, \Om}, \De^\# H_1^{loc} A_{\Om} ) -  \frac{a}{L^2} (QA_{0, \Om},  A_{\Om})
\end{equation}
If we replace   $H_1^{loc} $ by  $H_1$ we get zero.  Thus we can write this term as 
\begin{equation}
W_1 \equiv ( A_{0, \Om}, \De^\#( H_1^{loc} - H_1) A_{\Om} )
\end{equation}

The terms quadratic in  $A$ have the form 
\begin{equation}
   \frac{a}{2L^2}  (A,A) - \frac{a}{L^2}  (A, Q[H_1^{loc}A]_{\Om'}  )+
 \frac12 ( [H_1^{loc}A]_{\Om'} ,\De^\# 
[H_1^{loc}A]_{\Om'}) 
\end{equation}
We  replace  $A$ by   $A_{\Om}$.  The difference only comes from the first term
and it contributes to  $F^b$.  We also  insert  (\ref{shift}) and again the terms 
arising from   $[H_1^{loc}A_{\Om}]_{(\Om')^c} $ are all localized in   $\Om \cap (\Om'')^{c}$
and contribute to $F_b$.  We are left with    
\begin{equation}
  \frac{a}{2L^2} (A_{\Om},A_{\Om})  -   \frac{a}{L^2} (A_{\Om}, QH_1^{loc}A_{\Om}) 
 +\frac12 ( H_1^{loc}A_{\Om},\De^\#  H_1^{loc}A_{\Om})  
\end{equation}
\bigskip
If we replace  $H_1^{loc} $ by  $H_1$ we get   $\frac12(A_{\Om}, \tilde \De_1 A_{\Om})
= \frac12(A, \tilde \De_{1,\Om} A)$.  The difference is 
\begin{equation}
\begin{split}
W_2 \equiv   & - \frac{a}{L^2} (A_{\Om}, Q(H_1^{loc}-H_1)A_{\Om}  )  \\
 & +\frac12 ((H^{loc}_1- H_1)A_{\Om} ,\De^\#  H_1^{loc}A_{\Om}) +
 \frac12 ( H_1A_{\Om} ,\De^\# (H_1^{loc}- H_1)A_{\Om}) \\
\end{split}
\end{equation}
Now we have established the representation (\ref{bosontrans}) with   $W^b = W_1 + W_2$.

Let us analyze $W_1$.  It can be written  
\begin{equation}  \label{w1}
W_1 = (A_0,w_1 A)=  \sum_{ x,y \in \Om}  A_{\mu,0}(x) w_{1,\mu,\nu}(x,y) A_{\nu}(y)
\end{equation}
Now  $w_1$ depends on $H_1 -H_1^{loc}$ and hence on 
$C -C^{loc}$.   In appendix  \ref{A}, lemma \ref{coarse.loc}, we use a random walk expansion to 
 find a local expansion $C - C^{loc}  =\sum_{X} \de C(X)$ 
  in which  $ \de C_{\mu,\nu}(X,x,y)$  is  supported on  $X
\times X$ and only very large  $X$ contribute.  This leads to an expansion  
$w_1 = \sum_X w_1(X)$ with  $w_{1,\mu,\nu}(X,x,y)$ supported on $X \times X$
and satisfying  
\begin{equation}
|w_1(X,x,y)|   \leq   \cO(1)  e^{-  \beta  r(e_0) } e^{-  \beta M_1|X|_1 } e^{- \beta d(x,y)}
\end{equation}
Now we have   $W_1  =  \sum_{X}  W_1(X)$ where $W_1(X) = (A_0,w_1(X) A)$.
We have  the weak bounds   $|A_0| \leq \mu_0^{-1}p(e_0)$
and   
\begin{equation}
|A| \leq  |A-QA_0| + |QA_0| \leq \cO(\mu_0^{-1}p(e_0)) 
\end{equation}
The factors  $\cO(\mu_0^{-1}p(e_0)) $ are compensated by a factor $e^{- \frac 12 \beta r(e_0)}$.  We also 
get a volume factor  $|X|  = M_1^3|X|_1$ which is dominated 
by  a factor $e^{- \frac12 \beta M_1|X|_1 }$.  Overall we have 
\begin{equation}
|W_1(X)|   \leq   e^{-   \frac 12  \beta  r(e_0) } e^{- \frac12 \beta M_1|X|_1 }
\end{equation}

The analysis of  $W_2$   is similar.  This completes the proof.
\bigskip

At this point we have
\begin{equation}    \label{rep2}
\begin{split}
& \tilde  \rho_{1}(\Psi, A)  =c_0e^{ \de E_0}  \sum_{\Om}  
\int d \Psi_0 d A_0 \   \zeta_{0}( \Om^c, A,A_0)\  \chi_{0}( \Om,A, \tilde A_{\Om'} +A_0  ) \\
&\exp \left(-  \frac12  (A, \tilde  \De_{1,\Om}A) -\frac12(A_0, \De^\# A_0)  
- F^b
- W^b)
 \right) \\
&
\exp \left(- \frac{a}{L}| \Psi -Q_{e_0}( \tilde A_{\Om'}+A_0) \Psi_0|^2
- (\bPsi_0 ,( D_{e_0}( \tilde A_{\Om'} +A_0) + m_0 ) \Psi_0 ) -v(\de m_0)
\right) \
\\
\end{split}
\end{equation}

\subsection{new field restrictions}
Let us consider the characteristic function
$ \chi_{0}(\Om,A,\tilde A_{\Om'} +A_0) $
where we recall that  $A$ is a function on  $\bbT^1_{N+M}$
and that   $A_0, \tilde A =  H_1^{loc}A$ are functions on  $\bbT^0_{N+M}$. 
The characteristic function imposes that   on $\Om'$
\begin{equation} \label{inequality1}
\begin{split}
   |\pa  (\tilde A +A_0) | \leq  & \ \  p(e_0)  \\
  |\tilde A +A_0 | \leq  & \ \   \mu_0^{-1} p(e_0)\\
|A - Q (\tilde  A  +A_0) | \leq & \ \   p(e_0)  \\
\end{split}
\end{equation}
These inequalities imply inequalities on the fields   $A, A_0, \tilde A$
separately as follows

\begin{lem} Assume,  (\ref{inequality1}) holds  on $\Om'$.
Then there are constants $C_1,C_2 $ such that 
\begin{enumerate}
\item  
 $ |\pa  A | \leq   C_1 p(e_0)$  and 
$  | A | \leq    C_1 \mu_{0}^{-1} p(e_{0})$  on $\Om'$.
\item  
$ \tilde A= Q^T A  + \cO(p(e_0))$  on $\Om''$
\item  
$|A_0 | \leq  \ \  C_1 p(e_0) $ on $\Om''$
\item  
$|\pa  \tilde A  | \leq   C_2 p(e_0)  $  and 
 $ |\tilde A  | \leq  C_2  \mu_0^{-1} p(e_0)  $ on $\Om''$
\end{enumerate}
\end{lem}

\res  The bound  $\cO(p(e_0))$ on the fluctuation field   $A_0$ is 
a substantial improvement over  $\cO(\mu_0^{-1} p(e_0))$. 
Since we are on a unit lattice we have the
same bound for  $\pa A_0$.

The last bound  follows  directly  from  (1.) and (3.), a fact we use later on.
\bigskip

\pr  The proof follows  \cite{Bal82b}.
\begin{enumerate}
\item  
Let   $\check A  =  \tilde A  +A_0$.
We have for    $y , y + Le_{\mu} \in    \Om'$
\begin{equation}
\begin{split}
| (\pa_{\mu} A_{\nu})(y) | 
 =&L|A_{\nu}(y) - A_{\nu}(y + Le_{\mu})|  \\
 \leq & L|(Q \check A_{\nu})(y) - (Q \check A_{\nu})(y + Le_{\mu})|
+ \cO(p(e_0))  \\
\leq &  \cO( p(e_0)  )
\end{split}   
\end{equation}
where we have used  the  first and third  bounds  in  (\ref{inequality1}) .
This proves the bound on $\pa A$.
The bound on  $|A|$ follows similarly from second and third bounds in 
(\ref{inequality1}).
\item  
If  $x \in B(y) $ then  $(Q^TA)(x)  =  A(y)$.
Thus we must show that  for   $x \in B(y)  \subset \Om''$ that  
\begin{equation}  \label{key}
\tilde A  (x) = A(y)  + \cO(p(e_0))
\end{equation}
We write
\begin{equation}  \label{breakh}
\tilde A (x)  =  ( H_1^{loc} A_{\Om'} )(x) =
\left(H^{loc}_11_{\Om'})(x)A(y)  +(H^{loc}_1(A_{\Om'} - 1_{\Om'} A(y))\right)(x) 
\end {equation}
The second term  in (\ref{breakh})  is  $\sum_{y' \in \Om'}H^{loc}_1(x,y)(A(y') -  A(y))$.
This vanishes unless  $x$ and hence  $y$ are in the same component
of   $\Om'$ as $y'$.  Thus we can use the estimate on  $\pa A$ to get 
$|A(y') - A(y)| \leq  d(y,y') \cO(p(e_0))$.  The term is then    $\cO(p(e_0))$. 

For the first term  in (\ref{breakh}) we write 
\begin{equation}
(H^{loc}_11_{\Om'})(x)  =
(H_11)(x) - (H_11_{(\Om')^c})(x) + ((H^{loc}_1 - H_1)1_{\Om'})(x)
\end{equation}
For the last term here   we again use lemma
 \ref{coarse.loc} which has tiny  factors  $e^{-r(e_0)}$   to dominate the
 $\cO(\mu_{0}^{-1} p(e_{0})) $ bound on $A(y)$ and give   $\cO(p(e_0))$   (or better). 
For the second term  $H_1$ connects $x \in \Om''$ to points in  $(\Om')^c$
and thus also has tiny factors from the decay of  $H_1$. 
For the first term we have
\begin{equation} 
\begin{split}H_{1}1 =&  \frac{a}{L^2} CQ^T1 = C( \frac{a}{L^2} Q^TQ)1 \\
=& 1    -  C  (-\De_{\Om} + \mu_0^2)1  
 = 1 + \cO(\mu_0^2 )\\
\end{split}
\end{equation}
Hence $(H_11 )(x) A(y) = A(y) + \cO(p(e_0))$ and hence  the result.
\item 
Now  consider the bound on $A_0=\check A  -   \tilde A $.  For  
 $x \in B(y) \subset \Om''$ we have 
\begin{equation}   \label{HA=A}
 |A_{0}(x)| 
 \leq  | \check A(x) - (Q \check A)(y)| 
 +  |( Q\check A)(y) - A(y) | + |A(y)-  \tilde A(x)|  
\end{equation} 
The first term is  $\cO(p(e_0))$ by the first   bound  in  (\ref{inequality1}),
and the second term is   $\cO(p(e_0)$ by the third bound in  (\ref{inequality1}).
 For the
last term we  use (\ref{key} ).
\item 
This follows using the bounds (\ref{inequality1}) and the bounds on $A_0$.  
But we want to also show that it follows just from (1.) and (3.).  In fact if
(1.) holds then the  identity (\ref{key}) holds.  Hence the bound on $\tilde A$ follows from 
the bound on  $A$ and   the bound on  $\pa \tilde A$  follows from the bound on $ \pa  A$.
In the latter case   look at   $(\pa_{\mu}\tilde A_{\nu})(x)= \tilde  A(x+e_{\mu})- \tilde A(x)$
 separately in the cases where $x$ and $x+ e_{\mu}$ are or are not in the same $B(y)$. 
  This completes the proof.
\end{enumerate}
\bigskip

Now introduce new characteristic
functions  enforcing the conditions on  $A, A_0$ by defining 
\begin{equation}  \label{chidef}
\begin{split}
\tilde \chi_1 (\Om'',A)  = &  \prod_{x  \in \Om''}  \chi\left(\frac{ |\pa  A(x)| }{2  C_1p(e_0) } \right)
 \prod_{x  \in \Om''}  \chi\left(\frac{ |  A(x)| }{2  C_1 \mu_0^{-1}p(e_0) } \right)
\\  
 \chi^*(\Om'', A_0)  = &  \prod_{x  \in \Om''}  \chi\left(\frac{ |  A_0(x)| }{2  C_1 p(e_0) }
\right)\\  
\end{split}
\end{equation}
and inserting 
$\tilde \chi_1 (\Om'',A ) \chi^* (\Om'', A_0)$  under the integral sign in (\ref{rep2}).

Consider the old factor
\begin{equation} 
 \chi_{0}(\Om,A,\tilde A_{\Om'} +A_0  )
= \chi_{0}(\Om-\Om'',A, \tilde A_{\Om'}+A_0 )
 \chi_{0}(\Om'',A,\tilde A + A_0 )
\end{equation}
With the new characteristic functions in place  it is safe to break up  the second factor
by replacing $\chi$ by   $1 + (1-\chi)$ at each point.
We get an expansion of the form    
\begin{equation}
 \chi_{0}(\Om'',A,\tilde A + A_0 )=  \sum_{\tilde \La   \subset \Om'' }
\bar \zeta(\Om'' - \tilde \La, A, \tilde A + A_0)  
\end{equation}
The sum is over unions of  $R_0$ blocks $\tilde \La \subset \Om''$  and the function
 $ \zeta(\Om'' - \tilde \La, A, \tilde A + A_0) $ forces at least one of the inequalities
(\ref{inequality1}) to be violated in each block in  $ \Om''- \tilde \La $.

Let us collect some of the  the characteristic functions into 
\begin{equation}
\begin{split}
 \chi_{0,\Om, \tilde \La} (A,A_0) =&  \chi_{0}( \Om- \Om'',A, \tilde A_{\Om'} + A_0)
\ \chi^* (\Om''-\tilde \La'', A_0)\\
 \zeta_{0,\Om,  \tilde  \La} (A,A_0)= &\zeta_{0}( \Om^c, A,A_0)\
 \bar \zeta(\Om'' - \tilde  \La, A, \tilde A + A_0)\\
\end{split}
\end{equation}
The overall  expression is now 
\begin{equation}  
\begin{split}
& \tilde  \rho_{1}(\Psi, A)  =c_0e^{ \de E_0}  \sum_{\Om, \tilde \La}  
\int d \Psi_0 d A_0 \ 
  \zeta_{0,\Om, \tilde \La} (A,A_0) \  \chi_{0,\Om, \tilde \La} (A,A_0)\ \tilde \chi_1 (\Om'',A )\  \chi^*
(\tilde \La'', A_0) \\ &\exp \left(-  \frac12  (A, \tilde  \De_{1,\Om}A) -\frac12(A_0, \De^\# A_0)  
- F^b
- W^b)
 \right) \\
&
\exp \left(- \frac{a}{L}| \Psi -Q_{e_0}( \tilde A_{\Om'}+A_0) \Psi_0|^2
+ (\bPsi_0 ,( D_{e_0}( \tilde A_{\Om'} +A_0) + m_0 ) \Psi_0 ) -v(\de m_0)
\right) \
\\
\end{split}
\end{equation}
where the sum  is restricted by  $\tilde \La  \subset \Om''$.

\subsection{potential}

The quadratic  form  for   fermions has the external field  
 $ \tilde A_{\Om'} + A_0$. We want to take out the $A_0$,
  but only in the region where we have control.
 Let $\theta_0$ be the characteristic function of  $\Om''$ and write
$  \tilde A_{\Om'} + A_0  =    A^+  + \de A^+$  where
\begin{equation}
\begin{split}
  A^+  =&(1-  \theta_{0})( \tilde A_{\Om'} + A_0)
+   \theta_{0} \tilde A \\
\de A^+ =&
  \theta_{0} A_0    \\
\end{split}
\end{equation}
Then $\de A^+ =   \cO(p(e_0)) $ and on $\Om''$  we have
    $ \pa  A^+ = \pa \tilde A  =  \cO(p(e_0)) $.  
Now  write the 
quadratic form as 
\begin{equation}
\begin{split}
& \frac{a}{L}| \Psi -Q_{e_0}(  A^+  + \de A^+ ) \Psi_0|^2
+
 (\bPsi_0 ,( D_{e_0}(   A^+  + \de A^+) + m_0  ) \Psi_0 ) \\
=& \frac{a}{L}| \Psi -Q_{e_0}( A^+ ) \Psi_0|^2
+
 (\bPsi_0 ,( D_{e_0}( A^+) + m_0  ) \Psi_0 )  + V_0(\Psi, \Psi_0,A^+, \de A^+) \\
\end{split}
\end{equation}
(This is the same  $V_0$ as before with new arguments). 
We also want to include the mass  counterterm  and  part of 
the energy counterterm  defining
\begin{equation}
 V_0' =  V_0 +\de v_0 + \de \cE_0  L^{3(M+N)}
\end{equation}
We  need a  local version.
 
\begin{lem} \label{Vlemma} The potential   $ V_0'(\Psi, \Psi_0,A^+, \de A^+)$ 
has the  expansion
 $V_0' = \sum_X V_0'(X)$  where $X$ has one or two blocks and   $V_0'(X)$ only depends on  fields in 
$X$.  With  $\de A^+  =    \cO(p(e_0))$ we have 
\begin{equation}
\|V_0'(X) \|_h \leq   \cO(e_0  p(e_0)) 
 \end{equation}
\end{lem}

 \pr  We establish the result for each part separately.  Making the conjugate 
variables explicit we define for  $M_1$-blocks $\De, \De'$
\begin{equation}
V_0(\De, \De',\bPsi,\Psi,\bPsi_0,\Psi_0,A^+, \de A^+)   =   V_0(\bPsi,\Psi,1_{\De}\bPsi_0,
1_{\De'}\Psi_0,A^+, \de A^+)
\end{equation}
  This vanishes unless
$\De, \De' $ either coincide or have a common face.  For such  
$\De, \De' $  and $X = \De \cup \De'$ define $V(X) = V(\De , \De')$.
Define  $V(X) =0$ for any other polymer $X$.  Then
\begin{equation}
V_0  = \sum_{ \De, \De' }  V_0(\De, \De')  = \sum_{X}  V_0(X)
\end{equation}
(Since  $\de A^+$ vanishes away from  $\Om''$ we have that  $V_0(X)$  vanishes
for  $X$ away from   $\Om''$).
The bound on  $\de A^+$ gives   $|\exp(ie_0 \de A^+) -1| \leq  \cO( e_0p(e_0))$
and  this leads to     $\| V_0(X) \|_h \leq   \cO(e_0 p(e_0))$.

 Next   define 
\begin{equation}
\de v_0( \De )  =   \sum_{x, y \in \De} :\bPsi_0(x) \de m_0  \Psi_0(x) :_{\hat S_0}
\end{equation}
and then  $ \de v_0  =\sum_{\De} \de v_0 ( \De) \equiv \sum_{X} \de v_0 (X)$. 
We have   $  \|\de v_0(X)  \|_h \leq  \cO(e_0^2\cdot N) \leq  \cO(e_0^2 p(e_0))$.
   We also define  $ \de \cE_0(\De) = \de \cE_0 M_1^3$ where  $\de \cE_0  = \cO(e_0^2)$
is still to be specified.   Then   
$\de \cE_0 \ L^{3(M+N)} =  \sum_{\De}\ \de \cE_0 (\De)   \equiv   \sum_{X}\ \de \cE_0 (X) $.
Thus we have our expansion with $  V_0'(X) =  V_0(X) +\de v_0(X) +   \de \cE_0(X)$.

\subsection{fermion translation}

The quadratic form for fermions is now
\begin{equation}  \label{fermionform}
 \frac{a}{L}| \Psi -Q_{e_0}( A^+ ) \Psi_0|^2
+
 (\bPsi_0 ,( D_{e_0}( A^+) + m_0  ) \Psi_0 ) 
\end{equation}
 We introduce 
\begin{equation}   \label{Hloc}
H_1^{loc}( A^+) =
\begin{cases}    
 \frac{a}{L} \ \Ga^{loc}(A^+)  Q_{e_0}( - A^+)^T  &   \textrm{ on } \Psi \\
 \frac{a}{L} \  \Ga^{loc}(A^+) ^T    Q_{e_0}(  A^+)^T  &   \textrm{ on } \bPsi \\
\end{cases}
\end{equation}
where  $\Ga^{loc}(A^+)$ is defined in   (\ref{localprop}).  
This is not well-defined  everywhere since   $\pa A^+$ may be large.
Nevertheless we can use it in the  
 transformation 
\begin{equation}   \label{ftrans}
\Psi_0 \to    \Psi_0 +[H_1^{loc}( A^+) \Psi]_{\tilde \La'} \equiv   \Psi_0 + \tilde \Psi(A^+)_{\tilde \La'}  
\end{equation}
and similarly for  $\bPsi$.
This  only depends on  $A^+$ in $\tilde \La$ where  $A^+ = \tilde A$  and  we have control over  $\pa A^+ = \pa \tilde
A$. 

We also introduce 
\begin{equation}
\tilde D_1^{loc}(A^+)  =  \frac aL  - \frac {a^2}{L^2}Q_{e_0}(A^+) \Ga^{loc}(A^+) Q_{e_0}(-A^+)^T
\end{equation}
this is well-defined when restricted to $\tilde \La'$.

\begin{lem}  \label{wf} Under the transformation (\ref{ftrans})
the quadratic form becomes
\begin {equation}
 (\bPsi,[ \tilde D^{loc}_1( A^+)]_{\tilde \La'}\Psi)
 +(\bPsi_0,  D^\#(  A^+)\Psi_0)  +   F^f + W^f
\end{equation}
Here   $F^f=F^f(\Psi, \Psi_0,A,A_0)$ is localized in   $(\tilde \La'')^{c}$  and $W^f= W^f(\Psi,
\Psi_0,A,A_0)$ has the  local expansion
$ W^f =  \sum_X W^f(X)$
where   $W^f(X)$ has fields localized in  $X$ and satisfies
\begin{equation}   
\|W^f(X)\|_h \leq  \cO(e^{- \frac 12\beta r(e_0)})e^{-  \frac 12 \beta M_1|X|_1 }  
\end{equation}
\end{lem}

\re   In the course of the proof we will want to make further restrictions to  $\tilde \La$ such 
as
 $\Ga_{\tilde \La}(A^+) =   [D^\#(  A^+)]_{\tilde \La}^{-1} $ (first restrict, then inverse) and a local
version
 $\Ga^{loc}_{\tilde \La}(A^+)$ defined by the random walk representation as in (\ref{localprop}).
  These define restricted 
operators
 $ H_{1,\tilde \La}( A^+)$  and   $ H^{loc}_{1,\tilde \La}( A^+)$ just as in  (\ref{hhh})  (\ref{Hloc}).
 Defining
these operators as zero off $\tilde \La$ they are defined everywhere.
\bigskip

\pr
 The term quadratic in $\Psi_0$  remains
$(\bPsi_0,D^\#(   A^+ ) \Psi_0) $.  
The cross terms have the form
\begin{equation}
\begin{split}
&
 - \frac{a}{L}  (Q_{e_0}(-A^+) \bPsi_0,\Psi) + 
 (\bPsi_0, D^\#(  A^+) [H_1^{loc}( A^+) \Psi]_{\tilde \La'} )\\
&  - \frac{a}{L} (\bPsi,Q_{e_0}(A^+) \Psi_0) 
 +  ([H_1^{loc}( A^+) \bPsi]_{\tilde \La'} , D^\#( A^+)  \Psi_0) 
\\
\end{split}
\end{equation}
We focus on the first line. 
Replace   $\bPsi_0$ by  $\bPsi_{0, \tilde \La}$  and the difference contributes to  $F^f$.
Next replace   $D^\#(  A^+)$ by   $D^\#(  A^+)_{\tilde \La}$  at no cost. 
Then replace  $\Psi$ by     $\Psi_{\tilde \La'}$ and 
again the difference contributes to $F^f$.  Now in  $[H_1^{loc}( A^+) \Psi_{\tilde \La'}]_{\tilde \La'}$ 
the only paths which contribute avoid the boundary of $\tilde \La$ and so we can replace it 
by   $[H_{1,\tilde \La}^{loc}( A^+) \Psi_{\tilde \La'}]_{\tilde \La'}$.   
Finally we drop the outside restriction to $\tilde \La'$ at the cost of another contribution
to  $F^b$.
        At this point the
first line has become
\begin{equation}
 - \frac{a}{L}  (Q_{e_0}(-A^+) \bPsi_{0,\tilde \La},\Psi_{\tilde \La'}) + 
 (\bPsi_{0,\tilde \La'}, D^\#(  A^+)_{\tilde \La} H_{1,\tilde \La}^{loc}( A^+)\Psi_{\tilde \La'} )
\end{equation}
If we replace   $ H_{1,\tilde \La}^{loc}( A^+)$  by  $ H_{1,\tilde \La}( A^+)$ 
 we get zero.  The difference is 
\begin{equation} W_3  =  (\bPsi_{0,\tilde \La},
 D^\#(  A^+)_{\tilde \La} ( H_{1,\tilde \La}^{loc}( A^+)- H_{1,\tilde \La}( A^+) )\Psi_{\La'} )
\end{equation}

To estimate $W_3$ we use 
$
 \Ga_{\tilde \La}( A^+)
- \Ga_{\tilde \La}^{loc}( A^+) 
=  \sum_{X}   \de \Ga_{ \tilde \La ,X}(A^+) 
$  where  $ \de \Ga_{ \tilde \La ,X}(A^+)$   is supported on  $X \times X$
and 
\begin{equation}  \label{A2}
 | \de \Ga_{ \tilde \La ,X}(A^+) |  \leq  \cO(  e^{-\beta  r(e_0)}) e^{- \beta  M_1|X|_1 }    e^{- \beta
d(x,y)}   
\end{equation}
See appendix \ref{C}.
The same holds for  $ H_{1,\tilde \La}^{loc}( A^+)- H_{1,\tilde \La}( A^+) $
 and also breaking  $D^\#(  A^+)_{\La}$ into small polymers
(c.f. lemma \ref{Vlemma}  ) gives the local representation 
$W_3 =\sum_X W_3(X)$  with 
  $\|W_3(X)\|_h \leq  \cO(e^{- \beta r(e_0)})e^{-  \frac12 \beta M_1|X|_1 }   $.

 The term quadratic in  $\Psi$  is
\begin{equation}
\begin{split}
& \frac{a}{L}(\bPsi,\Psi) 
-\frac{a}{L}(\bPsi, Q_{e_0}(A^+) [H_1^{loc}(A^+) \Psi ]_{\tilde \La'})
-\frac{a}{L}( Q_{e_0}(-A^+) [H_1^{loc}(A^+) \bPsi ]_{\tilde \La'},\Psi)  \\
 &  + ( [H_1^{loc}(A^+) \bPsi ]_{\tilde \La'},\ D^\#( A^+)  [H_1^{loc}(A^+) \Psi]_{\tilde \La'}  )
\\
\end{split}
\end{equation}
We make the replacements   $D^\#( A^+) \to D^\#( A^+)_{\tilde \La}$ and  $\Psi \to \Psi_{\tilde \La'}$
and  $ [H_1^{loc}(A^+) \Psi]_{\tilde \La'}  \ \to H_{1,\tilde \La}^{loc}(A^+) \Psi_{\tilde \La'}$ . 
 Just as 
for the cross term the difference contributes to  $F^f$.
Now replace    $ H_{1,\tilde \La}^{loc}(A^+) \Psi_{\tilde \La'}$  by 
$ H_{1,\tilde \La}(A^+) \Psi_{\tilde \La'}$   and call the difference  $W_4$.  We are left with     
\begin{equation}
 \frac{a}{L}(\bPsi_{\tilde \La'},\Psi_{\tilde \La'}) 
  - \frac{a^2}{L^2} (\bPsi_{\tilde \La'},
 Q_{e_0}(A^+) \Ga_{\tilde \La}(A^+) Q_{e_0}( - A^+)^T \Psi_{\tilde \La'})
\end{equation} 
Next replace   $ \Ga_{\tilde \La}(A^+)$ by  $ \Ga^{loc}_{\tilde \La}(A^+)$  and call the difference  $W_5$.
Since they connect points  in  $\tilde \La'$  we can replace  $ \Ga^{loc}_{\tilde \La}(A^+)$
by $ \Ga^{loc}(A^+)$ at no cost and we get the announced term 
$(\bPsi_{\tilde \La'},  \tilde D^{loc}_1(A^+) \Psi_{\tilde \La'})$

A typical term in  $W_4$ is   
\begin{equation}
 (( H^{loc}_{1,\tilde \La}(A^+)- H_{1,\tilde \La}(A^+) )
  \Psi_{\tilde \La'},\ D^\#( A^+)_{\tilde \La} 
H_{1,\tilde \La}(A^+) \Psi_{\tilde \La'} )
\end{equation}
We insert  local expansions for   $ H^{loc}_{1,\tilde \La}(A^+)- H_{1,\tilde \La}(A^+) $
as above  and similarly for $ D^\#( A^+)_{\tilde \La} $ and $H_{1,\tilde \La}(A^+)$. The 
multiple sum over
polymers is rearranged into a single sum which gives the local expansion  for this part
of  $W_4$.
The estimates follow.  The other terms in  $W_4$ are
treated similarly as is $W_5$

This completes the proof of the lemma with  $W^f = W_3 + W_4 + W_5$.
\bigskip

The full expansion is now  with  $W = W^b +W^f$
\begin{equation}  
\begin{split}
& 
\tilde  \rho_{1}(\Psi, A)  =c_0e^{- \de E_1}  \sum_{\Om, \tilde \La}  
\int d \Psi_0 d A_0 \ 
  \zeta_{0,\Om, \tilde \La} (A,A_0) \  \chi_{0,\Om, \tilde \La} (A,A_0)\ \tilde \chi_1 (\Om'',A )\  \chi^*
(\tilde \La'', A_0) \\ &
\exp \left(-  \frac12  (A, \tilde  \De_{1,\Om}A) -\frac12(A_0, \De^\# A_0)  
- F^b) \right) \\
&
\exp \left(- (\bPsi, [\tilde D^{loc}_1( A^+)]_{\tilde \La'}\Psi)
 -(\bPsi_0,  D^\#(  A^+)\Psi_0)  -   F^f 
\right) \
\\
&
\exp \left(-V'_0(\Psi,   \Psi_0  + \tilde \Psi( A^+)_{\tilde \La'}   ,A^+, \de A^+)-W \right) \\
\end{split}
\end{equation}

\subsection {spacetime split}

Now let us focus on the fluctuation integral.  We consider 
the integral in two steps, first integrating over the fields
inside  $   \La   \equiv \tilde  \La''$ by conditioning on the fields outside, and then doing 
the remaining integral. The conditional expectation can be identified with a Gaussian integral
with non-zero mean depending on external variables.   The relevant
identities can be found in appendix   \ref{C}.    When we carry out this step only 
$V_0',W$ depend on variables inside $\La$.  Taking account also that  $A^+ = \tilde A$
in  $\La$  we obtain 
\begin{equation}  \label{5}
\begin{split} 
\tilde  \rho_{1}(\Psi, A)  =&c_0e^{- \de E_1}  \sum_{\Om,\tilde \La}  
\int d \Psi_0 d A_0 \ 
  \zeta_{0,\Om,\tilde  \La} (A,A_0) \  \chi_{0,\Om,\tilde \La} (A,A_0)\   \tilde \chi_1 (\Om'',A )\\
&
\exp \left(-  \frac12  (A, \tilde  \De_{1,\Om}A) -\frac12(A_0, \De^\# A_0)  
- F^b) \right) \\
&
\exp \left(- (\bPsi,[ \tilde D^{loc}_1( A^+)]_{\tilde \La'}\Psi)
 -(\bPsi_0,  D^\#(  A^+)\Psi_0)  -   F^f 
\right) \
\\
&
\Xi_{ \La} \left(\Psi,  \Psi_{0, \La^c}, A,A_{0,\La^c} \right)\\
\end{split}
\end{equation}
where
\footnote{  The expression $ \Xi_{ \La} $ might more   accurately be 
denoted   $ \Xi_{\Om,\tilde  \La} $.  Our notation reflects the fact
that the important dependence is on  $\La = \tilde \La''$.} 
\begin{equation}  \label{fluctuation}
\begin{split}
 \Xi_{ \La} \left(\Psi,  \Psi_{0, \La^c}, A,A_{0,\La^c} \right)
 =&
\int \ d\mu_{\st{C_{ \La}, \al_{\La}}} (A_{0,\La}) \ 
 d\mu_{\st{\Ga_{\La}(\tilde A),\beta_{\La}(\tilde A)}}(\Psi_{0,\La}) \chi^*(\La,A_0)  \\ 
& \exp
\left(-V'_0(\Psi,   \Psi_0  + \tilde \Psi( A^+)_{\tilde \La'}   ,A^+, \de A^+)-W(\Psi,\Psi_0,A,A_0)  
\right)   \\
\end{split}
\end{equation}
Here the Gaussian integrals have covariances 
and means 
\begin{equation}
\begin{split}
C_{\La}  = &[\De^\# _{\La}]^{-1}\\
\al_{\La}  = & [\De^\#_{\La}]^{-1}  \De^\#_{\La \La^c}A_{0,\La^c}\\
\Ga_{\La}(\tilde A)  =& [ D^\# (\tilde A )_{\La}]^{-1}\\
\beta_{\La}(\tilde A) = & [D^\#(\tilde A)_{\La}]^{-1}  [D^\#(\tilde A)]_{\La \La^c}\Psi_{0,\La^c}\\
\end{split}
\end{equation}
Here  $\beta_{\La}(\tilde A) $ is the mean for   $\Psi_{0,\La}$.  There is a similar 
 $\bar \beta_{\La}(\tilde A) $  as the mean for   $ \bPsi_{0,\La}$.  See the appendix.
We note that the integrand in (\ref{fluctuation}) contains extensive dependence 
on variables outside  $\La$.

\subsection{backing up}

Let us define   $ \Xi^W_{\La}$ to be the same as  $ \Xi_{\La} $ but with $V_0'=0$ and 
 $\chi^* =1$.
We will see that    $\Xi^W_{\La}$ is invertible, i.e. the no-fermion part is non-zero.
Thus we can write
\begin{equation}
\Xi_{\La} =\Xi^W_{\La}\ \left( \frac{ \Xi_{\La} }{\Xi^W_{\La}} \right)  \equiv \Xi^W_{\La}
\tilde   \Xi_{\La} 
\end{equation}
Now put this in  (\ref{5})
and then undo the conditioning  with the first factor $\Xi^W_{\La}$.  The second factor
$\tilde  \Xi_{\La} $ is
not affected since it only depends on variables in  $\La^c$.
This gives 
\begin{equation}  
\begin{split} 
\tilde  \rho_{1}(\Psi, A)  =&c_0e^{- \de E_1}  \sum_{\Om,\tilde  \La}  
\int d \Psi_0 d A_0 \ 
 \zeta_{0,\Om,\tilde  \La} (A,A_0) \ \chi_{0,\Om,\tilde  \La} (A,A_0)\   \tilde \chi_1 (\Om'',A )\\
&
\exp \left(-  \frac12  (A, \tilde  \De_{1,\Om}A) -\frac12(A_0, \De^\# A_0)  
- F^b-W^b \right) \\
&
\exp \left(- (\bPsi_{\La},[ \tilde D^{loc}_1( A^+)]_{\tilde \La'}\Psi_{\La})
 -(\bPsi_0,  D^\#(  A^+)\Psi_0)  -   F^f  -W^f 
\right) \
\\
&
 \tilde  \Xi_{\La}  \left(\Psi,  \Psi_{0, \La^c}, A,A_{0,\La^c} \right)\\
\end{split}
\end{equation}

We also undo the translations. 
We  make the inverse transformation
$\Psi_0 \to    \Psi_0 - \tilde \Psi( A^+)_{\tilde \La'} $.
By lemma   \ref{wf} the second  exponential above   is changed back to
to the exponential of the original fermion quadratic form (\ref{fermionform}).
We also make the inverse transformation   $A_0\to A_0 - \tilde A_{\Om'}$
By lemma   \ref{wb} the first     exponential above   is changed back to
to the exponential of the original boson quadratic form  (\ref{bquad}) .

In addition there the following changes   changes.
The characteristic functions become
\begin{equation}
\begin{split}   \chi'_{0,\Om,\tilde  \La} (A,A_0) \equiv &
 \chi_{0,\Om,\tilde  \La} (A,A_0- \tilde A_{\Om'})\\
 =&  \chi_{0}( \Om- \Om'',A,  A_0)
\ \chi^* (\Om''-\La, A_0 - \tilde A_{\Om'})\\
 \zeta'_{0,\Om,\tilde  \La} (A,A_0)  \equiv & \zeta_{0,\Om,\tilde  \La} (A,A_0- \tilde A_{\Om'})\\
= &\zeta_{0}( \Om^c, A,A_0)\ \bar
\zeta(\Om'' -\tilde  \La, A,  A_0)\\
\end{split}
\end{equation}
 The background field for  the fermions changes from  $A^+$ to  
\begin{equation}  A^*  \equiv
(1-   \theta_0) A_0
+   \theta_0  \tilde A  
\end{equation}
Finally the fluctuation part is transformed to
\begin{equation}
\tilde   \Xi'_{ \La} \left(\Psi,  \Psi_{0, \La^c}, A,A_{0,\La^c} \right)=
\tilde   \Xi_{ \La} \left(\Psi,  \Psi_{0, \La^c}- \tilde \Psi( A^*)_{\tilde \La'-\La} , A,A_{0,\La^c} - 
\tilde A_{\Om'-\La}    \right) 
\end{equation}

Thus our expression has become
\begin{equation}  
\begin{split} 
\tilde  \rho_{1}(\Psi, A)  =&c_0e^{- \de E_1}  \sum_{\Om,\tilde  \La}  
\int d \Psi_0 d A_0 \ 
  \zeta'_{0,\Om,\tilde  \La} (A,A_0) \  \chi'_{0,\Om,\tilde  \La} (A,A_0)\   \tilde \chi_1 (\Om'',A )\\
&
\exp \left(-  \frac{a}{2L^2}  |A-QA_0|^2 - \frac12  (A_0,   (-\De  +\mu_0^2 ) A_0  ) \right) \\
&
\exp \left(- \frac{a}{L}| \Psi -Q_{e_0}( A^* ) \Psi_0|^2
-
 (\bPsi_0 ,( D_{e_0}( A^*) + m_0  ) \Psi_0 ) 
\right) \
\\
&
\tilde   \Xi'_{\La} \left(\Psi,  \Psi_{0, \La^c}, A,A_{0,\La^c} \right)\\
\end{split}
\end{equation}

\subsection{scaling}

Now we scale to   $ \rho_{1} (\Psi_1, A_1) \equiv  \tilde \rho_1 ( \Psi_{1,L},  A_{1,L})$
where    $\Psi_1, A_1$  on   $\bbT_{N+M-1}^0$   scale up to   
 $\Psi_{1,L}, A_{1,L}$ is on   $\bbT_{N+M}^{1}$.
Define   $\cH_1^{loc} =  \sigma_L^{-1} H_1^{loc} \sigma_L$.    Then  
 $\tilde A =H^{loc}_1A$  
becomes  $H^{loc}_1A_{1,L}= \cA_{1,L}$
 where  $\cA_1 =\cH^{loc}_1A_1$  and 
the background field $A^*$ becomes  
\begin{equation} 
\hat A_1 =(1- \theta_0)A_{0}
+ \theta_0 \cA_{1,L}  
\end{equation}
We also define  $ \zeta_{1,\Om,\tilde  \La} (A_{1},A_0) = \zeta'_{0,\Om,\tilde  \La} (A_{1,L},A_0) $
and   $ \chi_{1,\Om,\tilde  \La} (A_{1},A_0) = \chi'_{0,\Om,\tilde  \La} (A_{1,L},A_0) $ and 
\begin{equation}
\begin{split}
\chi_1( L^{-1} \Om'', A_1)  = &
 \tilde \chi_1(\Om'', A_{1,L})\\
 \Xi_{1, \La} \left(\Psi_{1},  \Psi_{0, \La^c}, A_{1},A_{0,\La^c} \right)
=& \tilde \Xi'_{ \La} \left(\Psi_{1,L},  \Psi_{0, \La^c}, A_{1,L},A_{0,\La^c} \right) \\
\end{split}
\end{equation}
Then  we have the final expression
\begin{equation}  \label{finalexpression}
\begin{split} 
  \rho_{1}(\Psi_1, A_1)  =&c_0e^{- \de E_1}  \sum_{\Om,\tilde  \La}  
\int d \Psi_0 d A_0 \ 
  \zeta_{1,\Om,\tilde  \La} (A_{1},A_0) \  \chi_{1,\Om,\tilde  \La} (A_{1},A_0) \\ 
&
\exp \left(-  \frac{a}{2L^2}  |A_{1,L}-QA_0|^2 - \frac12  (A_0,   (-\De  +\mu_0^2 ) A_0  ) \right) \\
&
\exp \left(- \frac{a}{L}| \Psi_{1,L} -Q_{e_0}( \hat A_1 ) \Psi_0|^2
-
 (\bPsi_0 ,( D_{e_0}( \hat A_1) + m_0  ) \Psi_0 ) 
\right) \
\\
&
 \Xi_{1, \La} \left(\Psi_{1},  \Psi_{0, \La^c}, A_{1},A_{0,\La^c} \right)   \chi_1( L^{-1} \Om'', A_1)   \\
\end{split}
\end{equation}

\newpage

\section{Fluctuation integral}

\subsection{a local expansion}

We study the fluctuation integral   $\tilde \Xi   =\Xi_{\La}/ \Xi_{\La}^W$.
First we consider  the normalization factor  $ \Xi_{\La}^W$  which is given by 
\begin{equation}
\begin{split}
 \Xi^W_{ \La} \left(\Psi,  \Psi_{0, \La^c}, A,A_{0,\La^c} \right)
 =&
\int \ d\mu_{\st{C_{ \La}, \al_{\La}}} (A_{0,\La}) \ 
d\mu_{\st{\Ga_{\La}(\tilde A),\beta_{\La}(\tilde A)}}(\Psi_{0,\La}) \\
 & \exp \left(-W(\Psi,\Psi_0,A,A_0)   \right)   \\
\end{split}
\end{equation}
This is a tiny perturbation of a Gaussian and we have the local representation:

\begin{lem}   \label{xxiw}
 Assume  $|A| \leq \cO(\mu_0^{-1} p(e_0))$ on  $\Om$ and  $|A_{0}| \leq
\mu_0^{-1} p(e_0)$ on $\Om - \La$.  Then
\begin{equation}   \label{xiw}
\Xi^W_{\La} \left(\Psi , \Psi_{0, \La^c}, A,A_{0,\La^c} \right) 
= \exp \left(- \sum_X  W^*(X,\Psi , \Psi_{0, \La^c}, A,A_{0,\La^c}) \right)
\end{equation}
where  
\begin{equation}
\|  W^*(X)  \|_{h}  \leq   \cO(e^{-  \cO(1)  \beta r(e_0)})
e^{-\one \beta M_1|X|_1} 
\end{equation}
\end{lem}
\bigskip

\re   The assumed bounds on  $A,A_0$ are  needed for control over $W$ as we have seen in lemma \ref{wb}.  Let 
us note that they still follow from  the modified characteristic functions of the 
expansion (\ref{5}).  Indeed the bound on $A$ on $\Om''$ follows from the bounds of  
$\tilde \chi_1 (\Om'',A )$ while the bound on $A$ on $\Om-\Om''$ follows from the 
bounds of $ \chi_{0}( \Om- \Om'',A, \tilde A_{\Om'} + A_0)$.  The bound on  $A_0$ on $\Om''-\La$  follows 
from the bounds of  $ \chi^* (\Om''-\La, A_0)$  while the  bound on  $A_0$ on  $\Om - \Om''$
follows from the bound   $|\tilde A_{\Om'} + A_0|  \leq    \cO(\mu_0^{-1} p(e_0))$   on  $\Om - \Om''$
and the bound on  $A$.    Also note that we maintain control after the backing-up  transformation since 
$\Xi^W_{\La}$ and the characteristic functions undergo the same transformation.
\bigskip

\pr   The expression factors into a fermion piece and a boson piece.
We discuss the boson piece in detail; fermions are similar.
The boson piece is
\begin{equation}
\begin{split}
 \Xi^{W,b}_{ \La} \left( A,A_{0,\La^c} \right)
 =&
\int \ d\mu_{\st{C_{ \La}, \al_{\La}}} (A_{0,\La}) \ 
 \exp \left(-W^b(A,A_0)   \right)   \\
  =&
\int \ d\mu_{\st{C_{ \La}}} (A_{0,\La}) \ 
 \exp \left(-W^b(A,A_0  + \al_{\La})   \right)   \\
\end{split}
\end{equation}
 We recall from lemma \ref{wb} that  
$W^b(A,A_0) =  ( A_0, w_1   A)  + \frac12  ( A_0, w_2   A_0) $ so that  
\begin{equation}  
W^b(A,A_0  + \al_{\La}) =
 (  ( A_0 + \al_{\La}), w_1 A)  + \frac12    ( (A_0 + \al_{\La}), w_2   (A_0 + \al_{\La}))   
\end{equation}
We also have   $w_{1}  =  \sum_X w_{1}(X)$ and
\begin{equation}
|w_{1}(X,x,y)| 
  \leq   \cO(1)  e^{- \cO(1) \beta  r(e_0) } e^{- \cO(1)  \beta M_1|X|_1 } e^{-\cO(1)  \beta
d(x,y)}
\end{equation}
and similarly for $w_2$.  

The terms  $W_1^* =( \al_{\La} , w_1  A )  + \frac12 ( \al_{\La}, w_2   \al_{\La})   $
come outside the integral.   From lemma \ref{coarse} we have   $C_{\La}  =  \sum_X C_{\La}(X)$
 and hence  $\al_{\La}  =  \sum_X\al_{\La}(X)$ with the bound
$|\al_{\La}(X)| \leq  \cO(p(e_0) )  e^{- \cO(1)  \beta M_1|X|_1 }$.
Use  this together  with the expansions for $w_1,w_2$  and combine into to 
single expansion to get  $ W_1^* =   \sum_X W_1^* (X)$ where $ W_1^* (X)$ 
satisfies the bound of the lemma.
The point
is that the  strong   $e^{-\cO(1) \beta  r(e_0)}$ factor from  $w_1,w_2$ compensates
the weak  $\cO(p(e_0))$  bound in    $\al_{\La}$  and the   $\mu_0^{-1} p(e_0)$ bound on  $A$.

We are left with the expression
\begin{equation}
\int \ \exp \left( -  ( A_0,f ) - \frac12   ( A_0 , w_2   A_0 )  
\right)        d\mu_{\st{C_{ \La}}} (A_{0,\La}) \ 
 \end{equation}
where   $f =   w_1A  +  w_2  \al_{\La}$.   As above we have the local expansion  $f  =  \sum_X f(X)$ 
with   $|f(X)| \leq  \cO(e^{-\cO(1)  \beta r(e_0)})
e^{-\one \beta M_1|X|_1} $.  
Next break $A_0$ into $A_{0, \La^c}$ and $A_{0, \La}$.  The terms   
$W_2^* = (  A_{0, \La^c} ,f  )  + \frac 12 ( A_{0, \La^c} , w_2   A_{0,\La^c} )  $
again have a local expansion and we are left with 
\begin{equation}
\int \ \exp \left(  - (  A_{0, \La},f'  )- \frac12    ( A_{0, \La} , w_2  A_{0, \La})  
\right)        d\mu_{\st{C_{ \La}}} (A_{0,\La}) \ 
 \end{equation}
where   $f' = f + w_2  A_{0, \La^c} $.  Note that     $f'$ again has
a local expansion with the same bounds as $f$. This Gaussian integral can be
explicitly evaluated as  
\begin{equation}  \label{eval}
\det ( 1 + w_2 C_{ \La}) ^{-\frac12} \exp \left( \frac12  (f',  (w_2+C_{ \La}^{-1} )^{-1} f')\right)
\end{equation}

Consider the first factor in (\ref{eval}).   Since  $w_2 C_{ \La}$ has a small $\cL^2$-operator norm
we can write  
\begin{equation}
\begin{split}
\det ( 1 + w_2 C_{ \La}) ^{-\frac12} = & \exp \left( -\frac12\sum_{n=1}^{\infty} \frac {(-1)^{n+1}}{n} tr ( (
w_2 C_{
\La})^n) 
\right) =\exp \left(- \sum_Z W_3^*(Z)  \right)\\
\end{split}
\end{equation}
In the second step we have inserted  the local expansions for      $w_2 ,C_{\La}$ 
 and defined 
\begin{equation}
 W^{*}_3(Z) = \frac12  \sum_{n=1}^{\infty} \frac {(-1)^{n+1}}{n} 
\sum_{X_1,Y_1,\dots, X_n, Y_n \to  Z} tr (  w_2(X_1) C_{ \La}(Y_1) \dots  w_2(X_n) C_{ \La}(Y_n)      )  \
\end{equation}
The sum is over sets $X_1,Y_1,\dots, X_n, Y_n $ whose union is   $Z$.  The terms in 
this sequence must intersect their neighbors, else we get no contribution.   Thus $Z$ is 
connected.
We have the estimate 
\begin{equation}
|(w_2(X)C_{\La}(Y))(x,y)   |  \leq \cO(e^{-  \cO(1) \beta r(e_0)})e^{-\one \beta M_1(|X|_1 + |Y|_1)} 
e^{- \cO(1) \beta d(x,y)}
\end{equation}
Thus the operator norm satisfies   $\|w_2(X)C_{\La}(Y) \|  \leq \cO(e^{-  \cO(1) \beta r(e_0)})
e^{-\one \beta M_1(|X|_1+ |Y|_1)} $.  The Hilbert Schmidt norm  $\|w_2(X)C_{\La}(Y)   \|_2$  satisfies the
same bound
 with a negligible
factor of 
$|X | = M_1^3 |X |_1$ and the trace norm   $\|w_2(X)C_{\La}(Y)\|_1$ satisfies the same, again with
a negligible factor.  Now using  $|tr(AB)|  \leq  \|A\|_1 \|B\|$
we obtain
\begin{equation}
 |W^*_3(Z)| \leq   \sum_{n=1}^{\infty}\sum_{X_1,Y_1,\dots, X_n, Y_n \to  Z} 
  \prod_{i=1}^n\cO(e^{-  \cO(1) \beta r(e_0)})e^{-\one \beta M_1(|X_i|_1+ |Y_i|_1)}
\end{equation}
Next extract an overall factor  $ \exp  (- \cO(1) \beta M_1 |Z|_1)$.  The sum over  $|Y_n|$
 is  then  estimated
by  
\begin{equation}
\sum_{Y_n \cap X_n \neq  \emptyset}  e^{- \cO(1) \beta M_1 |Y_n|_1}  \leq  |X_n|_1  
\end{equation}
and the  $|X_n|$ is absorbed by   the factor   $e^{- \cO(1) \beta M_1 |X_n|_1}$.  Continue
estimating the sums in this fashion.  In the last step we get a factor  $|Z|_1$ which is 
also absorbed.  Thus we end with 
\begin{equation}
  |W^*_3(Z)|\leq   e^{- \cO(1)  \beta M_1|Z|_1 }  \sum_{n=1}^{\infty} (\cO(e^{-  \cO(1) \beta r(e_0)}) )^n 
\end{equation}
 which has the bound of the lemma.

Now consider the second  factor  in  (\ref{eval})  which we write
\begin{equation}
\begin{split}
& \exp \left( \frac12  (f',  (w_2+C_{ \La}^{-1} )^{-1} f')\right) =
  \exp \left( \frac12  (f',  C_{ \La} (1 + w_2C_{ \La})^{-1}f')\right) \\
= & \exp \left( \frac12  (f',  C_{ \La}\sum_{n=0}^{\infty}( w_2 C_{\La})^n  f')\right)  
=  \exp \left( \sum_Z W_4^*(Z)  \right)\\
\end{split}
\end{equation}
In the last step  we  have   inserted local expansions and grouped terms by their localization.  
The estimate   on  $W_4^*(Z)$  is similar to the estimate on  $W_3^*(Z)$, but    now only 
 operator norms enter.  This completes the proof.

\subsection{adjustments}

Insert the expression (\ref{xiw}) for   $\Xi_{\La}^W$ into  $\tilde \Xi   =\Xi_{\La}/ \Xi_{\La}^W$ and 
put the terms under the integral sign.
Also insert the local expansions for   $V_0'$ and   $W$  and  obtain 
\begin{equation}
\tilde  \Xi_{\La}
 =
\int \exp \left(  \sum_{X}  E_0 (X) \right)
 \chi^* (\La, A_0 )\ d\mu_{\st{C_{ \La}, \al_{\La}}} (A_{0,\La}) \  d\mu_{\st{\Ga_{ \La}(\tilde A),
\beta_{\La}(\tilde A)}}(\Psi_{0,\La})
\end{equation}
where   
\begin{equation}
\begin{split}
 E_0(X)   =& - V'_0(X,\Psi,   \Psi_0  + \tilde \Psi( A^+)_{\tilde \La'}   ,A^+, \de A^+)  \\
-&  W(X,\Psi,\Psi_0,A,A_0) +  W^*(X,\Psi,\Psi_0,A,A_0) \\
\end{split}
\end{equation}

This is the main object of our attention.  Before we attack it however
we want to make some  adjustments.
We concentrate on the case  $X \subset \La'$ in which case the potential can be 
 written  
$  V'_0(X,\Psi,   \Psi_0  + \tilde \Psi( \tilde A)  ,\tilde A, A_0)$
We break it into pieces by writing (at first without the translation)
\begin{equation}
V'_0(X,\Psi,   \Psi_0 ,\tilde A, A_0)
=V^Q_0(X,\Psi,   \Psi_0   ,\tilde A, A_0) + V^D_0(X,   \Psi_0   ,\tilde A, A_0) + \de V_0(X, \Psi_0)
\end{equation}
where  $V^Q $ is the part that comes from the variation 
of   $aL^{-1}|\Psi - Q_{e_0}(A) \Psi_0|^2$ and  $V_0^D$ is the part that comes from the variation of $(\bPsi_0
, D_{e_0} (A) \Psi_0)$ and $\de V_0$ are the counterterms.  Our goal is to 
 get rid of  the $\Psi$ dependence in  
$V^Q_0$ and make it a function of  $\tilde \Psi(\tilde A)  = H_1^{loc}(\tilde A) \Psi$ only.
This will be important later on.

To accomplish this we define  for any $A$  
\begin{equation}  \label{Mdefn}
M_{\La}(A) =  
\begin{cases}    
 \frac{L}{a} \  Q_{e_0}(  A) [ D^\#( A)]_{\La}      &   \textrm{ on } \Psi \\
 \frac{L}{a} \  Q_{e_0}( - A)  [ D^\#( A)]_{\La}  &   \textrm{ on } \bPsi \\
\end{cases}
\end{equation}
which satisfies 
\begin{equation}
M_{\La}(A)H_{1,\La}(A)  = I
\end{equation}
Then we write  with  $\de Q(\tilde A, A_0)=Q_{e_0} (\tilde A + A_0) - Q_{e_0} (\tilde A)$
\begin{equation}
\begin{split}
\sum_{X \subset \La'} V^Q_0(X,\Psi,   \Psi_0   ,\tilde A, A_0) 
=&- \frac{a}{L} (\bPsi, \de Q\Psi_{0,\La'})  + \dots\\
=&- \frac{a}{L}  (M_{\La}(\tilde  A)H_{1,\La}(\tilde A)  \bPsi, \de Q\Psi_{0,\La'}) + \dots\\
=& - \frac{a}{L} (M_{\La}(\tilde A) H^{loc}_{1,\La}(\tilde A)\bPsi,\de Q\Psi_{0,\La'})+ \cdots\\
&-  \frac{a}{L} (M_{\La_*}(\tilde A) (H_{1,\La}(\tilde A)-H^{loc}_{1,\La}(\tilde A))\bPsi, \de
Q\Psi_{0,\La'}) +
\dots\\
\equiv &
\sum_{X }  \tilde  V^Q_0(X,\tilde \Psi(\tilde A), \Psi_0,\tilde A, A_0) 
+   V^{\dagger}_0(X, \Psi,\Psi_0,\tilde A, A_0 ) \\
\end{split}
\end{equation}
Here  $ + \dots$ indicates a similar term with  $\bPsi_0 \Psi$ rather than   $\bPsi \Psi_0$.
In passing to the last line in the first term we have replaced $ H^{loc}_{1,\La}(\tilde Ax,y)$
by $ H^{loc}_{1}(\tilde A,x,y) $ which is allowed since  $x,y$ are  far from
$\pa \La$. Then we have  localized  the first term in  $\tilde \Psi(\tilde A), \Psi_0$
The  second term is  localized using a local expansion for 
$H_{1,\La}(\tilde A)-H^{loc}_{1,\La}(\tilde A)$, see lemma \ref{wf} for details in
a similar case.   Also it is tiny.  
We have the estimates 
$\|\tilde  V^{Q}_0(X)\|_h \leq   \cO(e_0p(e_0))e^{-  \one \beta 
M_1|X|_1} $
and  $\| V^{\dagger}_0(X)\|_h \leq   \cO(e^{- \one \beta r(e_0)})e^{-  \one \beta 
M_1|X|_1} 
$.

Having made this rearrangement in  $\La'$ the integrand is now 
$\exp( \sum_X  E'_0(X) )$  where $E_0'(X)= 
E_0'(X,\tilde \Psi(A^+),\Psi, \Psi_{0, \La^c},  \tilde A, A, A_{0,
\La^c} ) $ is given by 
\begin{equation}
E_0'(X) =  \begin{cases}
E_0(X)-V^\#(X)   &     X \cap  (\La')^c  \neq \emptyset   \\
-V_0^*(X) + R_0(X)  &   X \subset \La'  \\
\end{cases}
\end{equation}
where 
\begin{equation}
\begin{split}
V^\#(X)  = &[\tilde  V^Q_0(X)  + V^{\dagger}_0(X)]_{\Psi_0 =\Psi_0+ \Psi(\tilde A)}\\
V_0^*(X)   =& [ \tilde  V^Q_0(X)   +  V^D_0(X)  + \de V_0(X)]_{\Psi_0 =\Psi_0+ \Psi(\tilde A)}\\
R_0(X)  =&  [ -V^{\dagger}_0(X)]_{\Psi_0 =\Psi_0+ \Psi(\tilde A)} - W(X) +  W^*(X) \\
\end{split}
\end{equation}
The important term is    $V_0^*$ and we note that it is now a  
function  only  of the variables we want.

Before the translation we have bounds on all these functions.  The effect 
of the translation $\Psi_0  \to \Psi_0+ \Psi(\tilde A)$ is to lower
the value of $h=\cO(1)$ to $h'= \cO(1)$ satisfying  $h' (1 +  \|\cH_1^{loc} \|^{(2)}) \leq h$.
See lemma  \ref{cv} in appendix \ref{B}.  Thus we have the following bounds
\begin{equation}  \label{primebd}
\begin{split}
\|V_0^*(X)\|_{h'} \leq &  \cO(e_0 p(e_0))e^{- \one \beta M_1|X|_1} \\
\|R_0(X)\|_{h'} \leq &  \cO(e^{- \one \beta r(e_0)})e^{-  \one \beta M_1|X|_1} \\
\|E_0'(X)\|_{h'} \leq &  \cO(e_0 p(e_0))e^{- \one \beta M_1|X|_1}  \\
\end{split}
\end{equation}

\subsection{cluster expansion}

The fluctuation integral is now
\begin{equation}
\tilde  \Xi_{\La}
 =
\int \exp \left(  \sum_{X}  E_0' (X) \right)
 \chi^*(\La,A_0 )\ d\mu_{\st{C_{ \La}, \al_{\La}}} (A_{0,\La}) \  d\mu_{\st{\Ga_{ \La}(\tilde A),
\beta_{\La}(\tilde A)}}(\Psi_{0,\La})
\end{equation}
Our  cluster expansion expresses this as a sum of local parts, but only well inside 
$\La$.  Cluster expansions similar to ours appear in  \cite{GJS73}, \cite{BBIJ84}, \cite{Bal87}.

\begin{thm}  \label{one}
 Let $e_0$ be sufficiently small (depending on $L,M_0,M_1$).  Also let    $|\pa A| \leq \cO( p(e_0))$
on  $\Om''$ and     $ |A| \leq  \cO( \mu_0^{-1}p(e_0) ) $ on  $\Om$  and
  $|A_0|  \leq  p(e_0)$ on  $\Om-\La$.  Then
\begin{equation}  \label{full}
\tilde \Xi_{\La}
=
\sum_{\Theta \subset \La'}  \cT(\Theta^c)
\exp \left(\sum_{ X \subset   \Theta }  \tilde E (X) \right)
 \end{equation}
where 
\begin{enumerate}
\item  $\tilde E (X)  = \tilde E (X,\tilde \Psi(\tilde A), \Psi , \tilde A, A)$
only depends on the 
indicated fields   in  $X$ (and is independent of  $\Psi_{0,\La^c}, A_{0, \La^c}$).
There is a constant $h_1= \cO(1)$ and a universal constant  $\ka $ such that 
\begin{equation}  
\|  \tilde E (X)  \|_{h_1}  \leq    \cO(e_0 p(e_0)) e^{-\ka  |X|_1}  
\end{equation}
\item 
Let   $\{ \Theta^c_{\beta}\}$ be the connected components of $\Theta^c$. The  the sum over $\Theta$
is further
restricted by the constraint that each $\Theta^c_{\beta}$ contain a connected 
component of  $\La^c $. Also we  have the factorization
$ \cT(\Theta^c) = \prod_{\beta} \cT(\Theta^c_{\beta}) $
and the bound 
\begin{equation}
\| \cT(\Theta^c_{\beta})\|_{h_1}  \leq   
 e^{  \cO(1)|\Theta^c_{\beta} - \La' |_1}
 e^{-\ka   |\Theta^c_{\beta} \cap  \La'|_1} 
\end{equation}
\end{enumerate} 
\end{thm}
\bigskip
  
\res
\begin{enumerate}
\item   The assumed bounds  on $A,A_0$  still hold in our expansion.  See the remark
following lemma \ref{xxiw}.
\item
 We have retreated from  decay in the linear size $M_1|X|_1$ to just decay in  $|X|_1$.  Nevertheless the
decay  in $\cT(\Theta)$ will be  sufficient for convergence of the sum over $\Theta$.
\item  The function
$\cT(\Theta^c_{\beta}) = \cT(\Theta^c_{\beta},\tilde \Psi(A^+),\Psi, \Psi_{0, \La^c},  \tilde A, A, A_{0, \La^c}) $
only depends on the 
indicated fields   in $\Theta^c_{\beta}$. Both $\cT(\Theta)$ and $\tilde E(X)$ depend on
$\Om, \La$. 
\end{enumerate}
 \bigskip

\pr We suppress external variables throughout.
\bigskip

\noindent  \textbf{part I}:
If $X$ is contained in  $\La^c$ then $E'_0(X)$ does not depend on 
$\Psi_{0,\La}, A_{0, \La}$ and we can take these terms outside the integral.
 It remains to consider terms intersecting  $\La$.

We make a Mayer expansion and write
\begin {equation}
\begin{split}
&\exp ( \sum_{X: X \cap \La \neq \emptyset} E'_0(X) ) = \prod_X e^{E'_0(X)}  \\
 =&  \prod_X   \left( (e^{E'_0(X)}   -1) +1 \right) 
= \sum_{ \{ X_i\} }  \prod_i ( e^{E'_0(X)}  -1) =  \sum_X K_0(X)
\end{split}
\end{equation}
Here  the sum over $ \{ X_i\}$ is a sum  over collections of  distinct subsets intersecting  $\La$.  
In the last step we have grouped to together terms with the same  $X = \cup_i X_i$ and defined
\begin{equation}  \label{kzero}
K_0(X)  =  \sum_{ \{ X_i\}: \cup_i X_i = X }  \prod_i ( e^{E'_0(X_i)} -1) 
\end{equation}
Note that if  $\{ X_{\al} \}$ are the connected components of $X$, then   $K_0(X)  =  \prod_{\al}  K(X_{\al})$.

We  estimate  $K_0(X)$ for $X$ connected.
Under our assumptions on $A,A_{0,\La^c}$ together with the bound  
$|A_{0,\La}| \leq \cO(p(e_0))$ from  $\chi^*(\La, A_0)$ the bound (\ref{primebd})
on  $E_0'(X)$ holds.  Then for $X$ connected we have  
\begin{equation}  \label{k0}
\begin{split}
&\| K_0 (X) \|_{h'}  \leq  
 \sum_{ \{ X_i\}: \cup_i X_i = X }  \prod_i \| e^{E'_0(X_i)} -1\|_{h'} \\
& \leq  \sum_{ \{ X_i\}: \cup_i X_i = X }  \prod_i \cO(e_0 p(e_0))e^{- \one \beta M_1|X_i|_1} 
\leq   \cO(e_0 p(e_0)) e^{-\one \beta M_1|X|_1} \\
\end{split}
\end{equation}
In the last step the origin of the small factors is clear.  For the convergence of 
the sum we use that for collections  $\{X_i\}$ of connected  subsets of $X$, any $\al$,  and $\ka$ large enough
\begin{equation}   \label{productsum}
\begin{split}
&\sum_{ \{ X_i\} }   \prod_i \al e^{-\ka|X_i|}
\leq   \sum_n \frac{1}{n!} \sum_{(X_1, \dots, X_n)}   \prod_i \al e^{-\ka|X_i|} \\
&=  \exp(  \sum_{X' \subset X}\al e^{-\ka|X'|} )  \leq       \exp (\al |X|_1)\\
\end{split}
\end{equation}

We also remove the characteristic functions in  $\La-X$
 where they are not needed
We write
\begin{equation}
 \chi^*(\La,A_0)  =   \chi^*  (\La \cap X, A_0 ) \sum_{P \subset \La-X} \zeta^*(P,A_0) 
\end{equation}
Here   $\zeta^*(P,A_0) $ enforces that  every block in  $P$ has at least 
one point where the inequality  $|A_0|  \leq 4 C_1 p(e_0)$ is violated.
Explicitly 
\begin{equation}
\zeta^*(P,A_0) =  \sum_{Q:  \bar Q^{M_1} =  P} \ \ \  \prod_{x \in Q}
(\chi\left( \frac{|A_0(x)|}{2C_1p(e_0)} \right) -1)
\end{equation}
Here   $Q$ is a subset of lattice points in  P and    $\bar Q^{M_1}$ is the smallest 
union of  $M_1$ blocks containing $Q$.
The purpose of this  step is to arrange that every block  $\De$ either has nothing in
it, or something small.

Now we need to analyze:
\begin{equation}  \label{expression1}
\begin{split}
&  \sum_{X,P}  \int 
 K_0(X, \Psi_0, A_0)\ \chi^*(\La \cap X, A_0)\ \zeta^*(P,A_0) \ d\mu_{\st{C_{ \La}, \al_{\La}}} (A_0) \ 
d\mu_{\st{\Ga_{ \La}(\tilde A),
\beta_{\La}(\tilde A)}}(\Psi_0)\
 \\
\end{split}
\end{equation}

\noindent
\textbf{part II.}:  To  break up the integral we have to break up 
the covariances $C_{\La}$
and   $\Ga_{\La}(\tilde A)$.
This will 
be accomplished using the random walk expansion.

 We introduce a variable   $s=  \{ s_{\De}  \}$ with  $ 0 \leq s_{\De}  \leq 1$ for every  $M_1$ cube $\De$.
If  $\om  =(j_0,j_1,\dots, j_n)$ is a  path  in the $M_0$ lattice  with localization
domain $  ( \cO_{j_0}, \cO_{j_1}, \cdots, \cO_{j_n})$ 
we define 
\begin{equation}
s_{\om}  = \prod_{\De:  \De \cap  ( \cO_{j_1} \cup \dots \cup \cO_{j_n}) \neq \emptyset }
  s_{\De}
\end{equation}
Note that   $\cO_{j_0}$ is omitted. 
Hence if $\om$ is only the  single point  $\{j_0\}$  (i.e. $\ell(\om)=0$) the product is
empty and in this case  we set   $s_{\om} =1$.
Now we define 
\begin{equation}  \label{rws}
\begin{split}
C_{\La}(s) = &   \sum_{\om} s_{\om}  C_{\La,\om}\\
\Ga_{\La}(s,\tilde A)=& \sum_{\om}s_{\om}  \Ga_{\La,\om}(\tilde A)\\   
\end{split}
\end{equation}
When all variables $s_{\De}=1$ we recover the original operators  $C_{\La},\Ga_{\La}(\tilde A)$. 
If all the $s_{\De}=0$ we
have  only paths with  $\ell(\om)=0$  and have  the totally decoupled operators
\begin{equation}
\begin{split}
C_{\La}(0) = &   C^*_{\La} \equiv  \sum_j h_j  C_{\cO_j} h_j\\
\Ga_{\La}(0,\tilde A)=& \Ga^*_{\La}(\tilde A) \equiv \sum_j h_j  \Ga_{\cO_j}(\tilde A) h_j\\   
\end{split}
\end{equation}
Now we write
\begin{equation}
\begin{split}
C_{\La}(s) = &C_{\La}(0)+\de C_{\La}(s)  \\
\Ga_{\La}(s,\tilde A)=& \Ga_{\La}(0,\tilde A)  + \de \Ga_{\La}(s,\tilde A)\\   
\end{split}
\end{equation}
One can show  $ \De_{\cO}^\# \leq  \cO(L^3)$. It follows that  
 $\cC_{\cO }= [ \De_{\cO}^\# ]^{-1}\geq  \cO(L^{-3})$.
and hence  $C_{\La}(0) \geq   \cO(L^{-3})$.  On the other hand   
by the bounds of lemma  \ref{randomlem} in appendix  \ref{A}  we have  
 $\| \de C_{\La}(s)\|   \leq   \cO(M_0^{-1})$. 
Since $M_0$ is  assumed larger than  $L$ we see that  $C_{\La}(s)$ is positive.

We introduce the $s$ parameters into  (\ref{expression1}), replacing 
  $ C_{\La},\Ga_{\La}(\tilde A )$  by $C_{\La}(s),\Ga_{\La}(s,\tilde A)$.
This includes replacing   $\al_{\La}, \beta_{\La}(\tilde A)$ by
\begin{equation}
\begin{split}
\al_{\La}(s)  = & C_{\La}(s)  \De^\#_{\La \La^c}A_{0,\La^c}\\
\beta_{\La}(s,\tilde A) = & \Ga_{\La}(s,\tilde A)  [D^\#(A^+)]_{\La \La^c}\Psi_{0,\La^c}\\
\end{split}
\end{equation}
We have the expression for $s_{\De}=1$ and we study it by expanding around
$s_{\De}=0$,  but only for  $\De$ in 
$\La' - ( X  \cup P) $,  essentially the region with no contribution to the integrand.  
(We leave  $\La - \La'$  alone to avoid trouble with the fact that  $\al_{\La}$ is 
not small enough here.)  Now (\ref{expression1}) can be written
\begin{equation}  \label{expression2}
\begin{split}
&  \sum_{ X,P,Y  } \int ds_Y \frac{\pa }{ \pa s_Y}[ \int 
  K_0(X)
\chi^*(\La \cap X)   \zeta^*(P)  d\mu_{\st{C_{ \La}(s), \al_{\La}(s)}} (A_0) \  d\mu_{\st{\Ga_{\La}(s,\tilde
A),
\beta_{\La}(s,\tilde A)}}(\Psi_0)\ ]_{s_{Y^c} =0} 
 \\
& = \sum_Z  \tilde K(Z)  \\
\end{split}
\end{equation}
where   $s_Y  =  \{ s_{\De}\}_{\De \subset Y} $.
The sum is over  $Y \subset  \La'-(X \cup P)$ and $Y^c$ is the complement in this
set  so that  $s_{Y^c} =0$  is really  $s_{\La' - (X \cup P \cup Y)} =0$. 
 In the second step we have defined
$\tilde K(Z)$ to be given by the same expression,
 but with the sum  restricted to  $X \cup P \cup Y \cup  (\La - \La')= Z$

Now let $Z_{\ell}$ be the connected components of $Z$.  We claim that
 $\tilde K(Z) = \prod_{\ell}  \tilde K(Z_{\ell})$.
Since   $K_0(X)$ factors over the connected components of
$X$  it factors over the $\{ Z_{\ell}  \}$, and this is true for the entire integrand.   
Next consider the random walk expansions (\ref{rws}) for 
 $C_{ \La}(s), \Ga_{\La}(s,\tilde A)$.
Since  $s_{\La' - Z}=0$  
paths connecting different  $Z_{\ell}$ do not contribute.
 Hence  these operators do not connect different components.
Hence the Gaussian integrals factor over the connected components.
In each component  the measures can be taken  
as 
\begin{equation}
\begin{split} 
d \mu_{\La,Z_{\ell},s}  (A_0)
\equiv & d \mu_{\st{[C_{ \La}(s)]_{\La \cap Z_{\ell}}, [\al_{\La}(s)]_{\La \cap Z_{\ell}}}}
(A_{0, \La \cap Z_{\ell}})\\
 d \mu_{\La,Z_{\ell},s} (\Psi_0)
 \equiv &d  \mu_{\st{[\Ga_{\La}(s,\tilde A)]_{\La \cap Z_{\ell}},[\beta_{\La}(s,\tilde A)]_{\La \cap Z_{\ell}}}}
(\Psi_{0, \La \cap Z_{\ell}})\\ 
\end{split}
\end{equation}
The derivatives and integrals with respect to  $s_Y$ preserve 
the factorization since expressions like $[C_{ \La}(s)]_{\La \cap Z_{\ell}}$ only depend on  $s_Y$
for  $Y \subset Z_{\ell}$.  Finally the sum  over $X,P,Y$
factorizes as well.

To summarize our expression is 
\begin{equation}
\sum_Z \tilde K(Z)  =\sum_{\{ Z_{\ell} \}} \prod_{\ell}  \tilde K(Z_{\ell}) 
\end{equation}
where  the sum is over disjoint connected $\{ Z_{\ell}\}$ intersecting $\La $ and 
for any such connected  $Z=Z_{\ell} \neq  \emptyset$
\begin{equation}  \label{expression5}
\begin{split}
&\tilde K(Z)  = \sum_{ X,P,Y \to Z } \int ds_Y \frac{\pa }{ \pa s_Y}
[ \int  K_0(X)\ \chi^*(\La \cap X)\   \zeta^*(P)
\ d \mu_{\La,Z,s} (A_0) \  d \mu_{\La,Z,s} (\Psi_0)\ ] \\
\end{split}
\end{equation}
Here  $X,P,Y \to Z$ means 
$X \cap \La \neq \emptyset$ (unless $X = \emptyset$) and $P \subset (\La' -  X)$ and  
 $ Y  \subset  \La' - (X \cup P )$ and  $Z$ is the union of 
$X \cup P \cup Y$ and any connected components of  $\La-\La'$
touching this set.  (Or we could write  $P \subset (\La'\cap Z -  X)$, etc.) 

We note the following features:
\begin{itemize}
\item
In  $\tilde K(Z)$ all variables are localized in  $Z$.  
\item 
If $Z \subset \La'$ then  $Z$ contains no part of $\La-\La'$  and so  
$ [\al_{\La}(s)]_{ Z}=[\beta_{\La}(s,\tilde A)]_{ Z}=0$.
It follows that in this case   $\tilde K(Z) $ does not depend on
  $\Psi_{0, \La^c}, A_{0,\La^c}$.   
\item  We can rule out     $X \cup P = \emptyset$ since in this case $Y \neq \emptyset $
and we have   $ \pa/ \pa s_Y [1] = 0$.
\end{itemize}
\bigskip

\textbf{part III}.  We now digress to estimate  
$\tilde K(Z)  = \tilde K(Z, \tilde \Psi(A^+),\Psi, \Psi_{0, \La^c},  \tilde A, A, A_{0, \La^c} )$
in a series of  lemmas.   We break the analysis into three parts  
by writing
\begin{equation}
\begin{split}
\tilde K(Z) =&  \sum_{ X,P, Y \to Z }  \int ds_Y   \frac{\pa }{ \pa s_Y} F(X,P,s) \\
   F(X,P,s) =&
  \int  G(X,s)\ \chi^*(\La \cap X)\   \zeta^*(P)\ \ d \mu_{\La,Z,s} (A_0) \\
G(X,s) =& \int    K_0(X) 
 \  \  d \mu_{\La,Z,s} (\Psi_0)\
 \\
\end{split}
\end{equation}

\begin{lem}
The function  $G(X,A_0,s)$ is analytic in complex   $|A_0(x)| \leq  \one p(e_0)$ for  $x \in \La$ 
and satisfies in this domain
\begin{equation}
\|   \frac{\pa }{ \pa s_Y}  G(X,A_0,s)  \|_{h_1}  \leq
 M_0^{- \frac12|Y|_1} \prod_{\al}   \cO(e_0p(e_0))  e^{ -\cO(1) \beta M_1|X_{\al}|_1)} 
\end{equation}
where  $\{X_{\al}\}$ are the connected components of  $X$.
\end{lem}

\pr First we give the bound with  $Y = \emptyset$ and $A_0$ real. The  integral  be written 
in more detail as 
\begin{equation}
G(X,A_0,s) = \int   K_0(X,\ \Psi_0 + [\beta_{\La}(s,\tilde A)]_{\La \cap Z},\  A_0) 
 \  d\mu_{\st{[\Ga_{\La}(s,\tilde A)]_{\La \cap Z}}}(\Psi_0)  
\end{equation}
We have the bound   $|\Ga_{\La}(s,\tilde A, x,y)| \leq   \cO(1) exp(- \beta d(x,y))$ uniformly in $s$.  
This is proved in lemma  \ref{decaylem} in Appendix \ref{A} for  $s=1$ and the same proof holds
for general $s$.   Then   $\| \Ga_{\La}(s,\tilde A) \|^{(1)}$
 and $\| \Ga_{\La}(s,\tilde A) \|^{(2)}$  defined in  (\ref{1norm}) and (\ref{2norm}) are bounded 
uniformly in  $s$.
Now we have 
\begin{equation}
\begin{split}
\| G(X,A_0,s) \|_{h_1}  
 \leq   &    \| K_0(X,\ \Psi_0 + [\beta_{\La}(s,\tilde A)]_{\La \cap Z}, \ A_0)
\|_{h_1}\\
 \leq  & \| K_0(X,\Psi_0, A_0 ) \|_{h'}\\
\leq  &   \prod_{\al}   \cO(e_0p(e_0)  ) e^{ -\cO(1) \beta M_1|X_{\al}|_1} \\
\end{split}
\end{equation}
The first estimate follows from lemma  \ref{intest2} in Appendix \ref{B} and holds  provided we chose 
$h_1$ so that 
$ \sup_s \|\Ga_{\La}(s,\tilde A)\|^{(2)}  \leq
h_1^2$. The second estimate follows from    lemma  \ref{cv} in Appendix \ref{B} and holds provided 
$h_1( 1 + \sup_s\| \Ga_{\La}(s,\tilde A)  [D^\#(A^+)]_{\La \La^c}\|^{(1)}) \leq h'$, a condition 
on $h'$.
The last estimate is the bound (\ref{k0}).

The above estimates were carried out for real $s_{\De}$ with   $|s_{\De}|  \leq  1$ but we could 
have allowed complex $s_{\De}$ with    $|s_{\De}|  \leq  M_0^{\frac12}$.   In this case
we have   $|s_{\om}|  \leq  M_0^{(\ell(\om)+1)/2}$ instead of   $|s_{\om}|  \leq  1$   but this
does not spoil the estimates on   $\Ga_{\La}(s, \tilde A)$.  In fact $\Ga_{\La}(s, \tilde A)$  is analytic in 
this domain and  so is    $G(X,A_0,s)$.
The result for $Y \neq  \emptyset $ now follows by a Cauchy bound.

The analyticity in  $A_0$ holds for  $V,W$  hence for  $E_0,E_0', K_0$ 
and hence for  $G$.  The bounds and analysis are not affected.  This completes the 
proof.

\begin{lem}
The function $ F(X,P,s)$ satisfies 
\begin{equation}  \label{fest}
\| \frac{\pa }{ \pa s_Y} F(X,P,s)  \|_{h_0} 
\leq  M_0^{- \one |Y|_1} e^{-\one p(e_0)^2|P|_1}
 \left(  \prod_{\al}   \cO(e_0 p(e_0)) e^{-\cO(1) \beta M_1|X_{\al}|_1}\right) 
\end{equation}
with the first and third  $\cO(1)$ universal.
\end{lem}

\pr  First consider the bound with  $Y = \emptyset$.  The characteristic function  $\chi^*(X)$
ensures that the bounds of the previous lemma are applicable and we have  
\begin{equation}
 \|  F(X,P,s) \|_{h_1}   \leq   \left(  \int   \zeta^*(P)   \ d \mu_{\La,Z,s} (A_0)  \right) (\dots)
\end{equation}
where $(\dots)$ is the parenthetic expression in (\ref{fest}).
For  $\zeta^*(P)$ we  note 
that 
for any  $\gamma$  
\begin{equation}
(\chi\left( \frac{|A_0(x)|}{2C_1p(e_0)} \right) -1)
\leq   e^{  -16c_1^2\ga p(e_0)^2 + \ga  |A_0(x)|^2}  (\chi\left( \frac{|A_0(x)|}{2C_1p(e_0)} \right) -1)
\end{equation}
We can use this bound  at least once in each block of $P$ and  obtain
\begin{equation}  \label{zeta}
|\zeta^*(P)|\leq   e^{  - \one  p(e_0)^2|P|_1 + \ga  \|A_0\|_P^2}
\end{equation}
Thus we have the small factors we need and it remains to control 
\begin{equation}
\begin{split}
 \int   e^{  \ga  \|A_0\|_P^2}  \ d \mu_{\La,Z,s} (A_0)  =  &
 \int   e^{  \ga  \|A_0+ \al_{\La}(s)\|_P^2}      d \mu_{\st{C_{ \La}(s)} }(A_0) \\
\leq &    e^{ 2 \ga  \| \al_{\La}(s)\|_P^2}   
 \int   e^{ 2 \ga  \|A_0\|_P^2}      d \mu_{\st{C_{ \La}(s)} }(A_0) \\
\end{split}
\end{equation}
We first estimate $  \| \al_{\La}(s)\|_P^2$.
As in lemma  \ref{decaylem} in Appendix \ref{A}  we have the bound
 $|C_{\La}(s,\tilde A, x,y)| \leq   \cO(1) exp(- \beta d(x,y))$ uniformly in $s$.
Since we have been careful to separate  $P$ by a distance at least $r(e_0)$
from  $\La^c$  we  gain a factor   $e^{- \frac12 \beta r(e_0)}$ and thus in   
\begin{equation}
 \| \al_{\La}(s)\|_P^2  \leq   \cO(e^{- \frac12 \beta r(e_0)}p(e_0))|P|
\end{equation}
the dangerous factor $p(e_0)$  from  $A_0$ on   $\pa \La$ is controlled.
On the other hand  for  $\ga$ sufficiently small
\begin{equation}
 \int   e^{ 2 \ga  \|A_0\|_P^2}      d \mu_{\st{C_{ \La}(s)} }(A_0) \
=  \det  ( 1 -4\ga \chi_P C_{ \La}(s))^{-\frac12}  \leq  e^{ \cO(1) \ga |P|}
\end{equation} 
follows from  $\|4\ga \chi_P C_{ \La}(s)\| \leq  \cO(1) \ga$ and 
$\|4\ga \chi_P C_{ \La}(s)\|_1 \leq \cO(1) \ga |P|$.
Thus we have
\begin{equation}  \label{Pint}
 \int   e^{  \ga  \|A_0\|_P^2}  \ d \mu_{\La,Z,s} (A_0)
\leq  e^{  \cO(1) \ga    |P|} \leq  e^{ \cO(1) |P|_1}
\end{equation}
which can be absorbed by the tiny factor $e^{  - \one  p(e_0)^2|P|_1}$

Now consider  $Y \neq  \emptyset$.
 Unlike  the previous lemma we do
not have analyticity in  $s$  and cannot use a Cauchy bounds.
Instead we evaluate
the derivatives explicitly.
In general a Gaussian integral with covariance  $C(s)$ and mean $\al(s)$ satisfies
 \begin{equation} \label{Lidentity}
\frac{d}{ds}  \int f   \  d\mu_{\st{C(s), \al(s)}} 
=  \int ( \cL'(s)  f )  \ 
 d\mu_{\st{C(s), \al(s)}} 
\end{equation}
where  $\cL(s)$ is the  differential operator    
\begin{equation}
\cL(s)   = \frac12
\sum_{x,y} \frac{\pa} { \pa A_0(x) }    C(s,x,y)  \frac{\pa} { \pa A_0(y) }   
+  \sum_z     \al(s,z)\frac{\pa}{ \pa A_0(z) }  
\end{equation}
We apply this repeatedly and we also have derivatives acting on $G$.
Altogether we have     
\begin{equation}  \label{tough}
\begin{split}
&\frac{\pa }{ \pa s_Y} F(X,P,s)\\
= & \sum_{Y_0}\sum_n \sum_{ \{ Y_1, \dots, Y_n\}}  
\int \prod_{j=1}^n
\left( \frac {\pa \cL(s)}{\pa s_{Y_j}}\right) 
 [ \frac{\pa  G(X,s)  }{ \pa s_{Y_0}}\ \chi^*(\La \cap X)\   \zeta^*(P)] 
   \  d \mu_{\La,Z,s} (A_0)\
\end{split}
\end{equation}
where  the sum is over subsets  $Y_0 \subset Y$ and  partitions $\{ Y_1, \dots, Y_n\}$ of  $Y-Y_0$ .
Expanding the terms in  $\pa \cL(s)/\pa s_{Y_j}$  we get a sum over subsets  $\cN$ of   $(1, \dots, n)$  
and have 
\begin{equation}  \label{basic}
\begin{split}
= & \sum_{Y_0} \sum_n \sum_{ \{ Y_1, \dots, Y_n\}}  \sum_{\cN} \sum_{\{x_i,y_j\}_{j \in \cN}} 
\sum_{\{ z_j\}_ {j \notin \cN}}    
   \prod_{j \in \cN }  \frac12     \frac {\pa C(s)}{\pa s_{Y_j}}(x_j,y_j)
  \prod_{j \notin \cN}  \frac {\pa \al_{\La}(s) }{\pa s_{Y_j}}(z_j) \\
&
\int \prod_x \left( \frac{\pa}{ \pa A_0(x) } \right)^{n_x} 
  [ \frac{\pa  G(X,s)  }{ \pa s_{Y_0}}\ \chi^*(\La \cap X)\  
\zeta^*(P)]\ \ 
  \  d \mu_{\La,Z,s} (A_0)\
\end{split}
\end{equation}
Here $n_x$ is the number of times a variable $x$ occurs in $\{x_i,y_j\}_{j \in \cN}$
and   $\{ z_j\}_ {j \notin \cN}$.  We have   $n \leq \sum_x n_x \leq 2n$

We need to estimate the effect of  the derivatives with respect to  $A_0$.
To begin we have  for  $n \geq 0$
\begin{equation}
|\left( \frac{\pa}{ \pa A_0(x) }  \right)^n  \chi (\frac{A_0(x)}{2C_1 p(e_0)})|
 \leq  (n!)^2   \left( \frac { \one}{  p(e_0)} \right)^n 
\end{equation}
and hence   
\begin{equation}
|\prod_x \left( \frac{\pa}{ \pa A_0 }  \right)^{n_x}  \chi^* (\La \cap X)|
 \leq  \prod_x  (n_x!)^2     \left( \frac { \one}{  p(e_0)} \right)^{n_x} 
\end{equation}
One can obtain the same  estimate for derivatives of  $\zeta^*(P)$.
Since derivatives do not enlarge supports we can still extract decay 
in  $|P|_1$ as in  (\ref{zeta}).  Thus we have   
\begin{equation}
| \prod_x\left( \frac{\pa}{ \pa A_0(x) }  \right)^{n_x} \zeta^* (P) |  \leq 
 \prod_x [(n_x!)^2   \left( \frac { \one}{  p(e_0)} \right)^{n_x} ]\
 e^{  - \one  p(e_0)^2|P|_1 + \ga  \|A_0\|_P^2}
\end{equation}
By the analyticity result of the previous lemma and Cauchy bounds we also 
have 
\begin{equation}
\begin{split}
&|\prod_x \left( \frac{\pa}{ \pa A_0(x) } \right)^{n_x}   \frac{\pa  G(X,s)  }{ \pa s_{Y_0}}| \\
&\ \ \ \ \ \leq 
 \prod_x [n_x! \left( \frac { \one}{  p(e_0)} \right)^{n_x} ]\  M_0^{- \frac12|Y_0|_1}
 \prod_{\al}   \cO(e_0p(e_0))  e^{-\cO(1) \beta M_1|X_{\al}|_1)} \\
\end{split}
\end{equation}
 We combine the last three results and use  some elementary combinatorics to obtain
\begin{equation}   \label{toughestimate}
\begin{split}
&\prod_x \left( \frac{\pa}{ \pa A_0(x) } \right)^{n_x}   
 [ \frac{\pa  G(X,s)  }{ \pa s_{Y_0}}\ \chi^*(\La \cap X)\  \zeta(P)]\\
&\ \ \ \ \ \  \leq \prod_x  [  (n_x!)^2 ]  \left( \frac { \one}{  p(e_0)} \right)^{n}  
 e^{ \ga \|A_0\|_P^2}M_0^{  - \frac12 |Y_0|_1}  (\dots) \\
\end{split}
\end{equation}
where now  $(\dots)=   e^{  - \one  p(e_0)^2|P|_1 } \prod_{\al}   \cO(e_0p(e_0))  e^{ -\cO(1) \beta
M_1|X_{\al}|_1)} $ .

Now  using the  bound of lemma \ref{lastlem} from Appendix \ref{A} 
we obtain
\begin{equation}  \label{alsotough}
\begin{split}| 
&  \prod_{j \in \cN }       \frac {\pa C(s)}{\pa s_{Y_j}}(x_j,y_j)
  \prod_{j \notin \cN}  \frac {\pa \al_{\La}(s) }{\pa s_{Y_j}}(z_j) |  \\
\leq  & \prod_{j=1}^n \cO(1)M_0^{-\cO(1)|Y_j|_1}  \prod_{j \in \cN}   e^{- \beta  \cL(x_j,y_j, Y_j) }     
  \prod_{j \notin \cN}  e^{- \cO(1) \beta  \cL(z_j, Y_j) } \\
\end{split}
\end{equation}
where $\cL(x,y,Y)$ is the  length of the shortest tree through 
$x,y$ and the centers of the blocks in  $Y$.  Here we have used  
  $\exp( -\frac12 \beta d(Y, \La^c)) \leq  \exp
(-\frac12 \beta r(e_0))$  to suppress a factor    $p(e_0)$ in  $\al_{\La}(s)$.
 Next since the $Y_j$ associated with a particular $x$
are disjoint one can show that     
\begin{equation}
n_x! \leq \cO(1)^{n_x}  \exp \left( \sum_{j:\ x_j = x} d(x,Y_j)/M_1 +\sum_{j:\ y_j = x} d(x,Y_j)/M_1 +\sum_{j:\ z_j
= x} d(x,Y_j)/M_1\right)
\end{equation}
(see  \cite{GJS73}, \cite {BBIJ84}, \cite{DiHu00} for similar bounds)
and hence  
\begin{equation}  \label{nbd}
\prod_x n_x! \leq (\cO(1))^n  \exp \left( \sum_{j\in \cN} d(x_j,Y_j)/M_1 +\sum_{j\in \cN} d(y_j,Y_j)/M_1
 +\sum_{j \notin  \cN}d(z_j,Y_j)/M_1\right)
\end{equation}
Even after squaring this  factor can be dominated by  factors on the right side of  (\ref{alsotough}).
Combining   (\ref{toughestimate}), (\ref{alsotough}), (\ref{nbd})  we can now bound (\ref{basic}) by
\begin{equation}  
\begin{split}
 & \sum_{Y_0}\sum_n   \left( \frac { \one}{  p(e_0)} \right)^{n}   \sum_{ \{ Y_1, \dots, Y_n\}}  \sum_{\cN}
\sum_{\{x_i,y_j\}_{j \in \cN}} 
\sum_{\{ z_j\}_ {j \notin \cN}}  
  \\
&  
 \prod_{j=0}^n  M_0^{-\cO(1)|Y_j|_1}  \prod_{j \in \cN}   e^{-  \one \beta  \cL(x_j,y_j, Y_j) }     
  \prod_{j \notin \cN}  e^{- \one \beta  \cL(z_j, Y_j) }  
   \left( \int  e^{   \gamma\|A_0\|_P^2}
   \  d \mu_{\La,Z,s} (A_0)\ \right)  (\dots) \ \ 
 \end{split}
\end{equation}
We do the sum over $x_j,y_j$  and  $z_j$ 
 we    get a factor   $\prod_j \cO(1) |Y_j| e^{- \cO(1) \beta\cL(Y_j)}$.
The $|Y_j|$ can be absorbed by the  $ M_0^{-\cO(1)|Y_j|_1}  $.
 Also we have  
$ \prod_j M_0^{-\cO(1)|Y_j|_1}  =    M_0^{-\cO(1)|Y|_1} $.  This factor is 
incorporated into    $(\dots)$  which now has the form that we want for the lemma.  
 Nothing 
depends on  $\cN$ at this point and  the sum gives a factor  $ 2^n$. 
The  
Gaussian integral is bounded by 
(\ref{Pint}) and the resulting factor 
absorbed. 
Thus our expression is bounded by 
\begin{equation}  
   \left[ \sum_{Y_0}\sum_{n}   \al^n
 \sum_{ \{ Y_1, \dots,Y_n\}}    
 \prod_{j=1}^n   e^{- \cO(1)  \beta  \cL( Y_j) }  \right]  (\dots)   
\end{equation}
where   $\al \equiv \one p(e_0)^{-1}$ is small for $e_0$ small.
We dominate the sum over partitions by an unrestricted sum 
and obtain for the bracketed expression
\begin{equation}   \label{exp}
\begin{split}
[\dots] \leq  & \sum_{Y_0} \sum_{n} \frac{\al^n}{n!}     \sum_{(Y_1,\dots, Y_n)}
  \prod^n_{j=1}  e^{- \cO(1) \beta \cL( Y_j) }  \\
=  & \sum_{Y_0}  \exp\left(  \al \sum_{Y'\subset (Y-Y_0)}
   e^{- \cO(1)  \beta \cL( Y') } \right) \\
\leq     & \sum_{Y_0}  e^{\one \al|Y-Y_0|_1}  \leq  2^{|Y|_1} e^{\one \al |Y|_1} 
\leq      e^{|Y|_1}
\\
\end{split}
\end{equation}
The second inequality is standard.    The sum here  is over $Y'$ which are not connected
we definitely need the stronger  $ e^{- \cO(1)  \beta \cL( Y') } $ decay.   
Finally the factor    $ e^{|Y|_1}$  is 
 absorbed by the  $ M_0^{-\cO(1)|Y|_1} $ in  $(\dots)$ to complete the proof.

\begin{lem}  \label{kbound}
 For any $\ka >0$, $Z$ connected, and $Z \cap \La \neq \emptyset$ we have under  the hypotheses of the
theorem
\begin{equation}
  \|\tilde K(Z) \|_{h_1} \leq   \cO(e_0 p(e_0))  e^{|Z \cap (\La-\La')|_1} e^{-\ka|Z -(\La- \La')|_1}
\end{equation}
\end{lem}
\bigskip
 \re  If $Z \subset \La'$  this becomes
\begin{equation}  \label{goodbound}
  \|\tilde K(Z) \|_{h_1} \leq   \cO(e_0 p(e_0))  e^{-\ka|Z|_1}
\end{equation}

\pr 
Assuming  $M_0,M_1, p(e_0)$ are sufficiently large we have by the previous lemma
\begin{equation}
\begin{split}
\| \tilde K(Z)\|_{h_1} \leq & 
  \sum_{ X,P,Y \to Z }  M_0^{-\one |Y|_1} e^{-\one p(e_0)^2|P|_1}  \left(  \prod_{\al}   \cO(e_0 p(e_0)) e^{-\cO(1)
\beta M_1|X_{\al}|_1} \right)  \\
\leq & \sum_{ X,P,Y \to Z } \left(  \prod_{\al}  e^{-3\ka|X_{\al}|_1} \right) e^{-3\ka |P|_1}  e^{-3\ka |Y|_1}
   \\
\leq &  e^{-2\ka|Z -(\La- \La')|_1 } \left( \sum_{X\subset Z, X_{\al}  \cap  \La \neq \emptyset  }
 \prod_{\al} e^{-\ka|X_{\al}|_1}  \right)
(\sum_{P \subset Z\cap \La' }e^{-\ka|P|_1 } )
(\sum_{Y \subset  Z\cap \La' }e^{-\ka|Y|_1 } )  \\
\leq &  e^{-2\ka|Z -(\La- \La')|_1 }e^{|Z \cap \La|_1}e^{2|Z \cap \La'|_1} \\
\leq &  e^{-\ka|Z -(\La- \La')|_1 }  e^{| Z\cap(\La -\La')|_1 }  \\
\end{split}
\end{equation}
In the third inequality we use that   $X,P,Y$ are disjoint and $X \cup P \cup Y  \supset (Z - (\La-\La'))$
 and in
the last inequality we use   $(Z - (\La-\La'))
\supset  Z \cap \La'$.
Since it is not possible that   $X \cup P = \emptyset$  we can also extract the
factor  $\cO(e_0 p(e_0))$ from either the $X$-terms or the $P$ terms. 
This completes the proof of the lemma. 
\bigskip

\noindent
\textbf{part IV}: Now we return to the proof of the theorem.
At this point we have the expression
\begin{equation}  \label{expression}
\exp \left( \sum_{X: X \subset \La^c } E'_0(X) \right)\sum_{\{ Z_{\ell} \}:  Z_{\ell} \cap \La \neq \emptyset }
\prod_{\ell}  \tilde K(Z_{\ell}) 
\end{equation}
For each collection $\{ Z_{\ell} \}$ with   $Z_{\ell} \cap \La \neq \emptyset$
consider the $\{ Z_{\ell} \}$ touching   $(\La')^c$. Take the union of  $(\La')^c$
and  the $\{ Z_{\ell} \}$ touching   $(\La')^c$, add a corridor of  $M_1$-blocks and   
  call
this the   $\Theta^c$, written   $\{ Z_{\ell} \}  \to   \Theta^c$. (So $ (\La')^c  \subset  \Theta^c$
and  $  \Theta  \subset   \La'$ ). 
 We classify the terms in the sum by the 
$\Theta $ that they determine.  The remaining  $\{ Z_{\ell} \}$  are in 
   $\Theta$    and are otherwise unrestricted.
Thus the second factor in  (\ref{expression}) can be written 
\begin{equation}
\sum_{\Theta \subset \La'}
\sum_{\{ Z_{\ell} \} \to \Theta^c}
\prod_{\ell}  \tilde K(Z_{\ell}) \left( \sum_{\{ Z'_{\ell} \}:  Z'_{\ell} \subset    \Theta } 
\prod_{\ell}  \tilde K(Z'_{\ell})  \right)
\end{equation}
and  (\ref{expression}) itself can be written  
\begin{equation}   \label{special}
\sum_{\Theta \subset \La'}
\cT(\Theta^c)
\left( \sum_{\{ Z'_{\ell} \}:  Z'_{\ell} \subset    \Theta } 
\prod_{\ell}  \tilde K(Z'_{\ell})  \right)
\end{equation}
where 
\begin{equation}   
\cT(\Theta^c) =  \exp ( \sum_{X: X \subset \La^c } E'_0(X) ) 
\sum_{\{ Z_{\ell} \} \to \Theta^c}
\prod_{\ell}  \tilde K(Z_{\ell}) \
\end{equation}

Let  $\{ \Theta_{\beta}^c \}$ be the  connected components 
of $\Theta$ .   The sum over $\{ Z_{\ell}\}$ factors over  $\{ \Theta_{\beta}^c \}$
as do  the terms  in $  \sum_{X: X \subset \La^c } E_0(X)  $ since the $X$ must be connected
and so lie in a single connected component of $\La^c$. Thus we get 
$  \cT(\Theta_{\beta}^c)=  \prod_{\beta} \cT(\Theta_{\beta}^c)$
where  
\begin{equation}
\cT(\Theta^c_{\beta}) =  \exp \left(\sum_{X: X \subset \Theta^c_{\beta} \cap  \La^c} E'_0(X) \right)
\sum_{\{ Z_{\ell} \} \to \Theta^c_{\beta}  }
\prod_{\ell}  \tilde K(Z_{\ell}) \
\end{equation}

\begin{lem}
\begin{equation}
\|\cT(\Theta^c_{\beta}) \|_{h_1}\leq e^{\one |\Theta^c_{\beta} -  \La'|_1}  e^{-\ka|\Theta^c_{\beta} \cap 
\La'|_1}
\end{equation}
\end{lem}

\pr
Using the estimates (\ref{primebd})  and  (\ref{goodbound})
we have 
\begin{equation}
\|\cT(\Theta^c_{\beta}) \|_{h_1}\leq  e^{\cO(e_0 p(e_0)| \Theta^c_{\beta} \cap \La^c |} 
\sum_{\{ Z_{\ell} \} \to \Theta^c_{\beta}  } \prod_{\ell}  \cO(e_0 p(e_0))  e^{|Z_{\ell} \cap (\La-\La')|_1}
e^{-2\ka|Z_{\ell}- ( \La - \La')|_1}
\end{equation}
The first factor is less than   $ e^{\one |\Theta^c_{\beta} -  \La'|_1} $.
Now  a factor  $e^{-\ka|Z_{\ell}- ( \La - \La')|_1}$ is less than 
$e^{-\ka|Z_{\ell}  \cap  \La'|_1}$  and after the product over
$\ell$ this gives the   $e^{-\ka|\Theta^c_{\beta} \cap  \La'|_1}$ which is the 
decay factor we want.  The other  factor satisfies $e^{-\ka|Z_{\ell}- ( \La - \La')|_1} \leq 
e^{-\ka|Z_{\ell}|}   e^{\ka|Z_{\ell} 
\cap  (\La- \La')|_1}  $.  Then we use    $\prod_{\ell}  e^{(\ka+1)|Z_{\ell}  \cap  ( \La- \La')|_1} 
=    e^{(\ka+1)|\Theta^c_{\beta}  \cap  ( \La- \La')|_1} $ and this is less than 
  $ e^{\one |\Theta^c_{\beta} - \La'|_1} $ .  We are left with the sum
\begin{equation}
\sum_{\{ Z_{\ell} \} \to \Theta^c_{\beta}  } \prod_{\ell}  \cO(e_0 p(e_0))  
e^{-\ka|Z_{\ell}|_1}
\leq   \exp  (\cO(e_0p(e_0))  |\Theta^c_{\beta}  - \La'|_1)
 \end{equation}
For this estimate we enlarge the sum to $Z_{\ell}$ touching $\Theta^c_{\beta}-\La'$
and identify an exponential  as in  (\ref{productsum}).  This completes the proof.
\bigskip

\textbf{part V}:   
This bound on  $\tilde K$ in $\Theta$ is sufficiently small that we can 
exponentiate the expression in parentheses in  (\ref{special}). 
 See \cite{Sei82}, \cite{BBIJ84} for more details of this standard argument.
We first write
\begin{equation}
\sum_{\{ Z_{\ell} \}:  Z_{\ell} \subset   \Theta } 
\prod_{\ell}  \tilde K(Z_{\ell})  = 1+  \sum_{n=1}^{\infty} \frac{1}{n!} \sum_{(Z_1,\dots,Z_n)}
\{\prod_{i<j} \zeta(Z_i,Z_j)\}\tilde K(Z_1) \dots  \tilde  K(Z_n)
\end{equation}
where the sum  over  $(Z_1,\dots,Z_n)$ is now unrestricted, but  $\zeta(Z_i,Z_j) =0$ if the sets
touch and  
$\zeta(Z_i,Z_j) =1$  if they do not.  This can be rearranged as  
\begin{equation}
\sum_{\{ Z_{\ell} \}:  Z_{\ell} \subset    \Theta } 
\prod_{\ell}  \tilde K(Z_{\ell})= \exp (\sum_{X \subset     \Theta} \tilde E(X)) 
\end{equation}
where
\begin{equation}  \label{polyexp}
\tilde E(X)  =  \sum_{n=1}^{\infty} \frac{1}{n!} \sum_{\stackrel{Z_1,..., Z_n}{\cup_i Z_i =X} }
\rho^T(Z_1,..., Z_n) \tilde K(Z_1) \dots  \tilde K(Z_n)
\end{equation}
and where 
\begin{equation}
 \rho^T(Z_1,..., Z_n)  =  \sum_G \prod_{\{i,j\} \in G} (\zeta(Z_i,Z_j) -1)
\end{equation}
Here $G$ runs over the connected graphs on   $(1,...,n)$. 

One can use  the bound (\ref{goodbound}) on $\tilde K$ and a tree domination 
argument  to show 
\begin{equation} \label{nextgoodbound}
 \|  \tilde E (X)  \|_{h_1}  \leq   \cO(e_0 p(e_0)) e^{-\ka|X|_1}     
\end{equation}
This completes the proof
of the theorem.

\subsection{more adjustments}

We  make adjustments which will simplify the treatment of perturbation
theory in the small field region.

We first want to  resum the expression in the small field region.
Accordingly we define for any $\Theta  \subset \La'$ 
\begin{equation}
 Z^{\dagger}_{\Theta}
 =
\int \exp \left(  \sum_{X \subset \Theta}  E'_0 (X) \right)
 \chi^*(\Theta, A_0 )\ d\mu_{\st{C (1,\Theta)}}(A_0) \  d\mu_{\st{\Ga(1,\Theta,\tilde
A)}}(\Psi_0)
\end{equation}
where  for  $x,y \in \Theta$ 
\begin{equation} 
C(s,\Theta,x,y)   =  \sum_{\stackrel{\om: x \to y}{\bar \om \subset \Theta}} s_{\om}  \
C_{\om}(x,y)
\end{equation}
Here  $\bar \om$ is the localization of  $\om$ defined in Appendix \ref{A}.
This is positive definite.  There is a similar 
expression  for $\Ga(s,\Theta,\tilde A)$.  
If we repeat the steps of  theorem  \ref{one} we find that    
\begin{equation} Z^{\dagger}_{\Theta} = \sum_{\{ Z_{\ell} \}:  Z_{\ell} \subset \Theta}\prod_{\ell}  
K^{\dagger}(Z_{\ell}) 
 = \exp (\sum_{X\subset \Theta}
 E^{\dagger}(X)) 
\end{equation}
where   $ E^{\dagger}(X)$ is defined from  $ K^{\dagger}( Z)$ by (\ref{polyexp}), where 
\begin{equation} 
\begin{split}
& K^{\dagger}(Z)  = \sum_{ X,P,Y \to Z } \int ds_Y \frac{\pa }{ \pa s_Y}
[ \int  K_0(X)\ \chi^*( X)\   \zeta^*(P)
d\mu_{\st{C (s,\Theta)}}(A_0) \  d\mu_{\st{\Ga(s,\Theta,\tilde A)}}(\Psi_0)] \\
\end{split}
\end{equation}
and where $K_0$ is still defined  by  (\ref{kzero}). 
The sum is over disjoint $X,P,Y$ whose union is $Z$. 
We could as well write $C (s,Z)$ and  $\Ga(s,Z,\tilde A)$
for the covariances.

 The expression
for $ K ^{\dagger} (Z)$ is  identical with 
the expression for    $\tilde  K(Z)$ specialized to the case $Z\subset
\La'$. Indeed the measures have mean zero and the same covariance,
\footnote{When $Z \subset \La'$ and $s_{\De}=0$ for $\De$ around $Z$ we have
$[C_{\La}(s)]_Z =C(s,Z)$.}
 and the 
restrictions on the sums are the same. Hence also for 
$X
\subset
\Theta$ we have 
$ E^{\dagger}(X) =  \tilde  E(X)$ and so  
we have the resummation
\begin{equation}
 \exp (\sum_{X\subset   \Theta}
 \tilde E(X))  =   Z^{\dagger}_{\Theta}
\end{equation}

Next we work toward a more  translation invariant expression which means
eliminating  dependence on  $\Om, \La,  \Theta$. We first work on
the measure and define
\begin{equation}
 Z^{\#}_{\Theta}
 =
\int \exp \left(  \sum_{X \subset \Theta}  E'_0 (X) \right)
 \chi^*(\Theta, A_0 )\ d\mu_{\st{C^{loc}}}(A_0) \  d\mu_{\st{\Ga^{loc}(\tilde A)}}(\Psi_0)
\end{equation}
Now    $C^{loc}$ is a small perturbation   of  $C^*$  and hence 
is positive definite.  Thus the expression  is well-defined.

\begin{lem}
\begin{equation} Z^{\#}_{\Theta} = \sum_{\{ Z_{\ell} \}:  Z_{\ell} \cap \Theta  \neq
\emptyset}\prod_{\ell}   K^{\#}(Z_{\ell}) 
 = \exp (\sum_{X \cap  \Theta \neq \emptyset} E^{\#}(X)) 
\end{equation}
where   $K^\#, E^\#$ satisfy the same bounds  (\ref{goodbound}),(\ref{nextgoodbound}) as   $\tilde K, \tilde E$.
\end{lem}
\bigskip

\pr  This is another cluster expansion which follows the steps in Theorem \ref{one}
with some
differences.  
 $C^{loc}$ still has a broken  version  $C^{loc}(s)$ which is also
 a small perturbation   of  $C^*$ and positive definite.   
However  since there is no boundary for  $C^{loc}$ we now vary $s_{\De}$ for all  $\De$ in  $(X \cup P)^c$,  not
just 
$\Theta - (X \cup P)$.  We get polymer activities  $K^{\#}(Z)$  for  any connected  $Z$ 
intersecting  $\Theta$ given by     
\begin{equation} 
 K^\#(Z)  = \sum_{ X,P,Y \to Z } \int ds_Y \frac{\pa }{ \pa s_Y}
[ \int  K_0(X)\ \chi^*( X)\   \zeta^*(P)
d\mu_{\st{C^{loc} (s)}}(A_0) \  d\mu_{\st{\Ga^{loc}(s,\tilde A)}}(\Psi_0)] 
\end{equation}
where  $X \subset  \Theta$,  $P\subset  \Theta-X$,  $Y \subset (X \cup P)^c$, and 
$Z = X \cup P \cup Y$.  This is bounded as before and the exponential version follows
as before.
This completes the proof.
\bigskip

Now we compare   $\cZ^{\dagger}_{\Theta}  ,   \cZ^\#_{\Theta}$.
  Recall that  $\Theta$ is a union of $M_1$-blocks.  Take the $R_0$ blocks 
contained in  $\Theta$,  delete an $R_0$  corridor, and call the result 
$\Theta'$.  Deleting another corridor gives  $\Theta''$.   

\begin{lem}  \ \ \ {}
 \begin{equation} 
  \cZ^{\dagger}_{\Theta}    =   \cZ^\#_{\Theta}  \exp  \left( \sum_{X \subset ( \Theta'')^c,
X \cap \Theta \neq  \emptyset} B^\#(X)  +
\sum_{X  \cap  \Theta'' \neq  \emptyset}  R^\#(X)\right)
\end{equation}
where
\begin{equation}
\begin{split}
 \|B^\#(X)\|_{h_1}   \leq &  \cO(e_0 p(e_0)) e^{-\ka|X|_1}  \\   
 \|R^\#(X)\|_{h_1}   \leq &  \cO(e^{- r(e_0)/M_1}) e^{-\frac12 \ka|X|_1}  \\ 
  \end{split}
\end{equation}
\end{lem}
\bigskip

\pr    
We have   
 \begin{equation} 
  \cZ^{\dagger}_{\Theta}    =   \cZ^\#_{\Theta}  \exp  \left( \sum_{X \cap   \Theta  \neq  \emptyset}
   ( E^{\dagger}(X)-E^\#(X) )             \right)
\end{equation}
with the convention that  $E^{\dagger}(X) =0$ unless $X \subset \Theta$.
Define   $B^\#(X)$ and  $R^\#(X)$  to be the expression $
E^{\dagger}(X)-E^\#(X)$ when respectively   $X \subset  (\Theta'')^c$  
and when   $X$ intersects  $\Theta''$.  The bound on  $B^\#$ follows from
the  separate bounds on  $E^{\dagger}(X),E^\#(X)$.  
To bound  $R^\#$ first suppose that  $X$ also intersects  $(\Theta')^c$.
Then  $M_1|X|_1 \geq r(e_0)$ and  we use    $ e^{-\frac12  \ka|X|_1} \leq  e^{- r(e_0)/M_1} $
in each term  to obtain the bound.  It remains to consider  $R^\#(X)$ when   $X \subset
\Theta'$ and now we really look at the difference.

 We interpolate between $C(1,\Theta)$ and  $C^{loc}$.
with   
\begin{equation}
C(u)  = u\ C(1,\Theta) +(1-u)\  C^{loc} 
\end{equation}
which is  also positive definite.  Together with a similar $\Ga(u, \tilde A)$
we define $Z_{\Theta}(u)$. There are   also  broken versions  
$C(u,s)  = u\ C(s,\Theta) +(1-u)\  C^{loc}(s) $ and   $\Ga(u,s, \tilde A)$
which we use to expand    $Z_{\Theta}(u)$ in terms of  
\begin{equation} 
\begin{split}
& K(u,Z)  = \sum_{ X,P,Y \to Z } \int ds_Y \frac{\pa }{ \pa s_Y}
[ \int  K_0(X)\ \chi^*( X)\   \zeta^*(P)
d\mu_{\st{C (u,s)}}(A_0) \  d\mu_{\st{\Ga(u,s,\tilde A)}}(\Psi_0)] \\
\end{split}
\end{equation}
 and associated $E(u,X)$.  
We follow the setup of the previous lemma so  $Z$ is only required to intersect
$\Theta$.   However  $K(1,Z)$ is only non-zero 
if  $Z \subset \Theta$.

The derivative with respect to $u$ satisfies
\begin{equation}   \label{derivative}
\begin{split}
&  \frac{d}{du} K(u,Z)  = \sum_{ X,P,Y \to Z } \int ds_Y \frac{\pa }{ \pa s_Y} \\
& \{ \int  d\mu_{\st{C (u,s)}}(A_0)   \ \chi^*( X)\   \zeta^*(P)
 \frac{\pa }{ \pa u}[ \int  K_0(X)\ 
d\mu_{\st{\Ga(u,s,\tilde A)}}(\Psi_0) ]\ \\
&+\frac{\pa }{ \pa u'} \int  d\mu_{\st{C (u',s)}}(A_0)   \ \chi^*( X)\   \zeta^*(P)
 [ \int  K_0(X)\ 
d\mu_{\st{\Ga(u,s,\tilde A)}}(\Psi_0) ]_{u'=u}\}\ \\
\end{split}
\end{equation}

To estimate this expression  for  $Z \subset \Theta'$ we look at   $ \pa   C(u,s) / \pa
u=   C(s,\Theta) -C^{loc}(s)$.  We use that for  $x,y \in \Theta'$
  \begin{equation}
|C(s,\Theta,x,y)
-C^{loc}(s,x,y)|  \leq   \cO(e^{-\beta r(e_0) }) e^{-\beta d(x,y)}
\end{equation} 
To see this compare each term to  $C(s,x,y)$ and notice that only 
paths of length greater than $r(e_0)$ contribute.  There is a similar 
bound for  $\Ga(s,\Theta,\tilde A,x,y)
-\Ga^{loc}(s,\tilde A,x,y)$

  Because of the $\Gamma$ bound
the expression $\int  K_0(X) d\mu_{\st{\Ga(u,s)}}$ 
in  the first term in  (\ref{derivative}) is analytic
in $|u | \leq   e^{  \one \beta r(e_0)}$.  By a Cauchy bound the 
derivative in  $u$ gives a factor  $ e^{ -\one \beta r(e_0)}$
The rest of the estimate proceeds as in theorem \ref{one} and we have 
the bound  $ e^{ -\one \beta r(e_0)} e^{-\ka|Z|_1}$.  For the second 
term in  $(\ref{derivative})$ the derivative with respect to $u'$
is evaluated using the identity (\ref{Lidentity}).   This introduces
$ \pa   C(u,s) / \pa u$   gives us the factor   $e^{-\beta r(e_0)}$,
even when derivatives with respect to $s$ are piled onto it.
Again the estimate proceeds as in theorem \ref{one} with the same result.
Altogether then we have  for  $Z \subset \Theta'$
\begin{equation}  \label{goodbound2}
  \|  \frac{d}{du}    K(u,Z) \|_{h_1} \leq   \cO( e^{-\cO(1) \beta r(e_0)}) e^{-\ka|Z|_1}
\end{equation}
This leads to the bound for   $X \subset \Theta'$
\begin{equation}  \label{goodbound3}
  \|  \frac{d}{du}    E(u,X) \|_{h_1} \leq   \ \cO( e^{-\cO(1) \beta r(e_0)}) e^{-\ka|X|_1}
\end{equation}
and the same bound now holds for 
\begin{equation}
 E^{\dagger}(X) - E^\# (X)  =  E(1,X) - E(0,X)  = \int_0^1  
\frac{d}{ d u}   E(u,X)  \ du 
\end{equation} 
This is the bound on $R^\#(X)$ for  $X \subset \Theta'$, it is stronger than
we need, and the  proof is complete.  
\bigskip

Now we work on  $Z^\#_{\Theta}$.  We recall that in $\Theta \subset  \La'$ we have
$E_0'(X) = - V^*_0(X) + R_0(X)$   where $R_0(X)$ still has some
$\Om, \La$-dependence  but is tiny.  We drop this term and define 
\begin{equation} \label{zz}
Z^*_{\Theta}
 =
\int \exp \left(  \sum_{X \subset \Theta} - V_0^*(X) \right)
 \chi^* (\Theta, A_0 )\ d\mu_{\st{C^{loc}}} (A_0) \
  d\mu_{\st{\Ga^{loc}(\tilde A)}}(\Psi_0)
\end{equation}
As before follow theorem \ref{one} and obtain 
\begin{equation}   \label{three}
Z^*_{\Theta} = \sum_{\{ Z_{\ell} \}:  Z_{\ell} \cap  \Theta   \neq  \emptyset}\prod_{\ell}  
K^*(Z_{\ell}) 
 = \exp (\sum_{X\cap \Theta  \neq \emptyset}
 E^*(X)) 
\end{equation}
again with  $\|E^*(x) \|_{h_1}  \leq    \cO( e_0 p(e_0)) e^{-\ka|X|_1}$.

\begin{lem}  \ \ \ {}
 \begin{equation} 
  \cZ^\#_{\Theta}    =   \cZ^*_{\Theta}  \exp  \left( \sum_{X\cap \Theta \neq \emptyset}  R^*(X)
\right)
\end{equation}
where
\begin{equation}
 \|R^*(X)\|_{h_1}   \leq   \cO( e^{-\cO(1) \beta r(e_0)}) e^{-\ka|X|_1}     
\end{equation}
\end{lem}
\bigskip

\pr The proof is similar to the previous lemma, but easier.
 The formula holds with  $R^*(X)
=  E^\#(X) - E^*(X)$ so it suffices to establish   the bound for this object.    
To interpolate  between  $V_0^*(X) $ and $E_0'(X)$  we introduce  $E_0(u,X)  =  -V_0^*(X)  + u
R_0(X)$. This defines $K_0(u,X)$  by  (\ref{kzero}) and then  
\begin{equation} 
\begin{split}
& K(u,Z)  = \sum_{ X,P,Y \to Z } \int ds_Y \frac{\pa }{ \pa s_Y}
[ \int  K_0(u,X)\ \chi^*( X)\   \zeta^*(P)
d\mu_{\st{C^{loc}(s)}}(A_0) \  d\mu_{\st{\Ga^{loc}(s,\tilde A)}}(\Psi_0)] \\
\end{split}
\end{equation}
and from this   $E(u,X)$.   Since  $\| R_0(X) \|_{h'}   $   satisfies the bound  (\ref{primebd})  we
have that  $E_0(u,X)$ is analytic in   $|u| \leq   \cO( e^{\cO(1) \beta r(e_0)})$ and the
same holds  for   $K_0(u,X), K(u,Z), E(u,X)$.  Now from the bound $\|E(u,X)\|_{h_1}   \leq  
\cO(e_0 p(e_0)) e^{-\ka|X|_1} $ and a Cauchy bound we obtain 
\begin{equation}  
  \|  \frac{d}{du}    E(u,X  ) \|_{h_1} \leq   \ \cO( e^{-\cO(1) \beta r(e_0)})
e^{-\ka|X|_1}
\end{equation}
and the result now follows from  
\begin{equation}
 E^\# (X) -  E^{*}(X)  =  E(1,X) - E(0,X)  = \int_0^1  \frac{d}{du}   E(u,X)  \ du 
\end{equation}

\res  
\begin{enumerate}
\item  $E^*(X)  = E^*(X,  \tilde \Psi(\tilde A),  \tilde A)$   only depends
on the indicated variables in   $X$  (and in fact on  $\tilde \Psi(\tilde A)$
only in  $X \cap \Theta$).  We  can   also  consider
the kernel norm  of
$E^*(X)$ as defined in  (\ref{kernelnorm}).  We claim that in the kernel norm we have the  same bound:
\begin{equation}
|E^*(X) |_{h_1}  \leq    \cO( e_0 p(e_0)) e^{-\ka|X|_1}
\end{equation}
This follows by repeating the analysis of theorem \ref{one} 
working with a mixed norm which is kernel in  the external variable  $\tilde \Psi(\tilde A)$ and 
standard in the fluctuation variable $\Psi_0$. This bound will have better iteration
properties.
\item $E^*(X)$ is translation covariant.  If  $X \subset \Theta$ the $\Theta$ dependence 
drops out of the definition   of $E^*(X)$.  Hence if $X, X+a  \subset \Theta$  one can show 
\begin{equation}
 E^*(X+a,  \tilde \Psi(\tilde A)(\cdot -a),  \tilde A(\cdot -a)) = E^*(X,  \tilde
\Psi(\tilde A), 
\tilde A)
\end{equation}
Here we use  the translation covariance of  $V_0^*(X)$ and  $C^{loc}, \Ga^{loc}$.
\end{enumerate}

\newpage

\subsection{perturbation theory}

We study the perturbation theory for  $\cZ^*_{\Theta}$ defined in (\ref{zz}).
 In this expression we
have the potential
 $\sum_{X \subset \Theta}  V_0^*(X) \equiv V^*_{\Theta}  $.  This is essentially the original
potential restricted to  $\Theta$ and is given  explicitly by
\begin{equation}
\begin{split}
V^*_{\Theta}(\tilde \Psi(\tilde A), \Psi_0, \tilde A , A_0)    = &|  M(\tilde A)\tilde
\Psi(\tilde A) - 
 Q_{e_0}(\tilde A +A_0)[\tilde \Psi(\tilde A) + \Psi_0]_{\Theta}     |^2  -  \{A_0= 0\}\\
+ & \left([\tilde\bPsi(\tilde A) + \bPsi_0]_{\Theta},  D_{e_0}(\tilde A +A_0) 
[\tilde \Psi(\tilde A) + \Psi_0]_{\Theta}\right)   -  \{A_0 = 0\}\\
+ &  \sum_{x \in \Theta} :[\tilde \bPsi(\tilde A)+\bPsi_0](x) \de m_0 
[\tilde \Psi(\tilde A) + \Psi_0](x)  :_{\st{\hat S_0}}  +  \de \cE_0|\Theta|\\
\equiv &
V^{**}_{\Theta}(\tilde \Psi(\tilde A), \Psi_0, \tilde A , A_0)   
+  \de V_{\Theta}(\tilde \Psi(\tilde A), \Psi_0, )\\
\end{split}
\end{equation}
where    $M(\tilde A)$ is defined in (\ref{Mdefn}) (without the restriction to  $\La$
which is not needed here). 

For our perturbation expansion we introduce
\begin{equation}
V(t)=V^{**}_{\Theta}(\tilde \Psi(\tilde A), \Psi_0, \tilde A ,t A_0)   
+ t^2 \de V_{\Theta}(\tilde \Psi(\tilde A), \Psi_0, )
\end{equation}
so that   $V(1) = V^*_{\Theta}$ and $V(0)=0$.
We also define for  $0 \leq t \leq  1$
\begin{equation}  \label{gequation}
Z(t)
 =
\int \exp \left( - V(t) \right)
 \chi^* (\Theta, A_0 )\ d\mu_{\st{C^{loc}}} (A_0) \
  d\mu_{\st{\Ga^{loc}(\tilde A)}}(\Psi_0)
\end{equation}
and then  $Z(1)=Z^*_{\Theta}$.
If we restore a local expansion for $V(t)$ we can just as for  $Z^*_{\Theta}$   obtain 
 a local expansion
\begin{equation}  \label{tequation}
  \cZ(t) 
= \exp \left(\sum_{ X \cap   \Theta  \neq  \emptyset }  E (t,X) \right)  
\end{equation}
and   $ E (t,X) $ satisfies the same bounds  as $E^*(X)= E (1,X) $  uniformly in $t$.
\bigskip

\begin{figure} 
\includegraphics{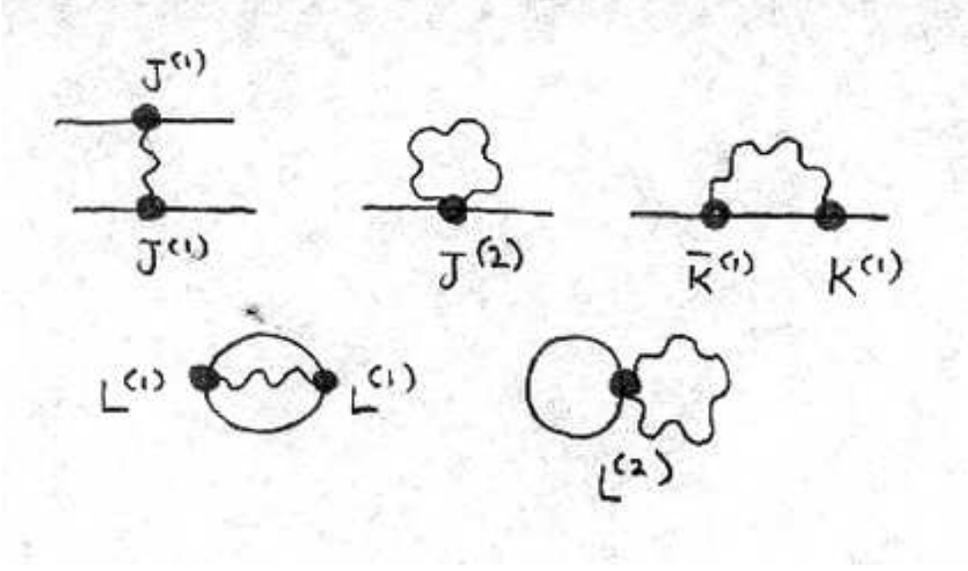}
\caption{second order perturbation theory}   \label{ptheory}
\end{figure}

Our perturbation theory is the expansion   $  \log \cZ(t) $
around  $t=0$ evaluated at $t=1$.  We are especially interested in second order. At
first  instead of   $\cZ(t)  $  let us consider
$\tilde\cZ(t)  $ defined just as  $\cZ(t)$ but with  the characteristic function  omitted.
Computing  some Gaussian integrals  we  find:

\begin{equation}    \label{pexp}
\begin{split}  P_{\Theta}(\tilde \Psi(\tilde A), \tilde A) & \equiv \frac{1}{2}(\log \tilde
\cZ)^{''}(0)\\
  =&  \frac12
\int \left( V'(0)^2 -  V''(0)\right)   \ \  d\mu_{\st{C^{loc}}}(A_0)
\ d\mu_{\st{\Ga^{loc}(\tilde A)}}(\Psi_0) \\
= &\frac12   \sum_{x,y \in \Theta}  J^{(1)}_{\mu}(x)
  C^{loc}( x,y)  J^{(1)}_{\mu}(y)
 -\frac12 \sum_{x \in \Theta}   J^{(2)}_{\mu,\mu} (x,x)
 C^{loc}(x,x) \\
 + &\frac12   \sum_{x,y,z,w \in \Theta}    \bar  K^{(1)}_{\mu}(z,x)
\Ga^{loc} (\tilde A;z,w)   K^{(1)}_{\mu}(w,y)\ C^{loc}(x,y)\\
 - &\frac12   \sum_{x,y,z,w \in \Theta}      K^{(1)}_{\mu}(z,x)
\Ga^{loc} (\tilde A;z,w)  \bar K^{(1)}_{\mu}(w,y)\ C^{loc}(x,y)\\
 - & \sum_{\stackrel{x,y,z,z'}{w,w' \in \Theta}}  tr\left(   L^{(1)}_{ \mu}( z,z',x)
\Ga^{loc}(\tilde A;z',w')  L^{(1)}_{ \mu}(w',w,y) \Ga^{loc} (  \tilde A;w,z)  \right)C^{loc}(x,y)   \\
  -\frac12 & \sum_{x,z,z' \in \Theta} 
 tr\left(   L^{(2)}_{ \mu}(z,z',x,x) \Ga^{loc}( \tilde A;z',z) 
\right)C^{loc}(x,x) \\   - & \sum_{x \in  \Theta}:\tilde  \bPsi (\tilde A ,x)  \de m_0  \tilde  \Psi
(\tilde A ,x):_{\hat S_0- \Ga^{loc}(\tilde A)} -\de \cE_0 |\Theta| \\
\end{split}
\end{equation}
These correspond to the diagrams of  figure  \ref{ptheory}  (except the counterterms).
The vertex functions are 
\begin{equation}
\begin{split}
&J^{(n)}_{\mu_1,\dots,\mu_n}(x_1,\dots,x_n,\tilde  \Psi (\tilde A ), \tilde A) 
=   \frac{\pa^n V^{**}_{\Theta} }{\pa A_{0,\mu_1}(x_1)\dots \pa A_{0,\mu_n}(x_n) } 
(   \tilde  \Psi(\tilde A),0, \tilde A,0)
\\
&\bar  K^{(n)}_{\mu_1,\dots,\mu_n}(z,x_1,\dots,x_n,\tilde  \Psi (\tilde A ), \tilde A) 
=    \frac{\pa^{n+1} V^{**}_{\Theta} }{\pa  \Psi_{0,}(z)\pa A_{0,\mu_1}(x_1)\dots \pa
A_{0,\mu_n}(x_n) }  (   \tilde  \Psi(\tilde A),0,\tilde A,0)
\\
&\bar  L^{(n)}_{\mu_1,\dots,\mu_n}(z,z',x_1,\dots,x_n,\tilde  \Psi (\tilde A ), \tilde A) \\
&  \ \ \ \ \ \ \ \ \ \ \ \ \ \ =  \frac{\pa^{n+2} V^{**}_{\Theta} }{\pa \bar \Psi_{0}(z) \pa   
\Psi_{0}(z')\pa A_{0,\mu_1}(x_1)\dots \pa A_{0,\mu_n}(x_n) }  (   \tilde  \Psi(\tilde A),0, \tilde
A,0)
\\
\end{split}
\end{equation} 
and  $ K$ has  $\pa/\pa \bar \Psi_0$ instead of  $\pa/\pa  \Psi_0$.  These vanish unless
$x_1=\cdots =x_n$ and $\mu_1 = \cdots = \mu_n$.
They are independent of $\Theta$ away from the boundary.

This differs somewhat from the  standard lattice perturbation theory.
There are  are no external photon lines corresponding to the  photon field  $\tilde A$. 
 Instead  $\tilde A$ appears as a background field in the propagators and vertices.   
Another difference is that the vertices are not entirely pointlike due to  the presence of
$Q$ terms in $ V_{\Theta}^{**} $

With this preliminary calculation out of the way we are now ready to state:
\begin{thm}
\begin{equation}
 \cZ^*_{\Theta } =  \exp \left( P_{\Theta} + \sum_{X \cap  \Theta \neq \emptyset }  E^{**}(X) \right) 
\end{equation}
The functions $E^{**}(X)$ are translation covariant inside $\Theta$ and 
 for any $\ep>0$ we have
\begin{equation}
\|E^{**}(X)\|_{h_1}   \leq   \cO(e_0^{4-\ep}) e^{-\ka|X|_1}  
\end{equation}
also in the kernel norm.
\end{thm}
\bigskip

\pr
We expand    $\log \cZ_{\Theta}^*  =  \log \cZ(1)$ by 
\begin{equation}
\log \cZ(1)  = \log  \cZ(0)     
 +   \frac{1}{2!}(\log  \cZ)^{''}(0)  +
+   \frac{1}{3!} \int_0^1   (1-t)^3    (\log \cZ)^{''''}(t) dt 
\end{equation}

 To treat  the remainder term
we use the  expansion $\log \cZ(t)  = \sum_{X} E(t,X)$ from 
(\ref{tequation}).  This gives a contribution 
$\frac{1}{3!} \int_0^1   (1-t)^3    E^{''''}(t,X)dt$ to  $E^{**}(X)$. 
To analyze this term we first
 claim  $E(t,X)$  is actually
analytic in complex
$|t| \leq  e_0^{-1+\ep}$ with a weaker bound.  This follows because $t$ enters  V(t)  either in the
combination  
$e^{ite_0A_0}-1$ for the main terms  or $t^2e_0$  for the counterterms.
In the first case  we have our characteristic 
function enforcing that  
$|A_0| \leq \cO( p(e_0))$  and thus this factor is  $\cO(e_0^{\ep}p(e_0))$.
The counterterms are   $\cO(e_0^{2\ep})$.   Altogether we have
that   $V(t)$   is     $\cO(e_0^{\ep}p(e_0))$.  This carries through
the proof of theorem  \ref{one}  and yields 
 $\| E(t,X)\|_{h_1} 
\leq   
\cO(e_0^{\ep}p(e_0)e^{-\ka|X|_1})$  for $|t| \leq  e_0^{-1+\ep}$
  Now returning to  $0 \leq t \leq 1$ and using a Cauchy bound we have  for the 
fourth derivative
$\| E^{''''}(t,X)\|_{h_1} 
\leq   
\cO(e_0^{4-3\ep}p(e_0)e^{-\ka|X|_1})$.
Since  $p(e_0) \leq e_0^{-\ep}$ and  $\ep$ is arbitrary we can take 
this to be  $\cO(e_0^{4-\ep}e^{-\ka|X|_1})$  as claimed.

Next consider the second order term.
This time we do not use   (\ref{tequation}) but instead use the global 
representation  (\ref{gequation}).   Thus we almost get the terms we 
computed in  (\ref{pexp}), but now we must deal with the characteristic functions.  We compute
\begin{equation}  \label{P}
\frac{1}{2}(\log  \cZ)^{''}(0)
=\frac12
\frac{\int \left( V'(0)^2 -  V''(0)\right)   \chi^*(\Theta,A_0) \  d\mu_{\st{C^{loc}}}(A_0)
\ d\mu_{\st{\Ga^{loc}(\tilde A)}}(\Psi_0) }{\int  \chi^*(\Theta,A_0) \  d\mu_{\st{C^{loc}}}(A_0) }
\end{equation}

The four-fermion part of this is 
\begin{equation}  \label{js}
\frac12   \sum_{x,y \in \Theta}  J^{(1)}_{\mu}(x)
    J^{(1)}_{\mu}(y) C^{loc}_{\chi}(x,y) 
  \end{equation}
where (no sum on $\mu$)
\begin{equation}
C^{loc}_{\chi}(x,y)=
\frac{\int  A_{\mu,0}(x)A_{\mu,0}(y)  \chi^*(\Theta,A_0) \  d\mu_{\st{C^{loc}}}(A_0)}
{\int  \chi^*(\Theta,A_0) \ d\mu_{\st{C^{loc}}}(A_0) }
\end{equation}
We claim that this  can be written as   $C^{loc}(x,y)$ 
plus something tiny with good decay.  This follows by another cluster expansion which 
we outline in Appendix \ref{D}. For us the most useful way to state it is  
\begin{equation}  \label{jt}
 C^{loc}_{\chi}(x,y)=  C^{loc}(x,y)  +   \sum_{X \ni x,y} \de K_{xy}(X)
\end{equation}
where   $|\de K_{xy}(X)|  \leq   \cO(e^{- \cO(1) p(e_0)^2}) e^{-2\ka |X|_1}$.  
Inserting this   in (\ref{js})   the first 
term contributes to $P_{\Theta}$.
The  second term  gives a contribution  to  $E^{**}(X)$  which    
is   
\begin{equation}
\sum_{x,y \in X \cap \Theta}  J^{(1)}_{\mu}(x)
    J^{(1)}_{\mu}(y) \de  K_{xy}(X)
\end{equation}
and this has the bound  $\cO(e_0^{4-\ep}e^{-\ka|X|_1})$  as claimed.

The treatment of the other  terms in   (\ref{P}) is similar in each case splitting into
a contribution to $P_{\Theta}$ and a contribution to  $E^{**}$.
The fermion propagator must also be localized with an expansion (see lemma \ref{coarse})
\begin{equation}
\Ga^{loc}(\tilde A;x,y)   =   \sum_{X \ni x,y}  \Ga_X^{loc}(\tilde A;x,y)  
\end{equation}
for its contribution to  $E^{**}(X)$.

Finally consider   the zeroth order term   $\log Z(0)$.  Again we use   (\ref{tequation}):
$\log \cZ(0)  =  \sum_X  E(0,X)$. 
Now there is no potential. The only contributions  to  $E(0,X)$ are from the characteristic functions.
This means lemma  \ref{kbound} can be modified to obtain a factor  $ \cO(e^{- \cO(1) p(e_0)^2})$
instead of  $\cO(e_0p(e_0))$. Hence  $E(0,X)$ is  $  \cO(e^{- \cO(1) p(e_0)^2}) e^{-\ka |X|_1}$.
  These terms contribute 
to  $E^{**}(X)$. 
\bigskip

\subsection{final adjustments}

Assembling the results of the previous two sections we have  
for the small field region
\begin{equation}
\cZ^{\dagger}  =  \exp  \left( P_{\Theta} + \sum_{X \cap  \Theta \neq \emptyset }  E^{**}(X) 
\sum_{X \cap \Theta \neq  \emptyset,X \subset ( \Theta'')^c} B^\#(X)  +
\sum_{X  \cap  \Theta \neq  \emptyset}  R'(X) \right)
\end{equation}
where     $R'(X)   =   R^\#(X)  + R^*(X)$.

It will be convenient to remove the $R'$ from the small field region.  It 
is possible because $R'$ is so tiny.  We make another Mayer expansion and 
write
\begin{equation}
\exp \left(
\sum_{ X  \cap  \Theta \neq  \emptyset }    R'(X) \right)
=   \sum_{\{X_i\}}  \prod_i (e^{R'(X_i)}-1)=   \sum_{S} 
U(S)
\end{equation}
 where  $\{ X_i\} $ are distinct connected sets and where
\begin{equation}
 U(S)  = \sum_{ \{ X_i \}: \cup X_i=S} \prod_i ( e^{R'(X_i)}
-1)
\end{equation}
The function  $U(S)$ factors over its connected components each one of 
which must intersect  $\Theta$.  From the bound on  $R'(X)$ we have  
for $S$ connected
\begin{equation}
\| U(S)  \|_{h_1} \leq  \cO(e^{-r(e_0)/M_1})  e^{- \cO(1)\kappa|S|} 
\end{equation}
This is sufficient to control the sum over   $S$.  Now we write 
\begin{equation}
 \sum_{S} U(S)  =\sum_{\Upsilon \subset \Theta} \left(  \sum_{S: \Theta - S = \Upsilon}U(S) \right)
\equiv \sum_{\Upsilon \subset \Theta}  \cU(\Upsilon^c)
\end{equation}
\bigskip

Let us return to the full fluctuation integral  (\ref{full}).  This can now 
be written 
\begin{equation}
\tilde \Xi_{\La}
=
\sum_{\Up \subset \Theta \subset \La'} \cG_{\Theta, \Upsilon}\ \  
\exp \left( P_{\Theta} + \sum_{X \cap  \Theta \neq \emptyset }  E^{**}(X)  \right)
 \end{equation}
where we have defined 
\begin{equation}
\cG_{\Theta, \Upsilon} = \cT(\Theta^c)\exp \left( \sum_{X \cap \Theta \neq  \emptyset,X \subset (
\Theta'')^c} B^\#(X) \right)
 \cU(\Upsilon^c) 
\end{equation}
We make  the inverse transformation   $A_0 \to   A_0 - [H_1^{loc}A]_{\Om'}$ in  $\La^c$ and 
similarly for fermions.  
This does not affect the exponential and changes $\cG_{\Theta, \Upsilon}$
to 
$\cG'_{\Theta, \Upsilon}$ whence 
\begin{equation}
\tilde \Xi'_{\La}
=
\sum_{\Up \subset \Theta \subset \La'} \cG'_{\Theta, \Upsilon}\ \  
\exp \left( P_{\Theta} + \sum_{X \cap  \Theta \neq \emptyset }  E^{**}(X)  \right)
 \end{equation}
Next we scale  replacing  $A$ by  $A_{1,L}$ , etc.       to obtain
\begin{equation}  \label{finalfluc}
 \Xi_{1,\La}
=
\sum_{\Up \subset \Theta \subset \La'} \cG_{1,\Theta, \Upsilon}\ \  
\exp \left( \cP_{1,L^{-1}\Theta} + \sum_{Y \cap L^{-1}\Theta \neq \emptyset }  E_{1}(Y)  \right)
 \end{equation}
where  $ \cG_{1,\Theta, \Upsilon}$  is the scaled version of  $ \cG'_{\Theta, \Upsilon}$. 
We have  also  defined
\begin{equation}  \label{growth}
\begin{split}
\cP_{1, L^{-1}\Theta}(\psi_1(\cA_1),  \cA_1 )  =&  P_{\Theta}( [\psi_1(\cA_1)]_L,  \cA_{1,L}  )  \\
 E_1 (Y,  \psi_1(  \cA_1), \cA_1 )
=   &
\sum_{X:  \bar X^L =  LY, X \cap \Theta \neq \emptyset}   E^{**}(X, [\psi_1(  \cA_1)]_L, \cA_{1,L} )\\
\end{split}
\end{equation}
Here  $ \bar X^L $ is the smallest union of  of  $LM_1$ blocks containing $X$.
Explicitly we have  for $S \subset  \bbT^{-1}_{N+M-1}$   
\begin{equation}    \label{pexprime}
\begin{split}&\cP_{1,S}(\tilde \Psi(\tilde A), \tilde A)\\
= &\frac12 L^{-6}  \sum_{x,y \in S}  \cJ^{(1)}_{\mu}(x)
   G^{loc}_1( x,y)  \cJ^{(1)}_{\mu}(y)
 -\frac12 L^{-4} \sum_{x \in S}   \cJ^{(2)}_{\mu,\mu} (x,x)
  G^{loc}_1(x,x) \\
 + &\frac12  L^{-6} \sum_{x,y,z,w \in S}    \bar  \cK^{(1)}_{\mu}(z,x)
   S^{loc}_{1} (\cA_1;z,w)   \cK^{(1)}_{\mu}(w,y)\  G^{loc}_1(x,y)\\
 - &\frac12  L^{-6} \sum_{x,y,z,w \in S}      \cK^{(1)}_{\mu}(z,x)
   S^{loc}_{1} (\cA_1;z,w)  \bar \cK^{(1)}_{\mu}(w,y)\  G^{loc}_1(x,y)\\
 - & L^{-6}\sum_{x,y,z,z',w,w' \in S}  tr\left(   \cL^{(1)}_{ \mu}( z,z',x)
   S^{loc}_{1}(\cA_1 ;z',w')  \cL^{(1)}_{ \mu}(w',w,y)    S^{loc}_{1} (  \cA_1;w,z)  \right)
G^{loc}_1(x,y)  
\\
- & \frac12 L^{-4}  \sum_{x,y,z,z' \in S}
 tr\left(   \cL^{(2)}_{ \mu}(z,z',x,x)    S^{loc}_{1}( \cA_1;z',z) 
\right) G^{loc}_1(x,x) \\ 
  - & L^{-3} \sum_{x \in  S}:  \bpsi_1 ( \cA_1 ,x) \ \de m_1 \  \psi
(\cA_1 ,x):_{\hat S_1- \cS_1^{loc}(\cA_1)} -\de cE_1' |S| \\
\end{split}
\end{equation}
Here  $\hat S_1=  (D(0) +m_1)^{-1}, m_1=Lm_0, \de m_1 = L\de m_0$, and $ \de \cE'_1 = L^3 \de \cE_0$.
The    vertices $\cJ^{(n)},\cK^{(n)},\cL^{(n)}$   are the scaled versions of 
 $J^{(n)},K^{(n)},L^{(n)}$.  For example
\begin{equation}
\cJ_{\mu}^{(1)}(x,  \psi_1 (\cA_1 ), \cA_1) 
=L^{5/2}J_{\mu}^{(1)}(Lx, [ \psi_1 (\cA_1 )]_L, \cA_{1,L})
\end{equation}
In general  $J^{(n)},K^{(n)},L^{(n)}$ have the scaling factors $L^{2+ \frac12 n},L^{1 + \frac12 n},
L^{\frac12 n}$.  With this choice and  $e_1 = L^{1/2} e_0$ we have that $J^{(n)},K^{(n)},L^{(n)}$
are 
$\cO(e_1^n)$.
    We have also introduced
\begin{equation}
\begin{split}
 G^{loc}_1(x,y)=& L C^{loc}(Lx,Ly) \\
  S^{loc}_{1} (\cA_1;x,y)   =&  L^2\Ga^{loc}_{0} (\cA_{1,L};Lx,Ly) \\
\end{split}
\end{equation}
which are local approximations to the propagators  $ G_1,S_1(\cA_1)$ defined earlier

\newpage

\section{Summary}
Now insert the expression (\ref{finalfluc}) for the fluctuation integral
 into our expansion  (\ref{finalexpression})  and
obtain the density on  $\bbT^0_{N+M-1}$
\begin{equation}  
\begin{split} 
  \rho_{1}(\Psi_1, A_1)  =&c_0e^{- \de E_1}  \sum_{\Om,\tilde  \La, \Theta,  \Upsilon}  
\int d \Psi_0 d A_0 \ 
  \zeta_{1,\Om,\tilde  \La}  \  \chi_{1,\Om,\tilde  \La} \  \cG_{1,\Theta, \Upsilon}\
 \\ 
&
\exp \left(-  \frac{a}{2L^2}  |A_{1,L}-QA_0|^2 - \frac12  (A_0,   (-\De  +\mu_0^2 ) A_0  ) \right) \\
&
\exp \left(- \frac{a}{L}| \Psi_{1,L} -Q_{e_0}( \hat A_1 ) \Psi_0|^2
-
 (\bPsi_0 ,( D_{e_0}( \hat A_1) + m_0  ) \Psi_0 ) 
\right) \
\\
&
\exp \left( \cP_{1,L^{-1}\Theta} + \sum_{X \cap L^{-1}\Theta \neq \emptyset }  E_{1}(X)  \right)
 \chi_1(L^{-1}\Om'',A_{1})
\\
\end{split}
\end{equation}
The sum is over   decreasing sets   $\Om \supset \tilde \La \supset \Theta \supset  \Upsilon$  
(which however are not the only restrictions).  
The factor $ \zeta_{1,\Om,\tilde  \La}  \  \chi_{1,\Om,\tilde  \La} \  \cG_{1,\Theta,
\Upsilon}$ supplies the convergence factors for the sums and is localized in $\Upsilon^c$. 
The function
$\cP_{1} +
\sum_{X  }  E_{1}(X)  $ is  the effective action for the small
field region with second order perturbation theory isolated.

 This completes the treatment of the first step.    In next paper we take
these expressions as a starting point and 
 iterate  the operations we have performed.   
After  $k$ steps we are on the smaller torus   $\bbT_{N+M-k}$ and we have 
an expansion  similar to the above  now with $4k$ decreasing sums over regions.  
In  the new small  field  we have  an effective action   $\cP_k + \sum_X  E_k(X)$.
The function  $\cP_k$ is second order in 
 a running  coupling constant   $e_k =  L^{-(N-k)/2}e$. It is 
given by the same diagrams as   $\cP_1$ but is expressed on the finer lattice   $\bbT^{-k}_{N+M-k}$ instead of 
$\bbT^{-1}_{N+M-1}$.

As $k$  gets large the ultraviolet  singularities begin to appear again.
In $\cP_k$ the effect 
is denied by explicit  renormalization cancellations. One can perhaps see how this 
will develop already in (\ref{pexprime}).
In higher orders the effect 
is reflected in the fact that every time we scale down we potentially 
gain a factor of $L^3$  (see  (\ref{growth})).  
This is handled  by the scaling properties of 
the fermion fields and  the allowed growth of the coupling constant.
In this respect our analysis is more like \cite{BDH95}.
The case of no fermion fields is special.  We take out the constant part with 
the energy counterterm and then locally 
we use gauge
invariance  to regard it as a function of the field strength  $dA$.  This has 
sufficiently good scaling properties to control the growth,  see \cite{DiHu92}.  
Thus the UV problems are handled.
The IR problems are also disappearing as $k \to
N$ since the volume is  shrinking. 
 Thus we will get an expression which is uniformly bounded in $N$.

\newpage

\appendix

\section{random walk expansions}  \label{A}

We quote/sketch some results on the covariance operators.
The original treatment for the Laplacian  is due to Balaban \cite{Bal83b}
and the treatment for the Dirac operator  can be found in Balaban, O'Carroll, and 
Schor \cite{BOS89}, \cite{BOS91}.  The operators on a unit lattice  $\bbT^0_N$ are  
\begin{equation}
\begin{split}
C_{\La}  = & [ \De^\#]_{\La}^{-1}  =   [- \De  + \mu_0^2 +\frac{a}{L^2} Q^TQ]_{\La}^{-1}\\
\Ga_{\La }(A)  =&[ D^{\#}(A) ]_{\La}^{-1}  = [  D_{e_0}( A) + m_0  + \frac{a}{L} Q_{e_0}(-A)^TQ_{e_0}(A)
]_{\La}^{-1}
\\
\end{split}
\end{equation}
Here $\La$ is a union of  $M_0$-cubes  centered on the   $M_0$ lattice  $\bbT^{M_0}_N$.
The operators   $[-\De]_{\La}$ and  $[D(A)]_{\La}$ are defined by restricting 
to bonds in $\La$.  For the Laplacian this means Neumann boundary conditions.

First  consider   $ C_{\cO},\Ga_{\cO }(A)$
 where   $\cO$  is one  of the sets $\cO_j$  defined 
for   $j \in T^{M_0}_N$ by 
\begin{equation}
\cO_j =  \{ x \in \La:  |x-j| \leq   M_0 \}
\end{equation}
Here   $|x-j| = \sup_{\mu}|x_{\mu} - j_{\mu}|$.
These overlap 
and cover  $\La$.
 For interior points  $\cO_j$ is a  $2M_0$  cube.  
 Let   $\|\pa A\|_{\cO}  =\sup_{\cO}|\pa A|$

\begin{lem} Let  $e_0 M_0 L^2 \|\pa A\|_{\cO}   $ be sufficiently small. 
Then 
$\Ga_{\cO}(A)$,   
$C_{\cO}$  exist and 
\begin{equation}
\begin{split}
\| \Ga_{\cO }(A) f \|_{2}  \leq &  \cO(L^2) \|f\|_{2} \\
\| C_{\cO} f \|_{2}  \leq & \cO(L^2) \|f\|_{2} \\
\end{split}
\end{equation}
\end{lem}

\pr  The bounds on   $ C_{\cO} $  follow from the   
 lower bound  $(f,[- \De  +   aL^{-2} Q^TQ]_{\cO} f) \geq  \cO(L^{-2}) \|f\|^2$. 
This in turn  follows by bounding it below by a sum over  $L$-squares with Neumann conditions.
On each $L$-square  $Q^TQ$ projects onto constants and on the complement of the constants
the Laplacian is bounded below by  $\cO(L^{-2})$.  See  \cite{Bal83b}. 
The bound for    $ \Ga_{\cO }(0) $
can be reduced to a bound on  the Laplacian.  
See  \cite{BOS91}, lemma VI.4 for general idea.

lf  $A_0$ is a constant (possibly large) we have on  $\cO $
that   $A_0 =  \pa \lambda_0$.  Then by the gauge covariance
\begin{equation}   \label{gauged}
 \Ga_{\cO}(  A_0)  =  e^{ie_0\la_0} \Ga_{\cO}( 0)  e^{-ie_0\la_0}
 \end{equation}
Hence  we have the result for constant background field.   

Now in general for a field  A   on $\cO$  we can write  
$A   = \bar A  + \de A $ where  $\bar A$ = constant is the average over  $\cO$ and   $|\de A| \leq   M \|\pa
A\|_{\cO}     
$.    Then put 
$V=     D^{\#}_{\cO}(\bar A + \de A) - D^{\#}_{\cO}(\bar A) $ 
This has the contribution  (\ref{pot2}) from    $D_{e_0}(A)$,  and also a contribution 
from  $ Q_{e_0}(-A)^TQ_{e_0}(A)$.  For   $ x \in B(y)$ 
we have the explicit formula
\begin{equation}
\left(Q(-A)^T Q(A) f\right)(x)   =  L^{-3} \sum_{\tilde x \in B(y)}  \exp\left( ie_0 A( \Ga_{xy}  \cup
\Ga_{y \tilde x}) \right) f(\tilde x)  
\end{equation}
Using these expressions we find 
\begin{equation}
\| V f\|_{2} \leq \cO(1) (e_0 M_0 \|\pa A\|_{\cO}) \|f\|_{2}
\end{equation}
 Under our assumptions the inverse exists and is given by      
\begin{equation}
\Ga_{\cO}( \bar A + \de A )  =\Ga_{\cO}( \bar A)   \sum_{n-0}^{\infty}
(- V \ \Ga_{\cO}( \bar A )  )^n 
\end{equation}
with the bound  
\begin{equation}
\| \Ga_{\cO}( \bar A + \de A )  f\|_{2}  \leq \cO(L^2) (1- e_0 M_0 L^2 \|\pa A\|_{\cO} )^{-1} \|f\|_{2}
 \leq \cO(L^2)\|f\|_{2}
\end{equation}
  This completes the proof.
\bigskip

We turn to the case of general  $\La$.
Now inverses are obtained by a random walk expansion.
  A path $\om$ in  $\La$
is a sequence of  lattice point $\om  =(j_0,j_1,\dots, j_n)$  which are neighbors in the sense that
$|j_{\al} - j_{\al+1}|=0$ or $M_0$ for any adjacent pair $(j_{\al},j_{\al+1})$.  The length of the path is 
$\ell(\om) = n$. Each path 
$\om$  also determines a sequence of cubes
$  ( \cO_{j_0}, \cO_{j_1}, \cdots, \cO_{j_n})$ as defined above.  
 A path connects points  $x,y$  in the unit lattice if  $ x \in  \cO_{j_0}$
 and  $y \in \cO_{j_n}$ in which case we write  $\om: x \to y$.

\begin{lem}   \label{randomlem}
 Let $L^2/M_0$  and     $  e_0 M_0 L^2 \|\pa A\|_{\cO}   $ be sufficiently small.
\begin{enumerate}
\item
   $\Ga_{\La }(A)$
 and    $C_{\La}$ exist. 
\item We have the random walk expansion
\begin{equation}  \label{random}
\begin{split}
\Ga_{\La }(A)  =&  \sum_{\om}  \Ga_{\La,\om}(A)\\
C_{\La } =&  \sum_{\om}  C_{\La,\om}\\
\end{split}
\end{equation}
 The kernels $\Ga_{\La, \om}(A,x,y)$ and  $C_{\La,\om}(x,y)$ vanish unless  $\om: x \to y$ and satisfy
for some constant $\al$
\begin{equation}
\begin{split}
|C_{\La,\om}(x,y)| \leq & \cO(L^2)(\frac{\al L^2}{ M_0})^{\ell(\om)}  \\
| \Ga_{\La, \om}(A,x,y)| \leq & \cO(L^2)(\frac{\al  L^2}{ M_0})^{\ell(\om)}  \\
\end{split}
\end{equation}
\item    $ \Ga_{\La,\om}(A)$  depends on $A$ only  in  $ \cO_{j_0} \cup \dots \cup    \cO_{j_n}$.
\end{enumerate}
\end{lem}
\bigskip

\pr We give the results for the Dirac operator.  First suppose  $\La$ is the whole torus $T^{0}_N $
and 
denote $\Ga_{\La }(A)$ as just  $\Ga(A)$. Choose smooth  $g$ with 
support in 
$\{ x:  |x|
\leq
\frac23\}
$    so that 
$g=1$ on 
  $\{ x:  |x| \leq \frac13\} $
and so that   $\sum_{ i \in \bbZ^3}  g^2(x-i) =1$.  Then for  $j \in T^{M_0}_N$ define
$h_j (x) =  g((x-j)/M_0)$.  Then $h_j$ is supported in  $\{ x:  |x-j| \leq \frac23 M_0\}\subset  \cO_j $ 
and  $h_j =1$ on  $\{ x:  |x-j| \leq \frac13 M_0\} $ and    
 $\sum_j h_{j}^2 =1$.     We
define a parametrix by 
\begin{equation}
\Ga^*(A) = \sum_j h_{j}\Ga_{\cO_j}(A)  h_{j}
\end{equation}
Here things are  arranged   to avoid  potential discontinuities of 
$\Ga_{\cO_j}  (x,y)$  for  $(x,y)$ on the  boundary  of  $ \cO_j$.
Since  $ D^\#(A) \Ga_{\cO_j}(A)=I$  inside  
$\cO_j$   we have  
\begin{equation}  
D^\#(A)\Ga^*(A)  =  I - \sum_j
R_j(A)  \Ga_{\cO_j}(A)  h_{j} \equiv I -R(A) 
\end{equation}
where  
\begin{equation}
 R_j(A) = - \left[D^\#(A), h_{j}\right]
\end{equation}
 The inverse of  $ D^{\#}(A) $  is now    
\begin{equation}   \label{Neumann}
\Ga(A) = \Ga^*(A) (I - R(A))^{-1} 
= \Ga^*(A) \sum_{n=0}^{\infty} R(A)^n  
\end{equation}
with convergence in $\ell^2$ provided   $\|R(A) \| <1$.

To see  this we develop some estimates.
For   $R_j(A)$ we 
note that  $\|[D_{e_0}(A), h_j]\| =  \cO(M_0^{-1})$  
since $|\pa h_j|    \leq  \cO(M_0^{-1})$.
Furthermore we compute  for  $x \in B(y)$  
\begin{equation}
\begin{split}
&\frac{a}{L}\left([Q(-A)^T Q(A),h_j] f\right)(x)  \\
&  L^{-3} \sum_{\tilde x \in B(y)}  \exp\left( ie_0 A(\Ga_{xy} \cup
\Ga_{y \tilde x}) \right) \frac{a}{L} (h_j(x) -h_j(\tilde x)) f(\tilde x)  \\
\end{split}
\end{equation}
Again this is bounded by  $|\pa h_j|$ 
and so 
$(a/L)\|[ Q(-A)^T Q(A), h_j]\| =  \cO(M_0^{-1})$. 
Combining these bounds we have  $\|R_j(A)  \|   \leq \cO( M_0^{-1})$.
Note also  that functions in the range of   $ R_j(A)$  have support
 in  $\cO_j$ since $L$ is much smaller than  $M_0$.

Now we have   since  $ h_i R_j(A)=0$ unless $i,j$ are neighbors
\begin{equation}
\| h_i R f\| \leq 
 \sum_j
\| h_i R_j(A)  \Ga_{\cO_j}(A)  h_{j}  f\|   \leq
\cO(L^2 M_0^{-1})   \sum_{j: |i-j| \leq M_0}\|  h_{j}  f\| 
\end{equation}
After a Schwarz inequality  we can replace the sum on the right by the expression
$( \sum_{j: |i-j| \leq M_0}\|  h_{j}  f\|^2)^{1/2}$.
Then 
\begin{equation}
\| R(A) f\|^2  = \sum_i \| h_i R(A) f\|^2  \leq  \cO(L^4 M_0^{-2})   \sum_{i,j: |i-j| \leq M_0}\|  h_{j} 
f\|^2  
\leq  \cO(L^4 M_0^{-2})  \|   f\|^2  
\end{equation}
 Hence   $\| R(A) \| \leq   \cO(L^2 M_0^{-1}) <1 $ and  (a.) is proved.

Now we can write
$\Ga(A)$ as a sum over random walks by inserting the definitions of  $\Ga^*(A)$ and  $R(A)$ into (\ref{Neumann})
and obtaining
\begin{equation}
\begin{split}
\Ga(A)=& \sum_{n=0}^{\infty}  \sum_{ j_0, j_1,...,j_n}
( h_{j_0}\Ga_{\cO_{j_0}}(A)  h_{j_0} )
( R_{j_1}(A)\Ga_{\cO_{j_1}}(A)  h_{j_1})
\cdots 
( R_{j_n}(A) \Ga_{\cO_{j_n}}(A)   h_{j_n})\\
\equiv & \sum_{\om}   \Ga_{\omega}(A)  \\
\end{split}
\end{equation}
Here in the last step we notice that we get zero unless   $j_{\al}, j_{\al+1}$ 
are neighbors.

By our previous estimates we have for some $\al = \cO(1)$
\begin{equation}
\|\Ga_{\om}(A)\| \leq   \cO(L^2)(\al L^2/ M_0)^{\ell(\om)} 
\end{equation}
Since we are on a unit lattice the same bound holds for  $|\Ga_{\om}(A,x,y)|$.
The restriction to  $\om: x \to y$ is obvious.  Thus (b.) is proved and (c.) follows by inspection.

Now suppose consider general  $\La$.  We replace $h_j$ by  $h^{\La}_j \equiv \chi_{\La}h_j$ 
and repeat the above argument with  $[D^\#(A)]_{\La}$ instead of $D^\#(A)$.
This completes the proof.   
\bigskip

\re If $\om$ has no points on $\pa \La$ then $C_{\La, \om}$ and   $\Ga_{\La, \om}(A)$ are independent 
of  $\La$.

\begin{lem}   \label{decaylem} Under the same assumptions with   $\beta = M_0^{-1}$
we have for   $x,y \in \La$.
\begin{equation}   \label{thedecay}
\begin{split}
|C_{\La} (x,y) |  &\leq  \cO(L^2)  exp(- \beta d(x,y))  \\
|\Ga_{\La} (A;x,y) |  &\leq  \cO(L^2)  exp(- \beta d(x,y))  \\
\end{split}
\end{equation}
Also 
\begin{equation}
|\Ga_{\La} (A; x,y)-\Ga(A;x,y) |  \leq   \cO(L^2) exp\left(- \beta(d(x,y)+ d(x,\La^c)  +d(y,\La^c))\right) 
\end{equation}
and similarly for   $C_{\La}(x,y)$.
\end{lem}

\pr  A crude estimate is
\begin{equation}
\begin{split}
\Ga_{\La}(A;x,y) \leq &   \cO(L^2)
  \sum_{\omega: x \to y}   (\frac{\al  L^2}{ M_0})^{\ell(\om)}\\
 \leq &   \cO(L^2)
  \sum_{n}   (\frac{\al  L^2}{ M_0})^{n}  | \{ \om:x \to y :  \ell(\om) =n    \} | \\
  \leq &   \cO(L^2)
  \sum_{n}   (\frac{27\al  L^2}{ M_0})^{n} 
\leq    \cO(L^2)
   \\
\end{split}
\end{equation}
 But the condition   $\om: x \to y$
means that    the sum is restricted to   $n \geq (d(x,y)/M_0)-2$.
Thus we can estimate   $ (\al  L^2M_0^{-1})^{n/2}$ by   $\cO(1) \exp ( -d(x,y)/M_0)$
and still have  another factor   $ (\al  L^2M_0^{-1})^{n/2}$ to estimate 
the sum as above.

For the second result  proceed as follows.  We write 
\begin{equation}
\Ga_{\La}(A;x,y) - \Ga(A;x,y)  =   \sum_{\om: x \to y}  \Ga_{\La,\om}(A;x,y)  -  \Ga_{\om}(A;x,y)
\end{equation}
But if  $\om$ has no  boundary   points  we have $ \Ga_{\La,\om}(A,x,y)  =  \Ga_{\om}(A,x,y)$
and hence no contribution.  Thus we can restrict to $\om$ with boundary points
and hence get the result for each term separately.
\bigskip

We will also need reblocked   random walk expansions on a scale 
$M_1$ larger than $M_0$.  We assume $\La$ is a union of $M_1$ blocks
centered  on $\bbT^{M_1}_N$ and we have 

\begin{lem}   \label{coarse}
Under the same assumptions 
\begin{equation}  
\begin{split}
C_{ \La}  = & \sum_{X \ni x,y}C_{ \La,X}\\
\Ga_{ \La} (A)  =&  \sum_{X \ni x,y}\Ga_{ \La,X}(A)\\
\end{split}
\end{equation}
where the sum is over connected unions of $M_1$-blocks $X$.
 The operator $\Ga_{ \La,X}(A)$ depends on $A$ only in X and the kernels 
  $C_{ \La} (x,y) ,\Ga_{ \La} (A,x,y) $ have support in  $X \times X$.
Furthermore 
\begin{equation}
\begin{split}
 \label{coarse.est}
|C_{ \La,X}(x,y)|   \leq &  \cO(1)  e^{-  \beta M_1|X|_1 }      e^{- \beta d(x,y)}  \\
|\Ga_{ \La,X}(A,x,y)|   \leq &  \cO(1)  e^{-  \beta M_1|X|_1 }      e^{- \beta d(x,y)}  \\  
\end{split}
\end{equation}
\end{lem}

\pr 
Define
\begin{equation}  \label{Xsum}
\Ga_{ \La,X}(A,x,y)  =  \sum_{\stackrel{\om: x \to y}{\bar \om = X}} \Ga_{ \La,\om}(A,x,y)
\end{equation}
where $\bar \om $ are the  $M_1$ blocks intersecting $\cO_{j_0} \cup \dots \cup  \cO_{j_n}$.
Now  $\bar \om$ cannot be very large without $\om$ itself covering some 
substantial distance and one can show  
\begin{equation}  \label{tricky}
M_0 \ell(\om) \geq 9 M_1(|\bar \om|_1-8) 
\end{equation}
Using this for a lower bound on  $\ell(\om)$ we extract  
the factor $ e^{-  \beta M_1|X|_1 }  $.
  The estimate now  proceeds as before.  This completes the proof.
\bigskip

We also want to consider local operators
$   C^{loc}_{\La} (x,y) $ and $\Ga^{loc}_{ \La} (A,x,y) $
defined by restricting the random walk expansion  (\ref{random}) to paths 
 which stay within  $r(e_0)/2$ 
of both $x$ and $y$.  Then we have

\begin{lem}   \label{coarse.loc}
Under the same assumptions 
\begin{equation} 
\begin{split}
C_{ \La} -  C^{loc}_{\La} = & \sum_{X }\de C_{ \La,X}\\
\Ga_{ \La} (A)-\Ga^{loc}_{ \La} (A)  =&  \sum_{X } \de \Ga_{ \La,X}(A)\\
\end{split}
\end{equation}
where  
\begin{equation}
\begin{split}
 \label{coarse.loc.est}
| \de C_{ \La,X}(x,y)|   \leq &  \cO( e^{-\beta r(e_0)}) e^{-  \beta M_1|X|_1 }      e^{- \beta d(x,y)} 
\\ | \de \Ga_{ \La,X}(A,x,y)|   \leq &  \cO( e^{- \beta r(e_0)}) e^{-  \beta M_1|X|_1 } 
     e^{- \beta d(x,y)}  \\  
\end{split}
\end{equation}
\end{lem}

\pr  $\de \Ga_{ \La,X}(A,x,y)$ has an expression like (\ref{Xsum}) in which 
  only paths with   $\ell(\om) \geq M_0^{-1}r(e_0)/2$ contribute. 
 This enables us  to extract the factor       $ e^{- \beta r(e_0)}$.
\bigskip

Finally we consider  operators of the form   $\pa  C_{\La}(s)/\pa s_Y  $ as defined in
the text in (\ref{rws})  
\begin{lem} \label{lastlem} Let  $\cL(x,y,Y)$ is the  length of the shortest tree through 
$x,y$ and the centers of the blocks in  $Y$.  Then with a universal constant  $\cO(1)$
we have 
\begin{equation}  \label{last}
\begin{split}
|\frac{\pa  C_{\La} }{\pa s_Y}(x,y)|
\leq &  \one  M_0^{-\cO(1)|Y|_1}  e^{- \beta \cL(x,y, Y) } \\ 
\end{split}
\end{equation}
and similarly for fermions. 
\end{lem}
 
\pr We can assume  $Y \neq  \emptyset$. We have    
  $C_{\La}(s,x,y)  =  \sum_{\om:x \to y}  s_{\om}  C_{\La, \om}(x,y)$.  
 After the  $\pa /\pa s_Y$  differentiation the only terms  $\om $ which survive are those with 
with  $\ell(\om) \geq 1$
and  $\bar   \om   \supset     \De_x \cup \De_y \cup  Y$ where  $\De_x$ is the $M_1$ block
containing $x$.  This leads to 
\begin{equation}
|\frac{\pa  C_{\La} }{\pa s_Y}(x,y)|
\leq    \cO(L^2)
  \sum_{\om}{}'  (\frac{\al  L^2}{ M_0})^{\ell(\om)}
\end{equation}
The prime indicates the restricted sum over paths.
Now   $\ell(\om) \geq  |\bar \om|_1/2  \geq  |Y|_1/2-1$  enables us 
to extract a factor   $M_0^{-\cO(1)|Y|_1} $.  (For the first inequality assume 
 $|\bar \om|_1 \geq 2$ and  use  $\ell(\om)
\geq |\bar\om|_1-1\geq  |\bar \om|/2$; check $|\bar \om|_1 =1$ separately).
On the other hand  by  (\ref{tricky}) and the fact that  $\bar \om$ is 
connected  we have  
\begin{equation}
M_0 \ell(\om)  \geq 9 M_1|\bar \om|- \one   \geq 9  \cL(\bar \om) 
- \one   \geq  \cO(1) \cL(\De_x \cup \De_y \cup Y)-\cO(1)  
\end{equation}
This enables us to extract a factor   $\one  e^{- \beta \cL(x,y, Y) } $.

\newpage

\section{fermion norms and integrals}  \label{B}

We consider the  Grassman algebra generated by   $\Psi_{\al} (x), \bPsi_{\al}(x)$
where  $(x,\al)$ are spacetime  and spinor  indices.
Let  $\xi$ stand for   $(0, \al, x)$ or $(1, \al , x)$ and define  $\Psi(\xi)$ by
$\Psi(0, \al, x) = \Psi_{\al}(x)$ and  $\Psi(1, \al, x) = \bPsi_{\al}(x)$.   Then the 
$\Psi(\xi)$ generate the algebra and any element can be uniquely written  
\begin{equation}  \label{element2}
 F(\Psi)= \sum_n  1/r! \sum_{\xi_1,...,\xi_r} f_r(\xi_1,..., \xi_r)
 \Psi(\xi_1) ...\Psi (\xi_r)  \end{equation}
where the coefficients  $ f_r$ are  totally antisymmetric 
functions.   We define a norm  $\| \cdot\|_h$ depending on a parameter $h>0$ by (c.f.  \cite{BEI92})
\begin{equation} \label{norm1}
\| F \| _h = \sum_r \frac {h^r} { r!}\| f_r \|_1 
\end{equation}
 where $\| f_r \|_1$ is the $\ell_1$ norm.

\begin{lem} $\| FG \|_h \leq \| F \|_h  \| G \|_h $ \end{lem}

\pr  $H=FG$ has the coefficients
 \begin{equation} h_r = 
\sum _ {s+t = r}  \frac {r!}{s!t!}\ \textrm{alt} (f_s \otimes g_t ) \end{equation}
and hence
\begin{equation} \| h_r \|_1 \leq \sum_{s+t=r}  \frac {r!}{s!t!}\ \| f_s \|_1 \| g_t \|_1 
 \end{equation}
Now multiplying by $ h^r / r! $ and summing over $r$ gives the result.
\bigskip

Next consider transformations of the form  $F'(\Psi)  =  F(A \Psi)$
where 
$(A\Psi)(\xi) =  \sum_{\xi'}  A(\xi,\xi') \Psi(\xi')$. We  can estimate 
the effect in terms of the norm
\begin{equation}  \label{1norm}
\|A\|^{(1)} = \sup_{\xi}  \sum_{\xi'}  |A(\xi, \xi')|  
\end{equation}

\begin{lem}  \label{cv}
 Let  $F'(\Psi)  =  F(A \Psi)$.  If   $h' \|A\|^{(1)} \leq h$ then  
\begin{equation}
\| F'\|_{h'}   \leq    \|F\|_h
\end{equation}
\end{lem}

\pr   $F'$ has coefficients  
\begin{equation}
f'_r(\xi'_1, \dots, \xi'_r)
= \sum_{\xi_1,\dots,\xi_r}   f_r(\xi_1, \dots ,\xi_n) \prod_{i=1}^r  A(\xi_i,\xi'_i) 
\end{equation}
Hence  
$\|f'_r\|  \leq   (\|A\|^{(1)})^r \|f_r\|$.    Now  multiply by    $(h')^r/r!$ and sum 
over $r$ to get the result. 
\bigskip

Now we distinguish between  $\Psi(x), \bPsi(x)$. 
 We use a  notation in which  the spinor indices are suppressed, i.e. $x_i$ really means the pair
$(x_i, \al_i)$.  The general element  (\ref{element2}) 
  can  then  be uniquely written  be in the form:
\begin{equation} 
\begin{split}
& F(\Psi, \bPsi) \\
 =& \sum_{n,m} \frac{1}{n!m!} 
 \sum_{\stackrel{x_1,...,x_n}{\bx_1,...,\bx_m}}
f_{n,m}(x_1,...,x_n,\bx_1,\dots, \bx_m) 
\Psi(x_1) \dots \Psi(x_n)\bPsi(\bx_1) \dots \bPsi(\bx_m) \\
 \equiv & \sum_{n,m} \frac{1}{n!m!}  \sum_{X_n, \bar  X_m}
f_{n,m}( X_n, \bar X_m)  \Psi(X_n)\bPsi(\bar X_m) \\
\end{split}
\end{equation}
where the  coefficients  $f_{n,m}$  are   antisymmetric  separately in $X_1 =(x_1,...,x_n)$ and  
$\bar X_m =(\bx_1,...,\bx_m)$.  We have in fact
\begin{equation}
  f_{n,m}(x_1,...,x_n,\bx_1,\dots, \bx_m) =f_{n+m}((0,x_1),...,(0,x_n),(1,\bx_1),\dots,(1, \bx_m)) 
\end{equation}
 We usually consider elements in which only terms with $n=m$ contribute.

Now  the norm (\ref{norm1}) can now be written  
\begin{equation}
 \| F \| _h = \sum_{n,m}  \frac{ h^{n+m}}{n!m!}  \| f_{n,m} \|_1 
\end{equation}

A fermion Gaussian "measure" with non-singular covariance $\Ga = D^{-1}$  can be  defined by 
\begin{equation}
 d\mu_{\st{\Ga}}( \Psi ,\bPsi)  = \frac   { e^{-(\bPsi,D\Psi)}  d\Psi d\bPsi} {  \int e^{-(\bPsi,D\Psi) } d\Psi
d\bPsi  }
\end{equation}
Then one can consider integrals of the form   $\int F d \mu_{\Ga}$.  Terms with  $n\neq m$ give 
zero while terms with  $n=m$ are integrated by  
\begin{equation}
\int  \Psi(x_1) \bPsi(\bx_1) \dots  \Psi(x_n) \bPsi(\bx_n)  \  d\mu_{\st{\Ga}}( \Psi ,\bPsi)
=   \det \left\{  \Ga(x_i,\bar x_j) \right\}
\end{equation}
For the proof see for example  \cite{FMRT94} whose conventions we have adopted.
This identity can also  be used to define   $d \mu_{\st{\Ga}}$  when  $\Ga$ is singular.

To estimate these integrals we introduce  
\begin{equation} \label{2norm}
\| \Ga \|^{(2)}  =  \left( \sup_x \sum_{\bar x}  |\Ga(x, \bx)|^2  \right)^{1/2}
\end{equation}

\begin{lem}  \label{intest1} If  $\sqrt{ \| \Ga \|^{(2)} } \leq h$ then 
\begin{equation}
 | \int F(\Psi, \bPsi ) d\mu_{\st{\Ga}}( \Psi ,\bPsi)|  \leq \| F \|_h
\end{equation}
\end{lem}

\pr  Let   $\si_n$ be the sign of the permutation that takes $\Psi(x_1) \dots \Psi(x_n)\bPsi(\bx_1) \dots
\bPsi(\bx_n)$   to   $ \Psi(x_1) \bPsi(\bx_1) \dots  \Psi(x_n) \bPsi(\bx_n)$.  
Then the integral is evaluated as 
\begin{equation}
 \int F d \mu_{\Ga}  = 
 \sum_{n} \frac{\sigma_n}{(n!)^2}   \sum_{\stackrel{x_1,...,x_n}{\bx_1,...,\bx_n}}
f_{n,n}(x_1,...,x_n,\bx_1,\dots, \bx_n)  \det \left\{  \Ga(x_i, \bx_j) \right\}
\end{equation}
By Hadamard's inequality  \cite{FKT00} we have  
\begin{equation}
| \det \left\{  \Ga(x_i, \bx_j) \right\}|  \leq   ( \|\Ga \|^{(2)})^n  \leq h^{2n}
\end{equation}
and so the result:
\begin{equation}
 |\int F d \mu_{\Ga} | \leq \sum_n \frac{h^{2n}}{(n!)^2} \|f_{n,n} \|_1  \leq  \|F\|_h
\end{equation}
\bigskip

Now suppose our Grassman algebra is generated by two sets of basis elements
  $\Psi(y), \bPsi(y)$ and   $\Psi'(x), \bPsi'(x)$. 
The general element has the form
\begin{equation}
\begin{split}
&F(\Psi, \bPsi,\Psi', \bPsi' ) \\
=&\sum_{n,m,\ell,k} \frac{1}{n!m!\ell!k!}  \sum_{ Y_n, \bar Y_m, X_{\ell}, \bar  X_{k}} 
f_{n,m,\ell,k}( Y_n, \bar Y_m, X_{\ell}, \bar  X_{k}) 
\Psi(Y_n)\bPsi(\bar Y_m) \Psi'(X_{\ell})\bPsi'(\bar X_{k})
\\
\end{split}
\end{equation}
where   $f_{n,m,\ell,k}$ is anti-symmetric separately in each of the four 
groups of variables.
The  norm is now
\begin{equation}
\|F\|_h =\sum_{n,m,\ell,k} \frac{h^{n+m+\ell+k}}{n!m!\ell!k!} 
\|f_{n,m,\ell,k}\|_1
\end{equation}
Gaussian integrals with respect to  $\Psi', \bPsi'$ only are defined in the 
obvious way.   They are estimated as follows:

\begin{lem}   \label{intest2}
If   $h', \sqrt{ \|\Ga\|^{(2)}  }  \leq h$
then  
\begin{equation}
\|\int  F(\Psi, \bPsi,\Psi', \bPsi' )  d\mu_{\st{\Ga}}( \Psi' ,\bPsi') \|_{h'}
\leq    \|F\|_h
\end{equation}
\end{lem}

\pr  The integral is evaluated 
as 
\begin{equation}
\begin{split}
&\sum_{n,m,\ell} \frac{\sigma_{\ell}}{n!m!(\ell!)^2}  \sum_{ Y_n, \bar Y_m, X_{\ell}, \bar  X_{\ell}} 
f_{n,m,\ell,\ell}( Y_n, \bar Y_m, X_{\ell}, \bar  X_{\ell})\
\det \{ \Ga( x_i, \bar  x_j)\}\ \Psi(Y_n)\bPsi(\bar Y_m)
\\
\end{split}
\end{equation}
By Hadamard's inequality the determinant is less than  $(\|\Ga\|^{(2)} )^{\ell} $  and this 
expression has $h'$-norm dominated by 
\begin{equation}
\sum_{n,m,\ell} \frac{(h')^{n+m}(\|\Ga\|^{(2)} )^{\ell} }{n!m!(\ell!)^2} \|f_{n,m,\ell,\ell}  \|_1 
\leq \|F\|_h \end{equation}

\section{spacetime split}  \label{C}

Let  $S$ be a finite set (e.g. one of our tori)  and let $\La$ be a 
subset  (e.g. a small field region).  Suppose we  have a Gaussian integral 
on  $\bbR^S$ defined by a positive self-adjoint operator  $T$ on $\bbR^S$. 
We want to carry out the integral over  $\bbR^{\La}$ first.
This is accomplished by:

\begin{lem} 
\begin{equation}  \label{condition}
\begin{split}
&\int_{\bbR^S} H(A_{\La^c} )  F(A)    \exp\left( - \frac12 (A,  T  A)\right) d  A =
\int_{\bbR^S} H(A_{\La^c} )  \tilde F(A_{\La^c})  \exp\left( - \frac12 (A,  T  A)\right) d A 
\\
\end{split}
\end{equation}
where
\begin{equation}
 \tilde F(A_{\La^c})   =  \int_{\bbR^{\La}}   F(A)     d\mu_{\st{C_{\La},\al_{\La}}}(A_{\La})\
\end{equation}
and  $\mu_{\st{C_{\La},\al_{\La}}}$  is the Gaussian measure    with
covariance 
  $C_{\La} =    T_{\La}^{-1} $ and  mean
$\al_{\La} =    - T_{\La }^{-1}T_{\La \La^c} A_{\La^c}$
\end{lem}

\re   $ \tilde F(A_{\La^c})$  is the conditional expectation of   $F(A)$ with 
respect to the variables $A_{\La^c}$. 
\bigskip

\pr  In  $(A,TA)$ we   write $T = T_{\La^c}+T_{\La^c \La}+ T_{\La \La^c}+ T _{\La}$
where $T_{\La^c \La}= \chi_{\La^c} T\chi_{\La} $, etc.  The cross terms are eliminated 
by the transformation
$
A_{\La} \to  A_{\La} + \al_{\La},  A_{\La^c} \to  A_{\La^c}
$  which we also write as   $A \to  A + \al_{\La}$.
Then the left side of (\ref{condition}) becomes 
\begin{equation}
\begin{split}
&
\int    d A_{\La^c}  H(A_{\La^c} )  \ \exp( - \frac12(A, R_{\La_c} A) )\
  \left(\int  F( A + \al_{\La})  
  \exp( -  \frac12 (A, T _{\La}A) )  d A_{\La}  \right)\\
\end{split}
\end{equation}
where   $R_{\La^c}  
 =  T_{\La^c} -  T_{\La^c \La}  T_{\La}^{-1}T_{\La \La^c}$.
If we divide by  $ \int \exp( -  (\Phi, T _{\La}\Phi) )  d \Phi_{\La}$
  the term in parentheses is identified as $
\tilde F(A_{\La^c})$  and we have  
\begin{equation}
\begin{split}
&
\int    d A_{\La^c}  H(A_{\La^c} ) \tilde F(A_{\La^c})  \ \exp( - \frac12(A, R_{\La_c} A) )\
  \left(\int
  \exp( -  \frac12 (A, T _{\La}A) )  d A_{\La}  \right)\\
\end{split}
\end{equation}
Now reverse the first step  to get the right side of  (\ref{condition}) . 
This completes the proof.
\bigskip

The above result is for bosons.  For fermions suppose 
we have Grassman variables  $\bPsi, \Psi$ indexed by $S$
and a  Gaussian measure   determined by an operator $T$ on $\bbR^S$. We
assume that  $T_{\La} = \chi_{\La}T  \chi_{\La}$ is non-singular.

\begin{lem}
\begin{equation}  \label{condition2}
\begin{split}
&\int H(\bPsi_{\La^c} ,\Psi_{\La^c})  F(\bPsi,\Psi)  
  \exp\left( -  (\bPsi,  T  \Psi)\right) d  \bPsi  d  \Psi  \\
=&
\int H(\bPsi_{\La^c} ,\Psi_{\La^c})  \tilde F(\bPsi_{\La^c},\Psi_{\La^c}) 
 \exp\left( -  (\bPsi,  T \Psi)\right) d \bPsi  d  \Psi
\\
\end{split}
\end{equation}
where
\begin{equation}
 \tilde F(\bPsi_{\La^c},\Psi_{\La^c})   =  \int   F(\bPsi,\Psi)    
d\mu_{\st{\Ga_{\La},\beta_{\La},  \bar \beta_{\La}}}(\bPsi_{\La}, \Psi_{\La})
\end{equation}
is the  the Gaussian integral  over $\bPsi_{\La}, \Psi_{\La}$ 
with covariance 
  $\Ga_{\La} =    T_{\La}^{-1} $
and  mean
$\beta_{\La} =    - T_{\La }^{-1}T_{\La \La^c} \Psi_{\La^c}$
and   $  \bar \beta_{\La} =    - (T_{\La }^{-1})^T(T_{\La^c \La})^T \bPsi_{\La^c}$.
\end{lem}

\pr  Follow the proof of the previous lemma.  The transformation is now 
$\Psi_{\La}  \to   \Psi_{\La} + \beta_{\La}$   and  $\bPsi_{\La}  \to   \bPsi_{\La} +  \bar \beta_{\La}$
and the 
 Gaussian integral is identified/defined as
\begin{equation}  \tilde F(\bPsi_{\La^c},\Psi_{\La^c})  =
\frac{\int  F(\bPsi  +\bar  \beta_{\La},\Psi + \beta_{\La})  
  \exp\left( -  (\bPsi,  T_{\La}  \Psi)\right) d  \bPsi_{\La}  d  \Psi_{\La} }{ \int  
  \exp\left( -  (\bPsi,  T_{\La}  \Psi)\right) d  \bPsi_{\La}  d  \Psi_{\La} } 
\end{equation}
The denominator is non-zero by the assumption that  $T_{\La}$ is non-singular.

\section{two point function}  \label{D}

We study the fluctuation two point function with characteristic functions present 
as defined by 
\begin{equation}
C_{\chi}(x,y)=
\frac{\int  A_{0,\mu}(x)A_{0,\mu}(y)  \chi^*(\Theta,A_0) \  d\mu_{\st{C}}(A_0)}
{\int  \chi^*(\Theta,A_0) \ d\mu_{\st{C}}(A_0) }
\end{equation}
where  $\chi^*$ is defined in  (\ref{chidef}) 
The following  results also hold with $C$ replaced by  $C^{loc}$

\begin{lem}  We have  with a universal  $\cO(1)$ in the exponent
\begin{enumerate}
\item $ |C_{\chi}(x,y)|  \leq    \cO(1) \exp( -  \cO(1) M_1^{-1}d(x,y))$
\item  $ |C_{\chi}(x,y) -C(x,y)|  \leq   \cO(e^{-\cO(1)p(e_0)^2}) \exp( -  \cO(1) M_1^{-1}d(x,y))$
\end{enumerate}
\end{lem}
\pr    We use a simpler version of the cluster expansion of 
theorem \ref{one} to which we refer for more  details.

Start with the numerator in  $C_{\chi}$.
Replace the characteristic function by   $\sum_{P  \subset  \Theta} \zeta^*(P)$
and then break up the measure by introducing the $s$-parameters in the complement of 
$\De_x \cup \De_y \cup  P$.  ($\De_x = M_1$-block containing $x$).
We find an expansion 
\begin{equation}  \label{aaa}
\int  A_{0,\mu}(x)A_{0,\mu}(y)  \chi^*(\Theta,A_0) \  d\mu_{\st{C}}(A_0)
=   \sum_{  \{Z_i\}}  \prod_i  K(Z_i) 
\end{equation}
where the sum is over collections of disjoint connected sets  $Z_i$ intersecting $\De_x \cup
\De_y \cup \Theta$. In fact the  points $x,y$ must be contained in a single $Z_i$,  otherwise the
contribution  vanishes since we are integrating an odd function.  Call this distinguished 
polymer  $Z^*$.  We have   
\begin{equation}
K(Z^*)  =  \sum_{P,Y \to Z^*}  
\int ds_Y \frac{\pa }{ \pa s_Y}
[ \int   A_{0,\mu}(x)A_{0,\mu}(y)  \  \zeta^*(P)
\ d\mu_{\st{C(s)}}(A_0) ]
\end{equation}
The sum is over disjoint $(\De_x \cup \De_y) \cup P \cup Y =Z^*$.  Given  $\ka$ 
this satisfies   $|K(Z^*)|  \leq  \cO(1)  e^{- \ka |Z^*|_1}$ provided  $M_0,M_1$
are large enough.
If $Z$ does not contain $x,y$ then $K(Z)$ is given by the same expression with
no $A_0-$fields.  Now  $P$ cannot be empty  and using (\ref{zeta}) we get    $|K(Z)|  \leq 
\cO(e^{-\cO(1)p(e_0)^2})  e^{-\ka |Z|_1}$. We separate off the distinguished polymer 
and write  for  (\ref{aaa})  
\begin{equation}
\sum_{Z^* \ni x,y}  K(Z^*)  \left(
 \sum_{  \{Z_i\}   \textrm{ in }  (Z^{**})^c   }  \prod_i  K(Z_i) \right)
=
\sum_{Z^* \ni x,y}  K(Z^*) \exp \left(
 \sum_{  X  \subset (Z^{**})^c   } E(X) \right)
\end{equation}
where  $Z^{**}$ is $Z^*$ enlarged by a corridor of $M_1$-blocks.
Here we take advantage of the small norm in  $(Z^{**})^c$ to exponentiate.  We have 
$|E(X)|  \leq  \cO(e^{-\cO(1)p(e_0)^2})  e^{- \ka |X|_1}$.

Now do the same thing without the fields and obtain
\begin{equation}
\int  \chi^*(\Theta,A_0) \ d\mu_{\st{C}}(A_0) =   
\exp \left(
 \sum_{  X    } E(X)\right)
\end{equation}

For the ratio  we have   
\begin{equation}  \label{bbb}
C_{\chi}(x,y)
=\sum_{Z^* \ni x,y}  K(Z^*) \exp \left(
 \sum_{  X   \cap  Z^{**}  \neq  \emptyset   } E(X)\right)  
\equiv \sum_{Z^* \ni x,y}  K'(Z^*) 
\end{equation}
Using   $|\sum_X E(X)|  \leq    \cO(e^{-\cO(1)p(e_0)^2})|Z^*|$  and 
the bound on   $K(Z^*)$  we get  $|K'(Z^*)|  \leq  \cO(1) e^{-\frac12\ka|Z^*|_1}$
This controls the sum over  $Z^*$ and also yields the  decay factor.  Thus 
the first result is established.

For the second result we make a cluster expansion for $C(x,y)$. 
Now  only polymers containing  $x,y$ contribute at all and we  have  
\begin{equation}  \label{ccc}
C(x,y)  =\int  A_{\mu,0}(x)A_{\mu,0}(y)   \  d\mu_{\st{C}}(A_0)
=   \sum_{Z^* \ni xy}    K^0(Z^*) 
\end{equation}
where
\begin{equation}
K^0(Z^*)  =  \sum_{Y \to Z^*}  
\int ds_Y \frac{\pa }{ \pa s_Y}
[ \int   A_{\mu,0}(x)A_{\mu,0}(y)  
\ d\mu_{\st{C(s)}}(A_0)  ]
\end{equation}
The difference is expressed as 
\begin{equation}
C_{\chi}(x,y) -C(x,y)  =   \sum_{Z^* \ni x,y}  \left(  K'(Z^*)-  K^0(Z^*) \right)
\equiv  \sum_{Z^* \ni x,y}   \de K(Z^*) 
\end{equation}
We claim that   
\begin{equation}
|\de K(Z^*)|   \leq     \cO(e^{-\cO(1)p(e_0)^2}) e^{- \frac12 \ka |Z^*|_1}
\end{equation}
which will give the result.
If we replace the exponential in  $K'(Z^*)$ by one, the difference is a term with this 
estimate.  Thus it suffices to look at 
\begin{equation}
 K(Z^*) -  K^0(Z^*)   =   \sum_{P,Y \to Z^*,  P \neq \emptyset}  
\int ds_Y \frac{\pa }{ \pa s_Y}
[ \int   A_{\mu,0}(x)A_{\mu,0}(y)  \  \zeta^*(P)
d\mu_{\st{C(s)}}(A_0)  ]
\end{equation}
Here we use  $\zeta^*(\emptyset)=1$.
The new condition  $P \neq \emptyset$ allows us to extract
a factor $ \cO(e^{-\cO(1)p(e_0)^2})$ and the rest of the bound goes as before.
This completes the proof.

\end{document}